\documentclass[twocolumn,trackchanges]{aastex701}
\usepackage[compatibility=false]{caption}
\usepackage{subcaption}
\usepackage{amsmath}
\usepackage{amssymb}
\usepackage{pifont}
\newcommand{\cmark}{\ding{51}}%
\newcommand{\xmark}{\ding{55}}%

\graphicspath{{./}{figures/}}

\begin{document}

\title{Can third- and fourth-order multipoles plus radial variation of iso-density ellipses explain the observed flux ratios in B1422$+$231? YES, and a lesson learned from a TNG100 lensing galaxy sample}

\author[sname=Feng,gname=Ruizhe,orcid=0000-0003-4463-389X]{Ruizhe Feng}
\affiliation{Department of Astronomy, Tsinghua University, Beijing 100084, People's Republic of China}
\affiliation{Dipartimento di Fisica e Astronomia ``Augusto Righi'', Alma Mater Studiorum Università di Bologna, Via Piero Gobetti 93/2, 40129 Bologna, Italy}
\email[show]{frz21@mails.tsinghua.edu.cn}

\author[sname=Xu,gname=Dandan,orcid=0009-0001-7628-0649]{Dandan Xu} 
\affiliation{Department of Astronomy, Tsinghua University, Beijing 100084, People's Republic of China}
\email[show]{dandanxu@tsinghua.edu.cn}

\author[orcid=0000-0001-6116-2095]{Dominique Sluse}
\email{dsluse@ulg.ac.be}
\affiliation{STAR Institute, Quartier Agora - All\'ee du six
Ao\^{u}t, 19c B-4000,  Li\'ege, Belgium}

\author[orcid=0000-0001-6150-4112]{Giulia Despali}
\affiliation{Dipartimento di Fisica e Astronomia ``Augusto Righi'', Alma Mater Studiorum Università di Bologna, Via Piero Gobetti 93/2, 40129 Bologna, Italy}
\affiliation{INAF-Osservatorio di Astrofisica e Scienza dello Spazio di Bologna, Via Piero Gobetti 93/3, 40129 Bologna, Italy}
\affiliation{INFN -- Sezione di Bologna, Viale Berti Pichat 6/2, 40127 Bologna, Italy}
\email{giulia.despali@unibo.it}

\author[orcid=0000-0002-5558-888X, gname=Anowar,sname='J. Shajib']{Anowar J. Shajib}
\altaffiliation{KICP Fellow}
\affiliation{Kavli Institute for Cosmological Physics, University of Chicago, Chicago, IL 60637, USA}
\affiliation{Department of Astronomy and Astrophysics, University of Chicago, Chicago, IL 60637, USA}
\affiliation{Center for Astronomy, Space Science and Astrophysics, Independent University, Bangladesh, Dhaka1229, Bangladesh}
\email{ajshajib@uchicago.edu}

\author[orcid=0000-0002-0901-9328]{Cai-Na Hao}
\affiliation{Tianjin Astrophysics Center, Tianjin Normal University, Tianjin 300387, People's Republic of China}
\email{hcn@bao.ac.cn}

\begin{abstract}

Flux ratio anomalies in multiply-imaged quasar lenses are a long-standing issue. Using a classical system B1422+231 as a case study, we investigate how typical non-clumpy perturbations beyond elliptical shapes -- multipoles $m_3,\,m_4$ and radial variations in $q,\,\phi_q$ -- can account for the observed image positions and flux ratios under different observational precisions. We extract these perturbations from a pre-selected strong-lensing galaxy sample from the TNG100 simulation. Smooth macroscopic models (\texttt{SIE}+$\gamma$, \texttt{EPL}+$\gamma$) are then fitted to the observed image positions alone and to both positions and flux ratios, with and without including the extracted perturbations. With astrometric uncertainty of $\sigma_{\rm p}=10$ mas, both macro-models alone can already successfully fit image positions within $3\sigma_{\rm p}$. At $\sigma_{\rm p}=2$ mas, however, ``astrometric anomalies'' appear if smooth macro-models alone are adopted. In this case, adding the extracted perturbations can explain the anomalous image positions. When both positions and flux ratios are adopted, the \texttt{SIE}+$\gamma$ model family already shows ``flux ratio anomalies'' at photometric uncertainty $\sigma_{\rm f}\leqslant 10\%$ (keeping $\sigma_{\rm p}=10$ mas). When \texttt{EPL}+$\gamma$ is used, the smooth model alone can simultaneously fit both positions and flux ratios with $\sigma_{\rm f}=10\%, \,5\%$, but not with $\sigma_{\rm f}=2\%$, where ``flux ratio anomalies'' appear. Adding all four types of extracted perturbations can rescue the macro-models and explain the observed anomalous flux ratios. We present important lessons learned regarding model flexibility and degeneracy. 

\end{abstract}



\section{Introduction}

In a number of strong lensing systems where a background quasar is quadruply lensed into point-like images, the flux ratios of images (measured at longer wavelengths) cannot be easily reproduced by commonly adopted smooth lens models, such as singular isothermal elliptical (\texttt{SIE}) or elliptical power law (\texttt{EPL}), plus an additional contribution from a spatially constant shear field $\gamma$ (e.g., \citealt{Jackson2000_JVASCLASS, Bradac2002_B1422_FA, Koopmans2003_JVASCLASS, DoblerKeeton2006, McKean2007_B2045, Nierenberg2014_B1422_FA}). In the past decades, many explanations to this so-called ``flux-ratio anomaly problem'' have been proposed. 

Such anomalous observations have been taken as evidence for the presence of clumpy matter perturbations, such as dark matter subhalos (e.g., \citealt{Mao1998, DalalKochanek2002_FACDMSub, Metacalf2002_FRDsub, Nierenberg2020_NLR_WFC3grism, Nierenberg2024_JWST_WarmDustFA}) or interloper halos along the line of sight (e.g., \citealt{Chen2003_FALOS, Despali2018LOS, Gilman2019_WD_withLOS}). Anomalies from such origins can also be identified through violations of certain asymptotic behaviors that hold universally for any smooth matter distribution of the lens. For example, a parity dependence of the induced anomalies is expected such that the brightness of the saddle image from a close triplet often tends to be suppressed (e.g., \citealt{Schechter_Wambsganss_2002_Parity, KochanekDalal2004_PropagationMacroM34, Bradac2004_Sim}). The ``cusp-relation'' (\citealt{SchneiderWeiss1992, Zakharov1995, Keeton2003_Cusp}) and the ``fold-relation'' (\citealt{BlandfordNarayan1986, Keeton2005_fold}) have also been constructed for the flux ratios of close triple- and pair-images, respectively, which asymptotically go to zero in systems where the lens mass distributions are smooth, and the image opening angles are small. Perturbations from clumpy mass substructures, either subhalos intrinsic to the lens (e.g., \citealt{MetcalfMadau2001_SimCDM, Bradac2004_Sim, Metcalf2012_FARcuspSimSys, Bradac2004_Sim, Xu2009, Xu2010, Xu2015}) or halos along the line of sight (\citealt{Xu2012, InoueTakahashi2012LOS, Inoue2016_FAlos}), would typically generate triplet and pair flux ratios of large values and violate these relations.

As a consequence, flux-ratio anomaly observations of quadruply-lensed quasar images have been widely used to measure properties of cold dark matter subhalos (e.g., \citealt{FadelyKeeton2012HE0435, Hsueh2020_SHARP-VII_DMFS, Gilman2020_CDMMC, Nierenberg2026_CDM_min}), constrain nature of dark matter, e.g., \citet{Inoue2015_FRWDM, Kamada2016PRD_FRWDM, Gilman2018_FR_DMNature, Harvey_2019_WDM, Gilman2020_WDM, Keeley2023_mixCDMWDM, Keeley2024_JWST_WarmDustFA, Keeley2025_JWST_Warm} for warm dark matter, \citet{Gilman2021SIDM, Gilman2023SIDM, hou2026fluxratioanomaliescuspquasars} for self-interacting dark matter, \citet{Chan2020_FRFuzzyDM, AlfredLim2023Nat, hou2026fluxratioanomaliescuspquasars} for fuzzy dark matter, and \citet{Wen_2024_PBH_B1422, Dike2023_PBH} for primordial black holes. 

In spite of such efforts, one shall note that flux-ratio anomalies as identified through large fitting residuals between the best-model predicted and the observed flux ratios may also arise due to inconsistency between the commonly adopted macroscopic density models and the true underlying distribution of the lens. For example, a number of studies showed that adding stellar disks to the standard \texttt{SIE}+$\gamma$ or \texttt{EPL}+$\gamma$ model can significantly improve fitting performances (e.g., \citealt{Moller2003DiskAffectRatio, SHARP-II_Hsueh2016_B1555, SHARP-IV_Hsueh0712_2017, Hsueh2018_IllustrisDisk, Gilman2017}). In addition, lower-order Fourier modes beyond quadrupole/ellipticity mode ($m=2$) have long been detected in isophotes of early-type galaxies, e.g., the fourth-order multipoles ($m=4$) can result in detectable disky or boxy morphologies in the light distributions (e.g., \citealt{Lauer1985BoxyDiscy, Jedrzejewski1987Discy, Bender1988, Bender1989, Rest2001, Pasquali2006, Hao2006, Kormendy2009, Krajnovic2013, Mitsuda2017, Shan2026_BarFA}). This effectively alleviates the observed flux-ratio anomaly problem (e.g., \citealt{EvansWitt2003, KochanekDalal2004_PropagationMacroM34, CongdonKeeton2005, Xu2015, Stacey2024, Cohen2024}), and thus effectively weakens the necessity of introducing clumpy mass perturbations to explain some of the anomalous observations. 

In this regard, many studies in the recent years have routinely taken into account the lower-order multipoles ($m=3, m=4$ modes) in model fitting (together with dark matter halo and subhalo populations), including the latest studies utilizing multi-band measurements of warm dust emission of multiply-imaged quasars from the JWST Lensed Quasar Dark Matter Survey (\citealt{Gilman2018_FR_DMNature, Nierenberg2024_JWST_WarmDustFA, Keeley2024_JWST_WarmDustFA, Keeley2025_JWST_Warm, Gilman2025_JWST_FAandArc, Nierenberg2026_CDM_min}), which has significantly increased the number of systems where flux ratios are affected by mass perturbation on sub-galactic scales instead of by micro-lensing or dust extinction. However, it is also worth noting that degeneracies may exist between the perturbation effects from the lower-order multipoles and those from lower-mass halos and subhalos. \citet{ORiordanVegetti2024} and \citet{Cohen2024} showed that the perturbation effects generated by lower-mass dark-matter subhaloes can be explained as due to $m=3,\,4$ multipole moments. Line-of-sight haloes may also act as multipole structures contributing to lensing perturbations (\citealt{Despali2018LOS, Amorisco2022}). Unless properly broken, such a degeneracy would jeopardize the reliability of using flux ratio anomalies to constrain the nature of dark matter. 

Another kind of deviation from the commonly adopted elliptical lens models is the radial variation in ellipticity and the twists of iso-density ellipses. Such behaviors have already been observed in the light distributions of massive early-type galaxies (e.g., \citealt{Michard1985CetrRndOutEllp, Jedrzejewski1987Discy, Vigroux1988TwistEC, FasanoBonoli1989Isophotwist, Bernstein1997Host0957TwistEC, Keeton2000Host0957TwistModel, Hao2006, Pasquali2006, Kormendy2009}, also \citealt{Stacey2024}). To account for such an effect in lens modeling, multiple components (for both baryonic and dark matter individually), each with different profile characteristics, may be adopted to account for ellipticity variations and iso-density twists (e.g., \citealt{BernsteinFischer1999DPL0957, Keeton2000Host0957TwistModel, Fadely2010DC0957}). Non-parametric mass models have also been invented for this purpose (e.g., \citealt{Liesenborgs2009NonParLens, Lefor2013LensModelReview, Lubini2014FreeModLens}), which can readily account for effects from such complicated azimuthal structures. \citet{Schramm1994LensingSheet} developed a method to create lens mass distributions using multiple elliptical slices, each with different ellipticity and orientation angle. Using such multi-slice technique, \citet{VdV2022_LackAF} investigated the impact of ellipticity change and iso-density twists on the accuracy of $H_0$ determination. Their potentially important consequence on causing observed flux ratio anomalies, however, has not been extensively investigated.

In this study, we take B1422+231 (or B1422 for short), one of the classical flux-anomaly systems (e.g., \citealt{Patnaik_1992_B1422, Impey_1996_B1422, Koopmans2003_JVASCLASS}), for a case study. In this system, a background quasar is quadruply imaged by a foreground lens, forming a close image triplet $A$, $B$ and $C$, and image $D$ on the other side of the lens (see Fig.~\ref{fig:img1422}). Image $A$ has been identified to be anomalous, possibly due to local density perturbation coming from of a $\sim 10^8M_{\odot}$ dark matter subhalo (see, e.g., \citealt{Bradac2002_B1422_FA, Chiba2005_B1422_FA, DoblerKeeton2006, Nierenberg2014_B1422_FA}). In this study, we ask the question whether macroscopic density perturbations (instead of clumpy ones such as dark matter halos and subhalos) can also explain the image configuration (positions and flux ratios) in this system.

To generate a realistic distribution of macroscopic (non-clumpy) perturbations, we extract the third- and fourth-order multipole moments as well as the radial variation in ellipticity and position angle of iso-density ellipses from a strong-lensing galaxy sample selected from the TNG100 simulation \citep{tng01,tng02,tng03,tng04,tng05,tngM01,tngM02}. We search for best-fit smooth lens models (both \texttt{SIE}+$\gamma$ and \texttt{EPL}+$\gamma$) that can well reproduce the observation data of B1422+231, in the presence of the macroscopic perturbations extracted from the simulation. In a way, this method can be viewed as a parameter-searching algorithm but with coarse-grained sampling for the highly-complex parameter space of the perturbation components. Through such an  experiment, we manage to address three questions: (1) under what fitting conditions lensing anomalies would appear; (2) whether the extracted macroscopic perturbations on top of the smooth elliptical mass distributions may account for the observed anomalies in this system; (3) how the results depend on the macro-model choice and on observational precisions. We find that typical macroscopic perturbations can indeed explain the observed image positions and flux ratios for B1422. Through the tests, we also see more clearly the difference lying between ``model explaining the data'' and ``model (physically) generating the data'', and we caution the potential danger of specific model assumptions artificially breaking hidden degeneracies and leading to biased inference results.

The paper is organized as follows: In Section \ref{sec:methods}, we demonstrate the main methodology used in this work; in Section \ref{sec:pert}, we show the properties of extracted macroscopic perturbations for our TNG100 galaxy sample; in Section \ref{sec:fit_position} and \ref{sec:fit_fluxratio}, we present results of B1422-analogs when fitting only image positions and positions plus flux ratios, respectively; and in Section \ref{sec:concl}, we summarize our conclusions and make discussions. Throughout the paper, the cosmology we used is a flat $\Lambda$CDM universe based on the \textit{Planck} results \citep{Planck_Collaboration_2016} with a total matter density of $\Omega_{\rm m} = 0.3089$, a baryonic matter density of $\Omega_{\rm b} = 0.0486$, and a Hubble constant of $h = H_0/(100\,{\rm km~s}^{-1}~{\rm Mpc^{-1}}) = 0.6774$, which is the same as that used in IllustrisTNG simulation. This cosmology is implemented as ``\texttt{Planck15}'' in the \texttt{astropy} \citep{astropy:2013,astropy:2018,astropy:2022} package.

\section{Methods} \label{sec:methods}

In this section, we present a brief introduction of observations towards B1422+231 (Section \ref{ssec:B1422intro}) and the main methods for selecting a lensing sample from the TNG100 simulation (Section \ref{ssec:samp_sel}), extracting lower-order perturbations (Section \ref{ssec:extract}), and generating B1422-analogues (Sections \ref{ssec:lens1422model} and \ref{ssec:mock}). The main codes implementing these methods have been made public at this repository\footnote{\url{https://github.com/asterg1122/B1422_pert}} and are deposited to Zenodo \citep{feng_2026_22537345}.

\subsection{The B1422+231 system}
\label{ssec:B1422intro}

\begin{figure}
    \centering
    \includegraphics[width=0.9\linewidth]{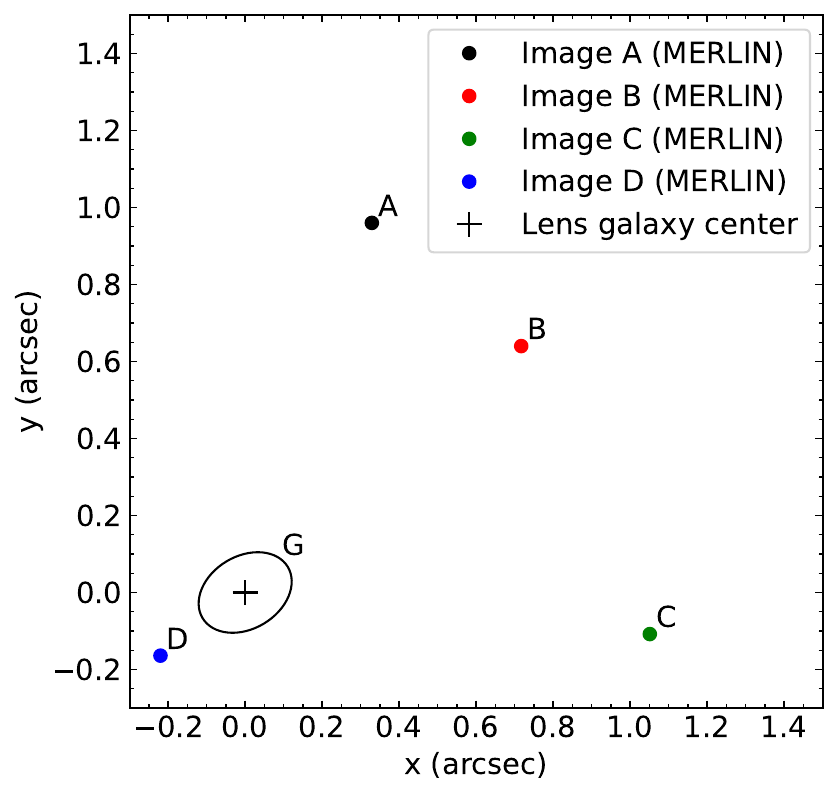}
    \caption{Image configuration of B1422+231 as in Table~\ref{tab:data1422}. Note that the origin point of the coordinate system has been shifted to the observed lens galaxy position. The black ellipse (semi-major axis, ellipticity, and position angle) is the photometry result of the lens galaxy in \citet{Impey_1996_B1422}.}
    \label{fig:img1422}
\end{figure}

B1422+231 is a quadruply-imaged quasar lensing system first found by \citet{Patnaik_1992_B1422, Lawrence1992B1422}, and is commonly known to suffer from flux ratio anomaly (e.g., \citealt{HB1994B1422, KSB1994B1422, Bradac2002_B1422_FA, Chiba2005_B1422_FA, DoblerKeeton2006, Nierenberg2014_B1422_FA}).  Its lens and source redshifts are $z_{\rm l}=0.34$ and $z_{\rm s}=3.62$, respectively. This system has been observed in various bands, including F480LP band with the Hubble Space Telescope \citep{Impey_1996_B1422}, $K$ band with KECK adaptive optics \citep{Nierenberg2014_B1422_FA}, mid-infrared band at 11.7 $\mu$m with the Subaru Telescope \citep{Chiba2005_B1422_FA}, 5-GHz band with MERLIN \citep{Patnaik_1992_B1422}, 8.4-GHz band with VLBA \citep{patnaik1999}, and 233-GHz band with ALMA \citep{Wen_2024_PBH_B1422}. Our astrometric and photometric references come from the MERLIN 5-GHz data (\citealt{Patnaik_1992_B1422, Impey_1996_B1422}), as listed in Table~\ref{tab:data1422} and shown in Fig.~\ref{fig:img1422}. We note that the adopted flux ratios here are consistent with the MERLIN long-time monitoring results (\citealt{Koopmans2003_JVASCLASS}) as well as the narrow-line flux ratios from the Infra-Red Imaging Spectrograph at KECK telescope (\citealt{Nierenberg2014_B1422_FA}).

\begin{deluxetable}{ccccc}
\tablehead{Image & $x$ & $y$ & $S_{\rm 5GHz}$ & Flux ratio \\ &(arcsec)&(arcsec)&(mJy)& }
\tablecaption{Observed image positions, fluxes and flux ratios of B1422+231 based on MERLIN 5-GHz image. Data are taken from Table 1 of \citet{Impey_1996_B1422} and \citet{Patnaik_1992_B1422}. Note that the origin point of the coordinate system has been shifted to the observed lens galaxy position (based on lens position w.r.t. image B, see Table 1 of \citet{Impey_1996_B1422} ).\label{tab:data1422} } 
\startdata
A & $0.329$ & $0.960$ & $216$ & $1$ \\
B & $0.717$ & $0.640$ & $221$ & $1.023$ \\
C & $1.051$ & $-0.108$ & $115$ & $0.532$ \\
D & $-0.220$ & $-0.164$ & $4.5$ & $0.021$ \\
Lens galaxy & $0.0$ & $0.0$ & & \\
\enddata
\end{deluxetable}

\subsection{Lensing galaxy sample selection}
\label{ssec:samp_sel}

We use galaxies from the IllustrisTNG simulation (see \citealt{tng01,tng02,tng03,tng04,tng05,tngM01,tngM02}), a suite of state-of-the-art magneto-hydrodynamic cosmological simulations. Specifically, our strong lensing galaxy sample is chosen from the TNG100-1 simulation, which has a box size of $110.7~\text{cMpc}$, a mass resolution of $m_{\rm DM}=7.5\times10^6M_\odot$ (for dark matter) and $m_{\rm baryon}=1.4\times10^6M_\odot$ (for baryon), and a gravitational soften length of $0.74~{\rm kpc}$.

\begin{figure*}
    \centering
    \includegraphics[height=0.4\linewidth,keepaspectratio]{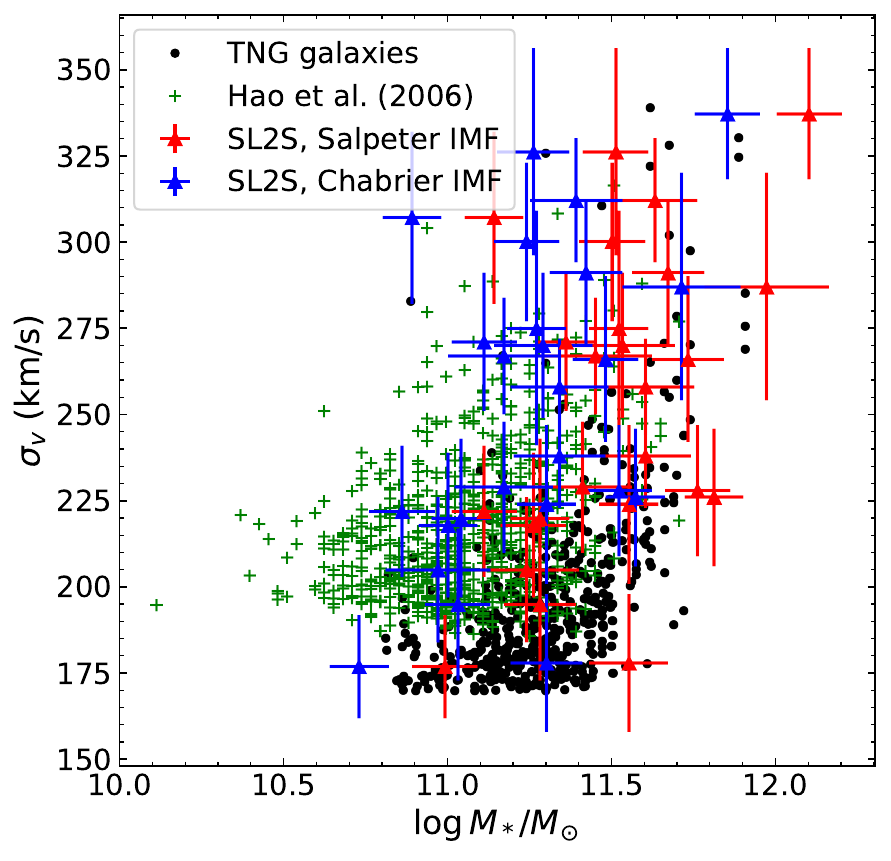}
    \qquad
    \includegraphics[height=0.4\linewidth,keepaspectratio]{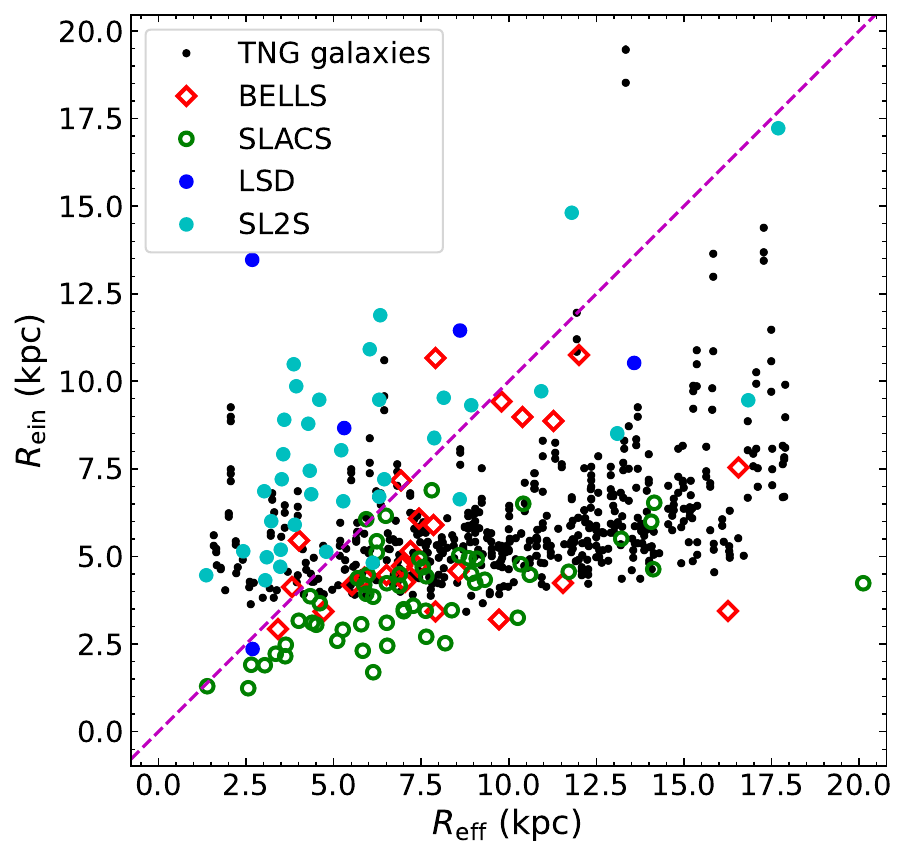}
    
    \caption{Properties of our lensing galaxy sample at $z=0.3$ (snapshot 078) from TNG100-1 simulation. Left: distribution of velocity dispersions and stellar masses of our sample, with results of SL2S \citep{Sonnenfeld2013a,Sonnenfeld2013b} and \citet{Hao2006} overlapped. The stellar masses of SL2S are estimated with two different initial mass functions (IMF). Note that the aperture of velocity dispersion in \citet{Hao2006} is one-eighth of the effective radius, and the data points in Figure~\ref{fig:SrcSelect} are corrected with the relation in \citet{Cappellari06ApertureCorr} to match our $0.5R_{\rm eff}$ aperture. We also note that the stellar masses of \citet{Hao2006} sample are taken from \citet{2005MNRAS.362...41G}, where they are calculated by the BC03 model \citep{bc03}, based on a Chabrier IMF. Right: distribution of Einstein radii and effective radii of our sample. The data points in Fig. 1 of \citet{Sonnenfeld2013a} are also plotted \citep{2010ApJ...724..511A,2012ApJ...744...41B,2004ApJ...611..739T}. The dashed line means $y=x$.}
    \label{fig:SrcSelect}
\end{figure*}

To build our sample, we first get all central galaxies in the TNG100-1 simulation at $z=0.3$ (snapshot 078), the closest to the lens redshift $z_{\rm l}=0.34$ of B1422+231. Since we only focus on projected lens mass distributions including perturbations, we create projections of these galaxies along the $x$-, $y$- and $z$-axis of the simulation box, and also treat them as independent cases to enlarge our sample size. We calculate properties of these projections and select the eligible ones according to their effective radius $R_{\rm eff}$ (in simulation, equivalent to the half-stellar-mass radius that encloses half of the total stellar mass of the galaxy), central velocity dispersion $\sigma_{\rm v}$ (measured within an aperture of $0.5R_{\rm eff}$), and stellar mass $\log M_*/M_\odot$. Motivated by existing strong lensing survey (see Figure~\ref{fig:SrcSelect}), we select galaxies that satisfy velocity dispersion $\sigma_{\rm v}\in[170,350]~{\rm km\,s^{-1}}$, stellar mass $\log M_*/M_\odot\in[10.8,12.2]$ and effective radius $R_{\rm eff}\leq18~\text{kpc}$. These criteria resulted in 536 galaxy projections in total, which, in particular, closely resemble strong lensing galaxies from Strong Lenses in the Legacy Survey (SL2S, \citealt{Sonnenfeld2013a,Sonnenfeld2013b}),  which has typical lens redshifts around that of B1422 and is also representative of strong lens-selected surveys.

In the left panel of Fig.~\ref{fig:SrcSelect} we show the distribution of velocity dispersions and stellar masses of our sample, compared to the results of SL2S and \citet{Hao2006}. As can be seen, our lensing galaxy sample is mostly constrained by central velocity dispersion $\sigma_{\rm v}$ instead of by stellar mass $M_*$. We also calculate the Einstein radius $R_{\rm ein}$ (assuming circular symmetry of the projected mass distribution) for each projection of our lensing galaxies and compare their distribution of $R_{\rm ein}-R_{\rm eff}$  with a number of strong lensing galaxy surveys. The result is given in the right panel of the figure. As can be seen, our galaxy sample is in general similar to typical lensing galaxies in terms of their size and lensing strength. We note that the selection criteria do not explicitly require that the sample be solely composed of early-type galaxies. Among the 536 galaxy projections in our sample, there are 42 ($\sim 8\%$) galaxies that have Sersic index $n_{\rm Ser}<2$, resembling typical light morphologies of late-type galaxies. The inclusion of these galaxies increases the morphological diversity and is also crucial in understanding the full origin of lensing anomalies.

\subsection{Extracting perturbations from TNG100 galaxies}
\label{ssec:extract}

To extract matter density perturbations from the simulation, we do not ray-trace through particle-based density field directly. Firstly, none of the overall mass distribution of the simulated galaxy projections (i.e., the mass perturbations together with their underlying elliptical density distributions) could ever happen to be the same as B1422. Secondly, even with moderately smoothed density distributions, shot noise can still successfully mimic the effect of true density perturbation and thus produce substantial amount of artificial lensing anomalies (see Figure 5 of \citealt{Xu2009}). For these reasons, we only utilize the best analytical models that describe the wanted {\it perturbations} in the particle-based density field (and add them to the smooth lens models that are automatically found by matching observations, see Section \ref{ssec:mock} for details). To do so, for each of the 536 sample galaxy projections, we first obtain the mass surface density map by assigning dark matter, stellar, and gas particles to a regular mesh with a projected smooth particle hydrodynamics (SPH) kernel \citep{1992ARA&A..30..543M,Xu2009}. The adopted smoothing scale is the radius of the sphere centered on each particle and enclosing the $64\pm1$ nearest dark matter particles.
The regular mesh covers a square from $-30~\text{kpc}$ to $30~\text{kpc}$ relative to each galaxy center and has a spatial resolution of $100~\text{pc}$ in either dimension. Similarly, we obtain the $r$-band light map by assigning the $r$-band luminosity of stellar particles to the same mesh with the same smoothing scale. The light map is used in comparison to the mass map in order to reveal differences between the distributions in light and total mass. 

To further extract perturbations from the above maps, we use the \texttt{Ellipse} function in the \texttt{photutils} \citep{photutils2.3.0} package. This is the \texttt{python} version that carries out similar tasks as the \textsc{iraf} program \textsc{ellipse} adopted in \citet{Hao2006}, and can perform isophotal fitting to a given map. 
Specifically, for each galaxy projection, we fit a given map (either mass or light) with a set of elliptical isophotes at different radii with a common center fixed at the center of the mesh, i.e., the galaxy center. Each isophote has its own semi-major axis $a$, axis ratio $q_a$, and position angle $\phi_a$ (with respect to the $x$-axis of the input map), collectively describing the variation in ellipticity and the twist of isophote as a function of radius. Each isophote also has its own Fourier coefficients $\alpha_m$ and $\beta_m$, which are linked to multipole amplitude
$a_m$ and orientation $\phi_m$ (at that radius) via Eq. A.4 of \citet{oh2026}:
\begin{eqnarray}
    a_m = {\rm sign}(\alpha_m)\sqrt{\alpha_m^2 + \beta_m^2}, \\
    \phi_m = \phi_a+\frac{1}{m}\arctan(\beta_m/\alpha_m).
\end{eqnarray}
In particular, $a_4>0$ corresponds to disky configurations, while $a_4<0$ boxy ones. At each radius we use the relative amplitude $a_m/a$ normalized by its semi-major axis $a$ to quantify a local perturbation strength from multipole $m$.  

Clearly, the extracted perturbations are radial dependent. As presented in Section \ref{ssec:lens1422model}, these extracted perturbations shall be implemented to the adopted elliptical mass models (i.e., \texttt{SIE} and \texttt{EPL}, plus external shear) in order to generate realistic mock lensing systems. This means that the extracted ellipticity variation, iso-density rotation, $m_3,\,m_4$ multipoles shall all be added with respect to some given elliptical shape specified by its own $q$ and $\phi_q$. To do so, for each galaxy projection in our sample, we first calculate some radially averaged axis ratio $\overline{q}$ and position angle $\overline{\phi}_{q}$ of the elliptical isophotes for a given map. These averaged quantities are used as references to derive the radial-dependent {\it relative} changes in axis ratio $\Delta q_a \equiv q_a - \overline{q}$ and in position angle $\Delta \phi_a \equiv \phi_a - \overline{\phi}_q$, which are then explicitly added on top of a given elliptical mass profile (described by its own $q$ and $\phi_q$) through a multi-slice technique (\citealt{Schramm1994LensingSheet}, see Section \ref{ssec:lens1422model} for details). 

For the multipole moments, it is important to note that radial variation in the multipole strength is clearly present in observed elliptical-galaxy isophotes, for example NGC6506 (see Figure 4 of \citealt{ORiordan2025_NGC6065}). For the simulated system, the relative multipole strengths $a_m/a$ also exhibit such radial variation extending to large radii. However, we do not explicitly incorporate their radial variations, primarily because no analytical formalism is available to describe them and incorporating such complexity would be computationally expensive: while the multi-slice technique would be the most straightforward route for implementation (\citealt{Schramm1994LensingSheet}), ensuring accurate derivatives of the lensing potential via complex derivative rules would become prohibitively expensive, since analytical solutions are only restricted to the potential and first-order derivatives for elliptical shapes (already implemented in \texttt{ElliSLICE} model in \texttt{lenstronomy} software). For this reason, we extract averaged $m_3$ and $m_4$ moments and add them as globalized multipole perturbations to a given elliptical mass profile.

To calculate all the averaged quantities above, we take a radial range between $0.8R_{\rm ein}$ and $1.2R_{\rm ein}$ of B1422, weight the values by $\frac{\rm intensity}{\rm rms}$ of the isophotes, in which both quantities are directly given by \texttt{photutils}. 
The weighted average axis ratio $\overline{q}$ and position angle $\overline{\phi}_{q}$ are then taken as the shape and orientation of the principal ellipse of that galaxy projection in our sample. The weighted average $m_3$ and $m_4$ are then taken as globalized multipole perturbation and added in the mock lens models. We note that the adopted radial range in fact extends beyond the radius of image $D$ for B1422. While the multi-slice technique is able to capture the exact relative change in ellipticity and rotation angle between the regions around $D$ and the Einstein radius, the multipole moments finally adopted at the position of $D$ are approximated by the averaged results. Consequently, our current approach does not allow us to evaluate the effects of radial variations in multipole strengths, despite their importance (see \citealt{ORiordanVegetti2024, ORiordan2025_NGC6065}). The impact of this additional complexity on the results, while computationally demanding, may be worth testing in the future.  

\begin{figure}
    \centering
    \includegraphics[width=0.9\linewidth]{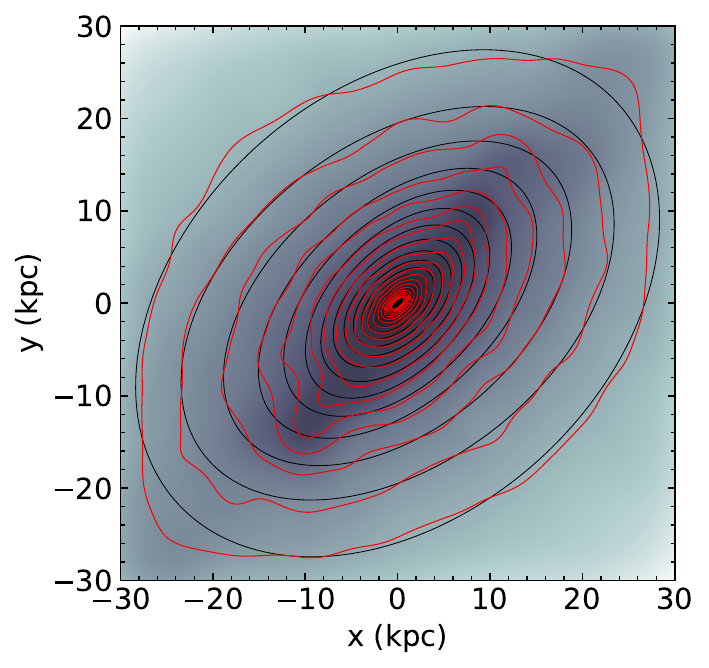}
    \caption{An example of isophote-fitting results of simulation galaxies (subhalo ID: 208612, projected along $z$-axis). Background grey-scale map: $r$-band light distribution; red lines: iso-density contours of total mass density distributions; black lines:  elliptical iso-density contours fitted to the mass map. }
    \label{fig:isovis}
\end{figure}

For the light maps, we perform the same procedures to extract the principal ellipses and the corresponding perturbations in light. These results are not used for generating lensing mocks but for comparisons between mass and light components, testing the mass-follow-light assumption commonly adopted in observations. As an example of visualization, Fig.~\ref{fig:isovis} shows one of our sample galaxies with its light map (grey scale), iso-density contours of total mass distribution (red), and {\it elliptical} iso-density contours fitted to the mass map (black). The latter two are obtained using the package of  \texttt{photutils}.

\subsection{Ingredients of the lens model}
\label{ssec:lens1422model}

In order to understand the effect of the investigated perturbations, we generate B1422-analogs under different combinations of smooth models and perturbations. In this subsection, we introduce the ingredients of the combined lens model that describe the overall mass density distribution. Specifically, we consider seven kinds of the perturbations listed below -- resulting in seven sets of B1422-analogs, which are referred to in the rest of the paper with the following acronyms:
\begin{itemize}
    \item \texttt{m3}: adding $m=3$ multipole alone
    \item \texttt{m4}: add $m=4$ multipole alone 
    \item \texttt{m3+m4}: adding both $m=3,\,4$ simultaneously
    \item \texttt{PAv}: adding radial variation $\Delta \phi_q$ alone
    \item \texttt{Qv}: adding radial variation $\Delta q$ alone
    \item \texttt{PAv+Qv}: adding both $\Delta \phi_q$ and $\Delta q$ simultaneously
    \item \texttt{m3+m4+PAv+Qv}: adding all four types of perturbations above, i.e., lower-order multipoles and radial variation of iso-density ellipses.
\end{itemize}

While searching for a best mass model to explain the observed data,  each time one set of the perturbations above is added to a given smooth underlying macro-model, for which we adopt two commonly used elliptical lens models, i.e., \texttt{SIE}+$\gamma$ and \texttt{EPL}+$\gamma$. We use the package \texttt{lenstronomy} \citep{lenstronomy01,lenstronomy02} and take parameter definitions there, i.e., for an \texttt{SIE} profile:
\begin{equation}
    \kappa(x,y)=\frac{1}{2} \left(\frac{\theta_{E}}{\sqrt{q x^2 + y^2/q}} \right),
\end{equation}
and for an \texttt{EPL} profile:
\begin{equation}
    \kappa(x, y) = \frac{3-s}{2} \left(\frac{\theta_{E}}{\sqrt{q x^2 + y^2/q}} \right)^{s-1},
\end{equation}
where $\theta_E$, $q$, and $s$ are the Einstein radius, axis ratio, and radial slope, respectively. Note that \texttt{SIE} is just a special case of \texttt{EPL} with $s=2$. A spatially constant external shear $\gamma$ is also added (\citealt{Keeton1997_ExternalShear, Wong2011_Shear}). The definitions for all model parameters are shown in Table~\ref{tab:defpri}.

For multipole-only perturbations, we do not explicitly account for their radial variations, but merely add globally averaged $m_3$ and $m_4$ moments to a given elliptical mass profile (see Section \ref{ssec:extract}). For this, we use the ready-made lens model \texttt{EPL\_MULTIPOLE\_M3M4} in \texttt{lenstronomy}. 

However, it is non-trivial to add radial variation of ellipticity and/or position angle onto a smooth model. To do so, we follow the method developed by \citet{Schramm1994LensingSheet} and adopted in \citet{VdV2022_LackAF}, which utilizes the \texttt{ElliSLICE} model in \texttt{lenstronomy}. A continuous \texttt{SIE}/\texttt{EPL} profile is discretized into 1000 elliptical-shaped slices, each described by the same $q$ and $\phi_q$ but with a gradually increasing semi-major axis ranging from 0.001\arcsec~to 5.5\arcsec ($\sim 30$ kpc) evenly spaced on a logarithmic scale. These settings are chosen to ensure that the slices are sufficiently dense to well capture the smooth macro-model profile when no perturbation is added. For a given perturbation of radial variations in ellipticity and orientation, the extracted relative changes $\Delta q_a \equiv q_a - \overline{q}$ and $\Delta \phi_a \equiv \phi_a - \overline{\phi}_q$ (with respect to the principal ellipse of the simulated galaxy projection) are first interpolated at the radii of the 1000 slices, and then added to $q$ and $\phi$ of the smooth macro-model. In order to calcuate image flux ratios, we take magnification $\mu$ at image positions under the point-source assumption. However, the default implementation of second derivatives for \texttt{ElliSLICE} uses numerical differentiation and may lead to a large error in some cases. Hence, we derive the second derivatives theoretically according to complex derivative rules (see Eq. 20--23 of \citealt{Schramm1994LensingSheet}) as follows: for any given point $(x,\,y)$ inside a slice,
\begin{equation}
    \begin{cases}
        \frac{\partial\alpha_{\rm in}}{\partial x}=(1-\epsilon\mathrm{e}^{2\mathrm{i}\varphi})\Sigma_0 \\
        \frac{\partial\alpha_{\rm in}}{\partial y}=\mathrm{i}(1+\epsilon\mathrm{e}^{2\mathrm{i}\varphi})\Sigma_0
    \end{cases}
\end{equation}
and for a point outside the slice,
\begin{equation}
    \begin{cases}
        \frac{\partial\alpha_{\rm ext}}{\partial x}=\frac{2ab}{f^2}\left(\mathrm{e}^{2\mathrm{i}\varphi}-\frac{\bar{z}\mathrm{e}^{3\mathrm{i}\varphi}}{\text{sign}(\bar{z}\mathrm{e}^{\mathrm{i}\varphi})\sqrt{\bar{z}^2\mathrm{e}^{2\mathrm{i}\varphi}-f^2}}\right)\Sigma_0 \\
        \frac{\partial\alpha_{\rm ext}}{\partial y}=-\mathrm{i}\frac{\partial\alpha_{\rm ext}}{\partial x}
    \end{cases}
\end{equation}
where $z=x+\mathrm{i}y$, and $\alpha=\alpha_x+\mathrm{i}\alpha_y$ is the deflection angle. The definitions of $\epsilon$, $\varphi$, $\Sigma_0$, $a$, $b$, $f$ and sign function follow those in \citet{VdV2022_LackAF}.

We note the reader that the methods of ensembling the lens model above allow us to self-consistently preserve all the relative changes in perturbation amplitudes and orientations. It is also worth noting that although both types of perturbations are extracted from the TNG100 simulation and explicitly added to the smooth macro-model \texttt{SIE}/\texttt{EPL}+$\gamma$, the way that $m_3,\,m_4$ are included eventually manifests a global perturbation effect from these lower-order multipoles, while the multi-slice technique allows us to explicitly incorporate the extracted radial variations in $\Delta q$ and $\Delta \phi_q$.

\subsection{Generating B1422-analogs}
\label{ssec:mock}

In this subsection, we explain how B1422-analogs are finally generated. When adding different perturbation components on top of a given underlying smooth macro-model, we request the final combined lens model shall satisfy two types of constraints: image positions alone and both positions and flux ratios, of the four images $A$, $B$, $C$ and $D$ of B1422. In principle, one can add the extracted perturbations to some existing best-fit macro-models, such as  $\texttt{SIE}$/$\texttt{EPL}$+$\gamma$ for B1422+231 (e.g., \citealt{Chiba2002, Wen_2024_PBH_B1422}). This, however, implicitly assumes that the adopted macro-models truly describe the underlying smooth matter distribution (and shear field) and the investigated perturbations only add small changes to the image configuration. Nevertheless, this may not be the case and the smooth models can be different in the presence of different perturbations. Hence we explicitly search for an appropriate set of underlying parameters of the adopted smooth macro-models in the presence of a given kind of perturbation.

The priors of the parameters are listed in Table \ref{tab:defpri}. Flat priors ($\mathcal{U}$) within commonly adopted ranges are taken for most of the model parameters, while Gaussian priors ($\mathcal{N}$) are applied to the axis ratio $q$ and position angle $\phi_q$ based on measurements from the light distribution of B1422. Indeed, the light and mass distributions can have different ellipticities, and observations seem to also suggest no correlation between the two (e.g., \citealt{Shajib2019UniMod13Quads}). Here in this study, we consider the Gaussian priors on $q$ and $\phi_q$ to be composed of two parts: one is from the light distribution of the lensing galaxy in \citet{Impey_1996_B1422}, where $q_{\rm light}=0.73\pm0.13$ and $\phi_{q,\,{\rm light}}=31^\circ\pm15^\circ$ (see the ellipse in Fig.~\ref{fig:img1422}); the other comes from the simulated misalignment between the mass and light elliptical shapes, which is given by $q_{\rm mass} - q_{\rm light} = 0.08 \pm 0.06$ and $\phi_{q,\,{\rm mass}} -\phi_{q,\,{\rm light}} = 0^{\circ} \pm 3.5^{\circ}$ (see Section~\ref{ssec:pert_ml} and Fig.~\ref{fig:qpa_mass_light}). The final priors combine the two Gaussian distributions, resulting in $q=0.81\pm0.14$ and $\phi_q=31.0^{\circ}\pm 15.4^{\circ}$. In addition, we have further adopted an absolute cut to the prior on $q$ such that $q\in [0.4,1.0]$. The Gaussian priors are crucial in particular when finding $\texttt{EPL}$+$\gamma$ parameters with image positions alone, where the model degree of freedom increases and the observational data start losing constraining power.

\begin{deluxetable}{ccc}
    \tablehead{Parameter & Definition & Prior}
    \tablecaption{Definitions and priors of macro-model parameters\label{tab:defpri}}
\startdata
$\theta_E$ & Einstein radius (\arcsec) & $\mathcal{U} (0.4,\,1.2)$ \\
$q$ & Axis ratio & $\mathcal{N} (0.8,\,0.14^2)$ \\
$\phi_q$ & Position angle ($^\circ$) & $\mathcal{N}(31,\,15.4^2)$ \\
$s$ & Radial density slope (for \texttt{EPL}) & $\mathcal{U} (1.5,\,2.5)$ \\
$\gamma$ & Shear amplitude & $\mathcal{U} (0.0,\,0.3)$ \\
$\phi_\gamma$ & Shear orientation angle ($^\circ$) & $\mathcal{U} (0,\,180)$ \\
$x_s$ & Source position $x$ (\arcsec) & $\mathcal{U} (-1,\,1)$ \\
$y_s$ & Source position $y$ (\arcsec) & $\mathcal{U} (-1,\,1)$ \\
\enddata
\end{deluxetable}

The observed B1422 image configuration is used to find a best smooth macro-model in the presence of the investigated perturbations. For this, we adopt two different sets of astrometric uncertainties at $\sigma_{\rm p}=10~{\rm mas}$ and $2~{\rm mas}$ when only positions are used as constraint. When both positions and flux ratios are adopted, we fix the astrometry uncertainty at $\sigma_{\rm p}=10~{\rm mas}$ and assume three levels of photometric uncertainty, i.e., $\sigma_{\rm f}=10\%$, 5\% and 2\% for the flux ratios of $B/A$ and $C/A$, and one fixed flux ratio uncertainty at $15\%$ for $D/A$. In general, an astrometric uncertainty $\sigma_{\rm p}\sim 10\,{\rm mas}$ is of typical precision for radio quasar images from MERLIN data (e.g., \citealt{Impey_1996_B1422, Koopmans2003_JVASCLASS}), $\sigma_{\rm p}=5-10\,{\rm mas}$ for KECK- and HST-based position measurements of quasars' narrow line regions and warm dust emissions (e.g., \citealt{Nierenberg2014_B1422_FA,  Shajib2019UniMod13Quads, Schmidt2023STRIDES, Nierenberg2024_JWST_WarmDustFA}) and $\sigma_{\rm p} \lesssim 2\,{\rm mas}$ for currently the highest resolution radio quasar images from VLBA and VLBI data (e.g., \citealt{patnaik1999, Spingola2018, Spingola2019VLBA, Spingola2020_B0712_VLBI}). Here the level of uncertainties with $\sigma_{\rm p}=10\,{\rm mas}$ and $\sigma_{\rm f}\lesssim 2\%$ for image $B/A$ and $C/A$, and $10-15$\% for $D/A$ is broadly consistent with existing measurements of this system (see \citealt{Patnaik_1992_B1422, Impey_1996_B1422, Bradac2002_B1422_FA, Nierenberg2014_B1422_FA}).

While searching for the macro-model parameters, we always fix the lens center position at $(0,0)$, i.e., explicitly assuming the center of the lens is the same as as the center of the simulated galaxies from which perturbations are extracted. When adding multipole-only perturbations, we directly solve the lens equation and minimize the $\chi^2$ value (of image positions and/or flux ratios) in the image plane. However, in the case of radial variations in ellipticity and position angle, solving the lens equation and minimizing the $\chi^2$ value in the image plane (for the 1000 elliptical slices and through complex derivative rules, see Section \ref{ssec:lens1422model} for details) is much more computationally expensive. Hence, when adding these perturbations to a given set of macro-model parameters, we follow the practice of \citealt{keeton2001}, ray-shoot the observed image positions back to the source plane and carry out the $\chi^2$ minimization therein. 

As this procedure is nothing more than conducting a standard minimum-$\chi^2$ search, there will always be a model corresponding to the minimum $\chi^2$ found. However, we require that only those (minimum-$\chi^2$ models) that can reproduce the observed image positions within $3\sigma_{\rm p}$ and the observed flux ratios within $3\sigma_{\rm f}$ shall be regarded as a {\it successful} case - a B1422-analog. One can think of this method as a cost-effective shortcut for exploring a large set of lens models with both smooth and perturbative parameters. Instead of explicitly traversing the vast, intricate parameter space of the perturbative terms, the method relies on simulations to sample over that space in a ``coarse-grain'' fashion.

As we shall see in Sections \ref{sec:fit_position} and \ref{sec:fit_fluxratio}, very often there can be more than one successful cases (out of a total of 536 projections, where perturbations features are extracted from) under the same perturbation type (e.g., \texttt{m4} or \texttt{PAv+Qv}). We note the reader, as long as one successful model can be obtained for a given perturbation type, then it is regarded as a solution to this system, which means the observed configuration of B1422 can be explained by the combination of the smooth macro-model and the investigated macroscopic perturbation, without the inclusion of any other components, e.g., dark matter subhalos. On the contrary, if no such a solution is found, it does not necessarily mean the combined mass model family cannot generate the observed data; it may simply reflects the fact that our sample from the simulation is not sufficiently sampling the huge parameter space of the perturbation feature, which is particularly possible for the radial variations in  ellipticity and position angle (can be clearly seen in Section \ref{sec:fit_fluxratio}). As for the questions how likely such a scenario can explain a sample of observed systems, this would require more than just one B1422 to address. We defer the reader to a parallel study that evaluates the probabilities for this TNG100 strong lenses to account for existing flux ratio observations (Feng et al. in prep). In Sections \ref{sec:fit_position} and \ref{sec:fit_fluxratio}, we present the detailed results under different (combinations of) investigated perturbations and see how they vary as the adopted macro-models change from \texttt{SIE}+$\gamma$ to \texttt{EPL}+$\gamma$, and as the assumed observational uncertainties progressively decrease to smaller values.

\section{Macroscopic perturbations}
\label{sec:pert}

In this section, we present the general differences between the mass and light distributions (Section \ref{ssec:pert_ml}) and the perturbation properties of the total mass distributions (Section \ref{ssec:prop_mass}) for our lensing galaxy sample as described in Section~\ref{ssec:extract}.

\subsection{Macroscopic shapes between mass and light}
\label{ssec:pert_ml}

\begin{figure*}
    \centering
    \includegraphics[height=0.37\linewidth,keepaspectratio]{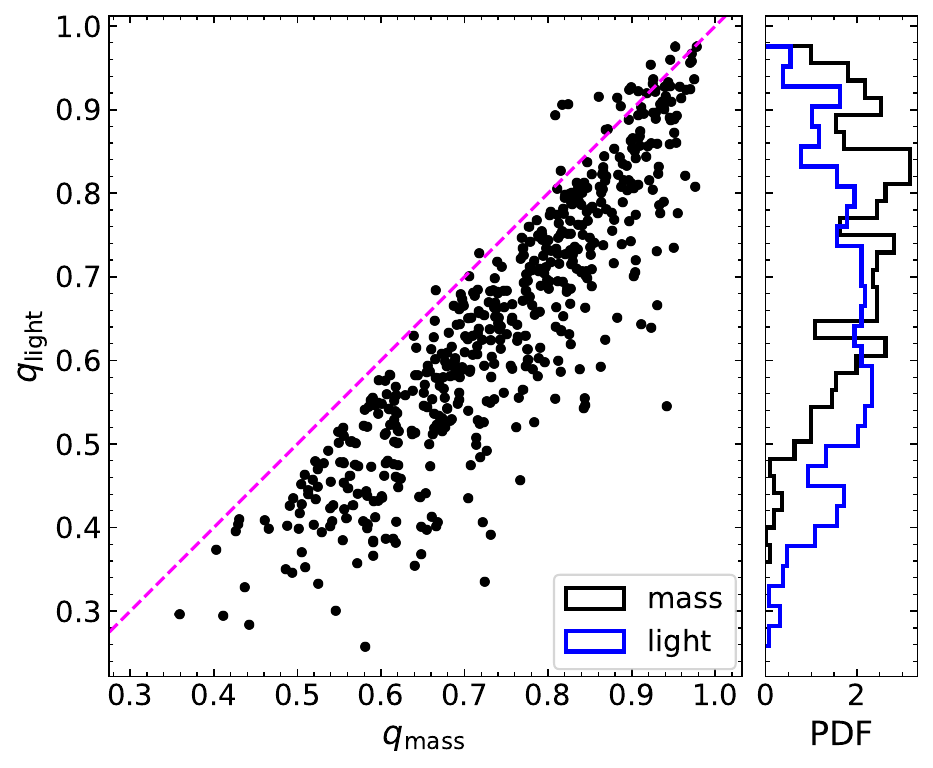}
    \qquad
    \includegraphics[height=0.37\linewidth,keepaspectratio]{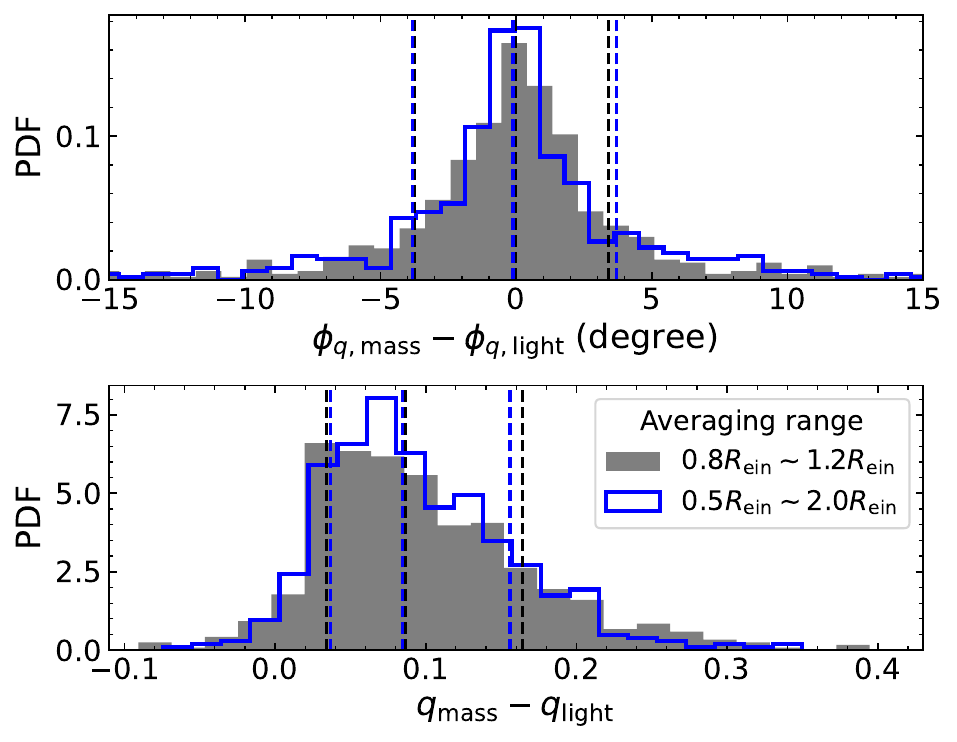}
    \caption{Difference in axis ratio $q$ and position angle $\phi_q$ between mass and light. Left: principal values of axis ratio $q$ for mass and light maps of our galaxy sample, with both histograms added on the top panel. The magenta dashed line means $y=x$. Right: difference of averaged $\phi_q$ and $q$ (within two annuli) between mass and light maps for our sample. Dashed lines indicate 16\%, 50\%, and 84\% quartiles.} \label{fig:qpa_mass_light}
\end{figure*}

We first compare the differences between the general shapes of the mass and light distributions, which is not possible in observations but feasible in simulations where the true mass distributions are known. The left panel of Fig.~\ref{fig:qpa_mass_light} shows the principal axis ratios $q$ extracted from both mass and light maps, averaged between $0.8R_{\rm ein}$ and $1.2R_{\rm ein}$. As can be seen, the two quantities are well correlated with each other, indicating a general validity of the \emph{mass-follow-light} assumption. However, the axis ratios of light maps are overall smaller than those of mass maps. The right two panels of Fig.~\ref{fig:qpa_mass_light} show the difference of principal $\phi_q$ and $q$ between mass and light maps: $q_{\rm mass} - q_{\rm light} = 0.08 \pm 0.06$ and $\phi_{q,\,{\rm mass}} -\phi_{q,\,{\rm light}} = 0^{\circ} \pm 3.5^{\circ}$. The mass axis ratios are systematically larger than those for the light -- the total matter profile is systematically rounder than the light distribution. The position angles of the two are very well consistent, with a small scatter of $\sim3.5^\circ$. These differences are adopted in determining the prior on the lens ellipticity when fitting the mock lenses (see Section~\ref{ssec:mock}). We also note that the comparison results in the shape properties do not change as the annulus varies from $0.8\sim1.2 R_{\rm ein}$ (shaded histogram) to $0.5\sim2.0 R_{\rm ein}$ (blue histogram).

\begin{figure}
    \centering
    \includegraphics[width=\linewidth]{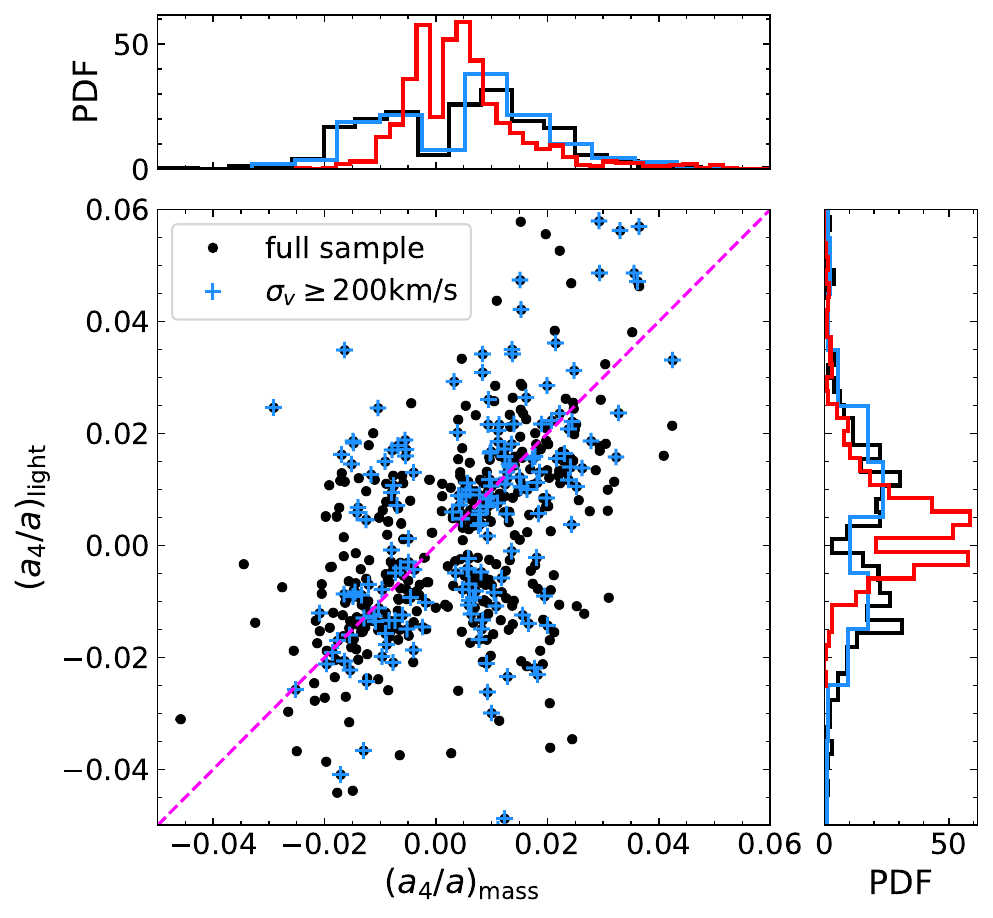}
    \caption{Comparison of $a_4/a$ between the mass and light distributions, for our full galaxy sample (black points and histograms) and a subsample with velocity dispersion $\sigma_{\rm v}\geqslant200~\text{km/s}$ (blue points and histograms). The magenta dashed line means $y=x$, and the red histogram is the distribution from \citet{Hao2006}. }
    \label{fig:a4_mass_light}
\end{figure}

Fig.~\ref{fig:a4_mass_light} shows the comparison of the averaged $a_4/a$ between the mass and light distributions. It can be seen that the two $m_4$ moments are, by and large, correlated with each other, though with substantial scatter. On average the amplitude $|(a_4/a)_{\rm mass}| \approx |(a_4/a)_{\rm light}| = 0.01\pm 0.01$. The difference $(a_4/a)_{\rm mass} - (a_4/a)_{\rm light}$ is centered around zero with a scatter of 0.015. In the figure, the histogram of $(a_4/a)_{\rm light}$ from \citet{Hao2006} calculated for the light distribution of a sample of massive early-type galaxies is also shown in red histograms. We note that the observed sample from \citet{Hao2006} has a larger velocity dispersions after aperture correction. We therefore also investigate the comparison for a subsample of the simulated galaxies with $\sigma_{\rm v}\geqslant200~\text{km/s}$ (187 galaxies, blue). As can be seen, the subsample has generally the same trend as the full sample; the histograms of $a_4/a$ in both light and mass of the simulated galaxies generally agree with but are markedly broader than that of \citet{Hao2006}\footnote{It is worth noting that \citet{Hao2006} averaged multipole strengths within an annulus that typically covers smaller radii than ours: the inner bound of adopted by \citet{Hao2006} is twice the seeing radius (and outer bound is $1.5R_{\rm eff}$). The difference in annulus radial coverage may partially account for a smaller scatter in their sample than in ours.}. Between the disky ($a_4>0$) and boxy ($a_4<0$) shapes, there are 66\% of the observed galaxies are disky in \citet{Hao2006}. Such a higher fraction towards disky galaxies is also seen in our subsample with $\sigma_{\rm v}\geqslant 200~\text{km/s}$: disky galaxies take a fraction of 64\% according to their mass distributions and 59\% according to their light distributions. For the full sample with lower velocity dispersion limit (at $\sigma_{\rm v}\geqslant 170~\text{km/s}$), the fractions drop slightly to 60\% according to mass and 51\% according to light. We note that the numerical values presented here are obtained based on 536 galaxy projections from one single snapshot of the TNG100 simulation. In a parallel work (Feng in prep), we adopt a larger sample of over 1500 lensing galaxy projections from three snapshots at $z=0.3$, 0.5, 0.7 from the same simulation and generally confirm our numerical findings given in this subsection. We will present a more detailed comparison between the light and mass distributions in that study.

\subsection{Perturbations in total matter distribution}
\label{ssec:prop_mass}

\begin{figure*}
    \centering
    \includegraphics[width=0.45\linewidth]{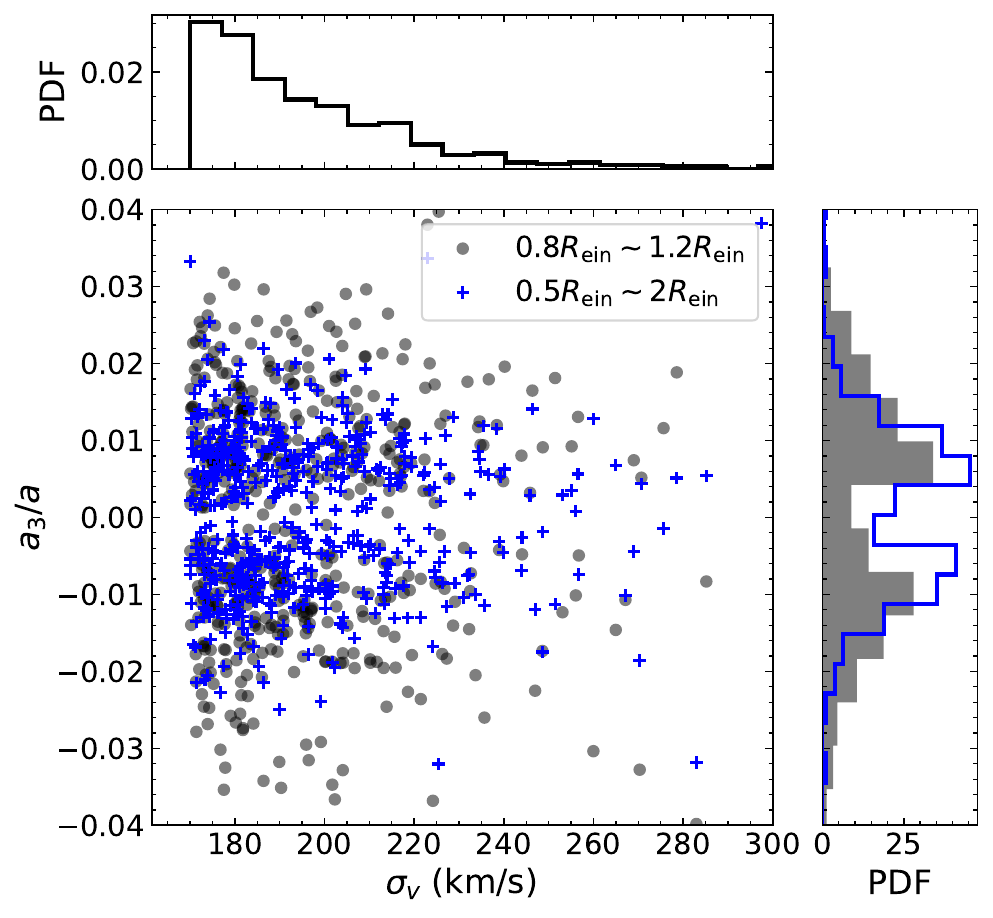}
    \quad
    \includegraphics[width=0.45\linewidth]{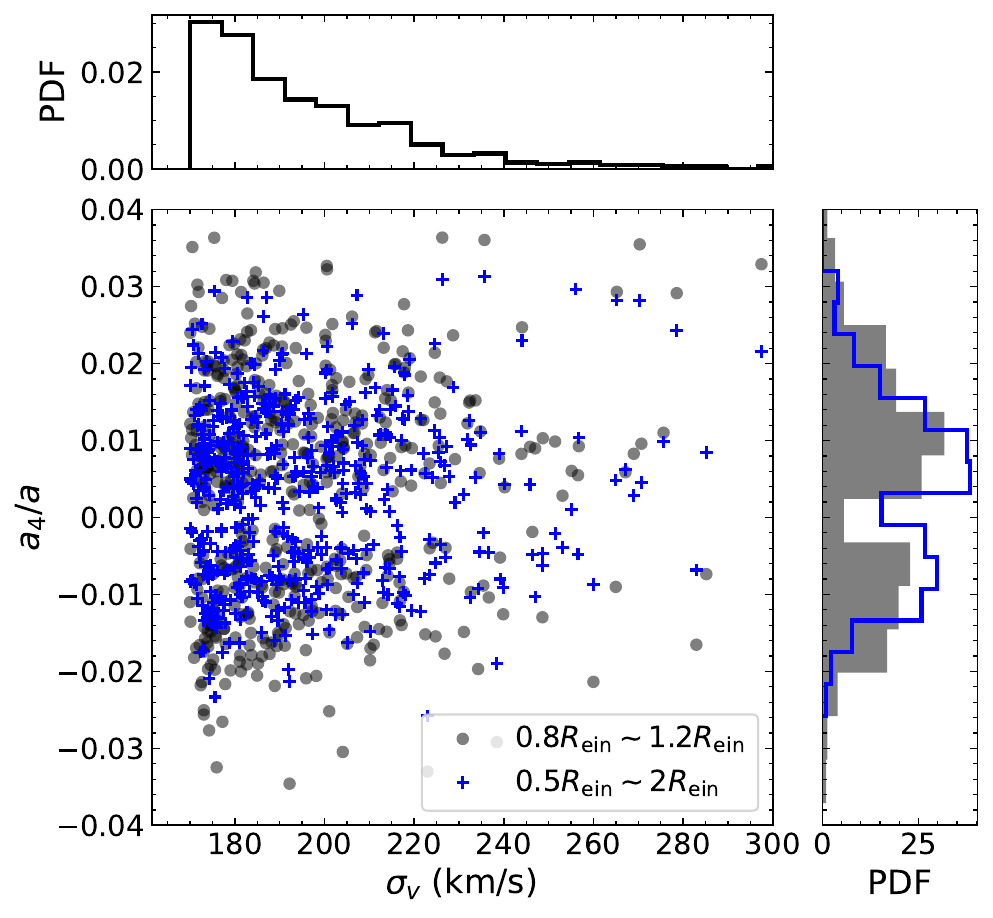}

    \caption{Relations between multipole strengths from total mass maps and velocity dispersion. Left: $a_3/a$ VS $\sigma_{\rm v}$; right: $a_4/a$ VS $\sigma_{\rm v}$. For comparison, the multipole strengths are averaged within 2 different annuli (as the right panel of Figure~\ref{fig:qpa_mass_light}): $0.8\sim1.2 R_{\rm ein}$ (grey points and shaded histogram) and $0.5\sim2.0 R_{\rm ein}$ (blue points and histogram).
    }
    \label{fig:a3a4withV}
\end{figure*}

The lensing effect is purely determined by the total mass distribution instead of light, in this subsection, we focus on the perturbation properties of the mass distribution. Fig.~\ref{fig:a3a4withV} shows the relation between the multipole strengths $a_3/a$ (left panel), $a_4/a$ (right panel), averaged within $0.8\sim1.2 R_{\rm ein}$, as a function of velocity dispersion $\sigma_{\rm v}$. In general there is little dependence of the multipole strengths on velocity dispersion $\sigma_{\rm v}$. On average the amplitude $|a_3/a|=0.01\pm 0.01$, same as $|a_4/a|$. As the averaging annulus changes from $0.8\sim1.2 R_{\rm ein}$ (shaded histogram) to $0.5\sim2.0 R_{\rm ein}$ (blue histogram), i.e., extending to more inner radii, the distributions of the averaged multipole strengths $a_m/a$ become slightly narrower. 

\begin{figure}[htbp]
    \centering
    \includegraphics[width=0.95\linewidth]{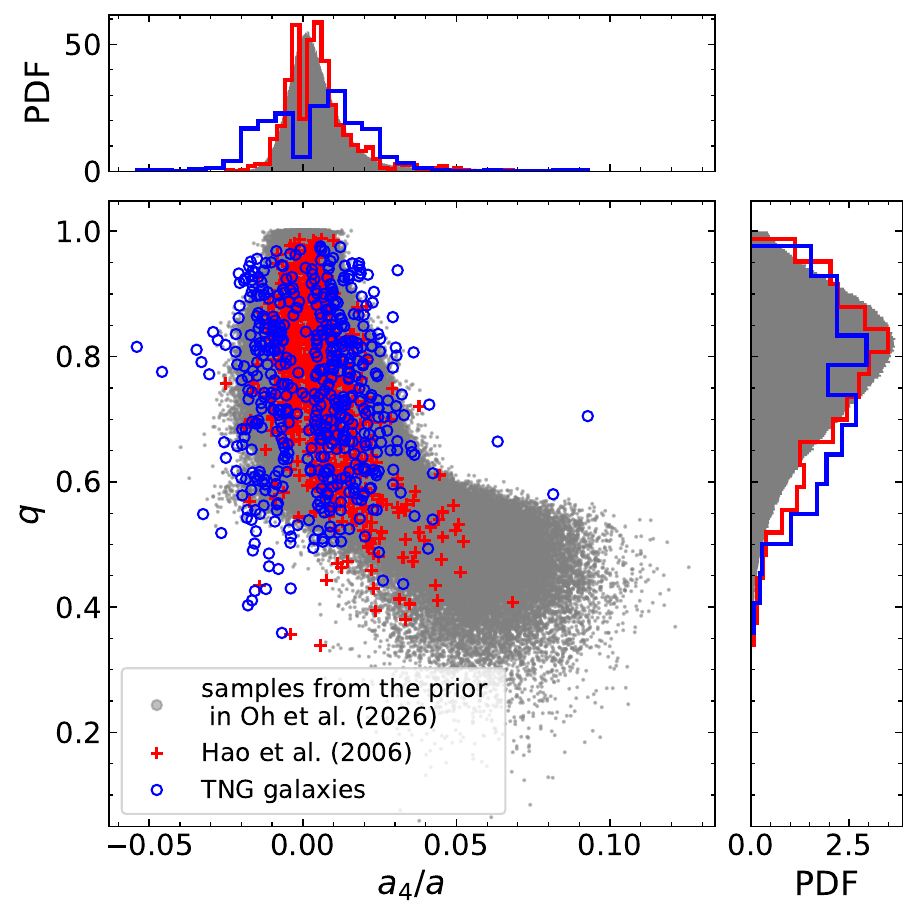}
    \caption{Relation between $m_4$ strength $a_4/a$ and axis ratio $q$. Blue points and histograms: our sample; red points and histograms: \citet{Hao2006}; grey points and shaded histogram: random sample generated by prior in \citet{oh2026}. See also Figure 6 in \citet{hou2026fluxratioanomaliescuspquasars}.}
    \label{fig:a4q}

\end{figure}

Fig.~\ref{fig:a4q} shows the relation between $a_4/a$ and axis ratio $q$ for our sample (blue), compared to the results from \citet{Hao2006, Hao2006_Erratum} (red). As can be seen, the two distributions are generally consistent, although the simulation sample has a wider distribution in $a_4/a$ than the observation. In addition, at $q \lesssim 0.55$ (a minor fraction at the tail of $q$), the simulated galaxies exhibit a slight bias towards boxy galaxies ($a_4<0$), while \citet{Hao2006, Hao2006_Erratum} towards the disky ones ($a_4>0$). 
For comparison, we also plot a distribution that samples the prior of $a_4/a-q$ from \citet{oh2026} (grey points). \citet{oh2026} fit a prior of multipole parameter values as a function of $q$ based on data from \citet{Hao2006}. We note the reader that the extra tail in grey to the bottom right of the overall distribution come from the usage of \citet{Hao2006} without taking into account the $\sqrt{q}$-factor correction (see \citealt{Hao2006_Erratum} for details).

\begin{figure*}[htbp]
    \centering
    \includegraphics[width=0.45\linewidth]{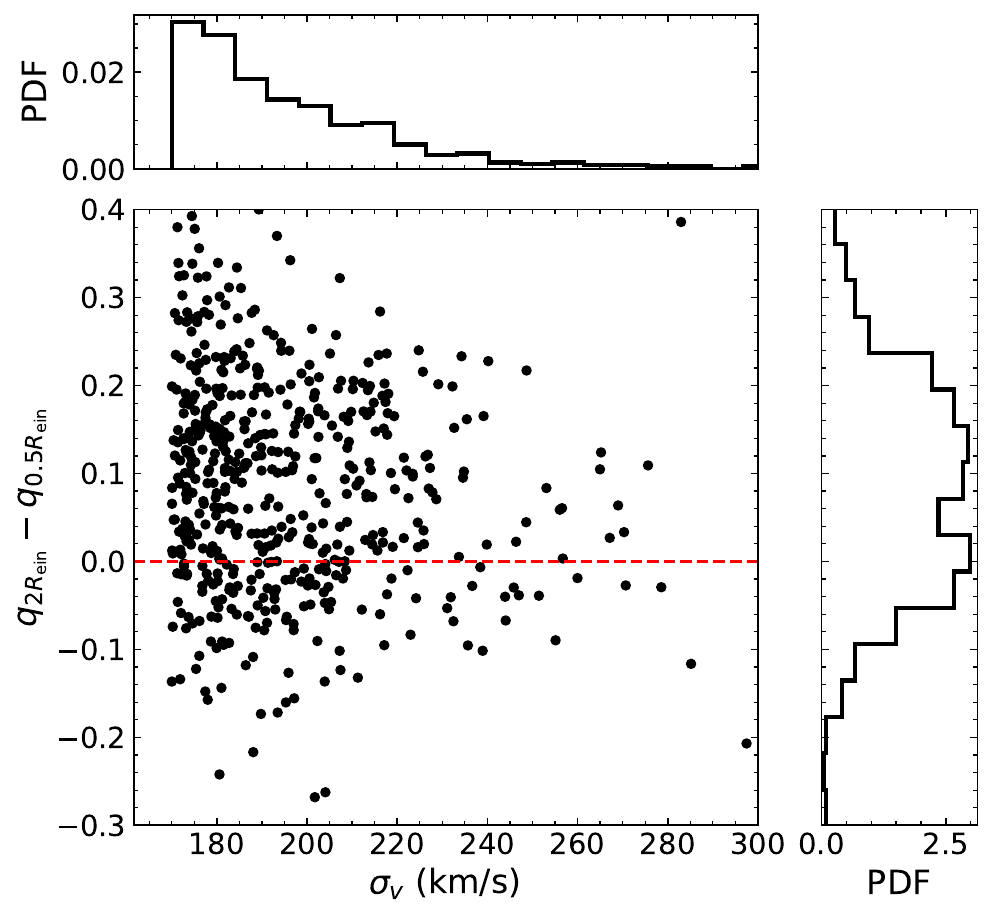}
    \qquad
    \includegraphics[width=0.45\linewidth]{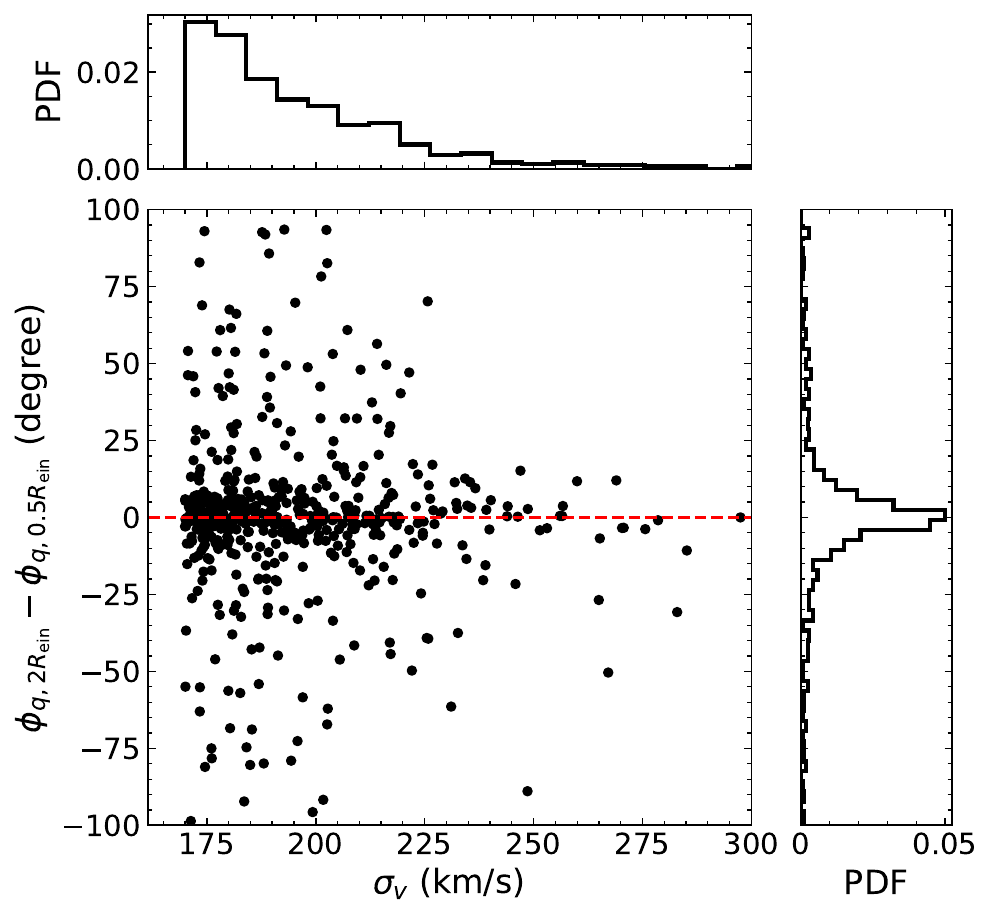}
    \caption{Relations between $\Delta q$, $\Delta \phi_q$ (difference of $q$ and $\phi_q$ between $2R_{\rm ein}$ and $0.5R_{\rm ein}$) and velocity dispersion $\sigma_{\rm v}$. Left: difference in $q$; right: difference in $\phi_q$. The red dashed line indicates zero difference.}
    \label{fig:DPADqwithV}
\end{figure*}

Fig.~\ref{fig:DPADqwithV} presents radial variations of the axis ratio and the position angle, $\Delta q \equiv q_{2R_{\rm ein}}-q_{0.5R_{\rm ein}}$ (left) and $\Delta \phi_q \equiv \phi_{q,\,2R_{\rm ein}}-\phi_{q,\,0.5R_{\rm ein}}$ (right) calculated over a radial range between $0.5R_{\rm ein}$ and $2R_{\rm ein}$, as a function of velocity dispersion $\sigma_{\rm v}$. The scatter in both quantities is seemingly bigger at lower $\sigma_{\rm v}$ than at higher $\sigma_{\rm v}$, due to less samples in the more massive end. Quantitatively, for galaxies with $\sigma_{\rm v}<220\, {\rm km\,s^{-1}}$, $\Delta q =0.10 \pm 0.12$ and $\Delta \phi_q =0^{\circ} \pm 15^{\circ}$. At $\sigma_{\rm v}\geqslant 220\, {\rm km\,s^{-1}}$, $\Delta q=0.04 \pm 0.10$ and $\Delta \phi_q=-1^{\circ}\, ^{+12^{\circ}}_{-21^{\circ}}$. There are slightly more galaxies with $\Delta q>0$ than those with $\Delta q<0$, indicating that the iso-density ellipses tend to be more circular at larger radii compared to smaller radii. Instead, the distribution of $\Delta \phi_q$ is more symmetric around zero -- no preferred direction for the twist/rotation. Across this radial range, the averaged scatter of position angle variation $\Delta \phi_q$ is at a level of $\sim 15^{\circ}$.

\begin{figure*}[htbp]
    \centering
    \includegraphics[width=0.49\linewidth]{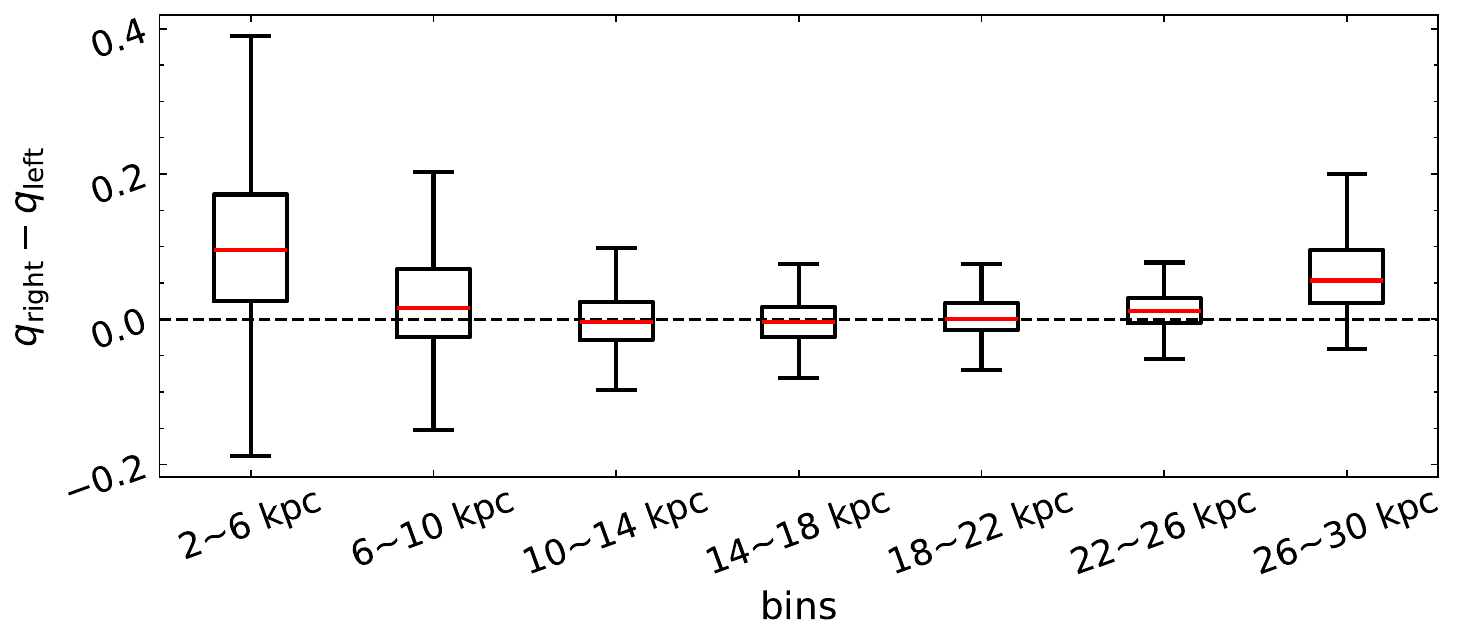}
    \hfill
    \includegraphics[width=0.49\linewidth]{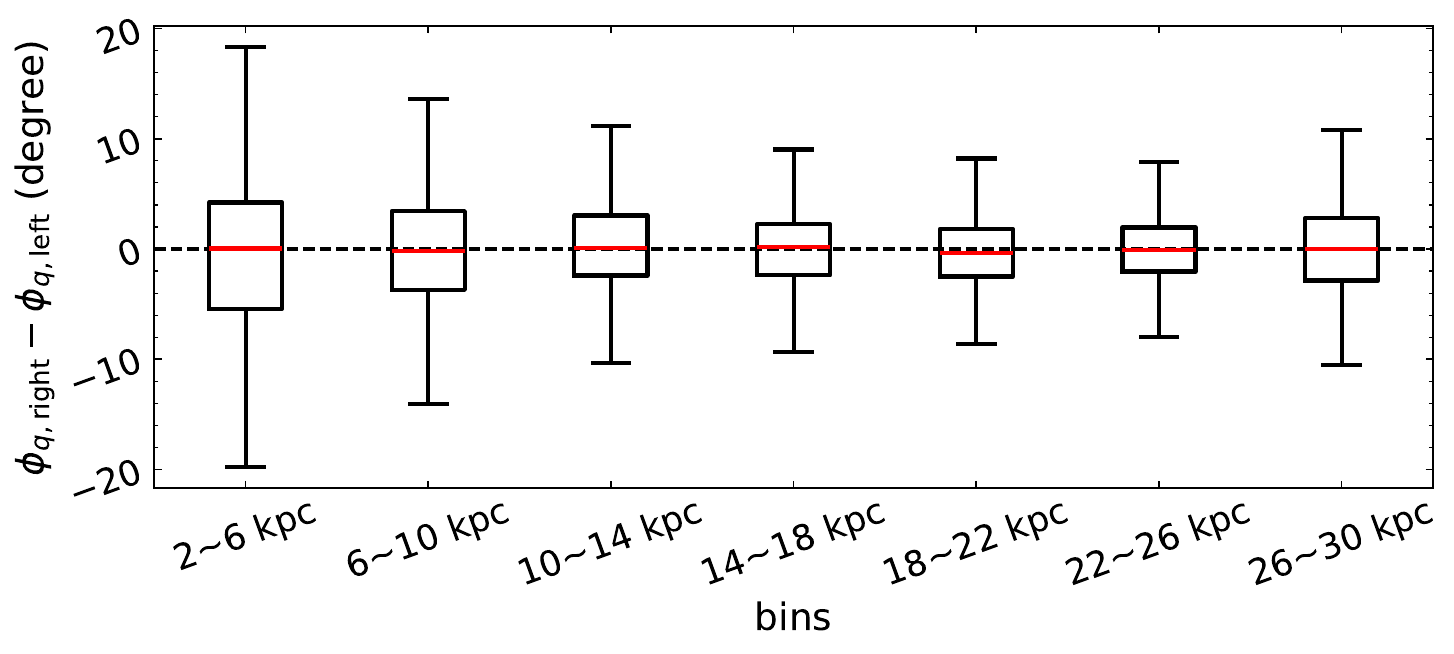}
    
    \caption{Box plot of variations in $q$ and $\phi_q$ (value at the right side minus that at the left side) in different radial bins with a width of 4 kpc. Left: difference in $q$; right: difference in $\phi_q$. For each bin, the red line means median, the box spans from $Q1$ (first quartile) to $Q3$ (third quartile), and the whiskers extend from the box to the smallest and largest values within $1.5\times IQR$ (interquartile range). The dashed line indicates zero difference.}
    \label{fig:qpa_range}
\end{figure*}

Fig.~\ref{fig:qpa_range} shows the variations in $q$ (left) and $\phi_q$ (right) as a function of radius in more detail. We divide radii (semi-major axes) from 2 kpc to 30 kpc into 7 bins with a width of 4 kpc, calculate the difference in $q$ and $\phi_q$ for each galaxy by subtracting the values at smaller radius from those at larger radius in each bin, and then get the statistics of all galaxies in each bin. It can be seen that for both ellipticity variation and position angle twist, the effect is most significant at innermost radii (below 6 kpc), where a bigger averaged $\Delta q$ and a larger scatter in $\Delta \phi_q$ are observed, indicating a tendency of more flattened and more twisted mass distributions as moving inwards towards the galactic center.

\section{Fitting the observed image positions}
\label{sec:fit_position}

In this section, we present results from fitting two families of macro-models (i.e., \texttt{SIE}+$\gamma$ and \texttt{EPL}+$\gamma$), in the presence of the investigated perturbations extracted from the simulation (i.e., $m_3,\,m_4,\,\Delta q,\,\Delta \phi_q$), to the observed image {\it positions alone} of B1422. For model fitting, we assume two sets of astrometric uncertainties of $\sigma_{\rm p}=10$ mas and 2 mas. We require that the best model produce image positions within $3\sigma_{\rm p}$ from the observed true positions, for all four images. We adopt the priors as listed in Table \ref{tab:defpri} (see Section \ref{ssec:mock} for details).

Table \ref{tab:ratepos} presents the numbers and percentages of successful cases. As can be seen, with the astrometric uncertainty as large as $\sigma_{\rm p}=10$~mas, both the adopted smooth models (\texttt{SIE}+$\gamma$ and \texttt{EPL}+$\gamma$), either alone or in combination with the investigated perturbation, can all successfully fit the observed image positions, i.e., no ``astrometric anomaly'' would be reported in this case, consistent with existing observations.

When adopting $\sigma_{\rm p}=2$\,mas, the observed image positions provide stronger constraints on the model. Now, the same macro-model families alone can no longer reproduce image positions -- observed ``astrometric anomalies'' indeed start appearing at such high astrometric precision (\citealt{Sluse2012_AAduetoSatellite, Spingola2018}).

Adding the investigated perturbations to these macro-models can improve the fitting performance at different levels, and more so for \texttt{EPL}+$\gamma$ than for \texttt{SIE}+$\gamma$. In particular, when all four types of perturbations are simultaneously included (last column), best-fit \texttt{EPL}+$\gamma$ models are found in two thirds of all cases, and one third for \texttt{SIE}+$\gamma$, indicating that the presence of lower-order multipole $m_3, \,m_4$ and radial variation $\Delta q$ and $\Delta \phi_q$ can explain the emerged astrometric anomalies at a $\sigma_{\rm p}=2$\,mas uncertainty level.

\begin{deluxetable*}{ccccccccc}
\tablehead{Setup & No-pert & \texttt{m3} & \texttt{m4} & \texttt{m3+m4} & \texttt{PAv} & \texttt{Qv} & \texttt{PAv+Qv} & \texttt{m3+m4+PAv+Qv}}
\tablecaption{Number and percentage of successful cases (out of a total of 536), which find a best-fit model that predicts image {\it positions} within $3\sigma_{\rm p}$ around observed true image positions for all four images, given different perturbation complexity and baseline model setups. ``No-pert'' is short for ``no perturbation'', similarly hereinafter. \label{tab:ratepos}}

\startdata
$0.01^{\prime\prime}$\texttt{-SIE} & \cmark & $535~(99.8\%)$ & $536~(100\%)$ & $536~(100\%)$ & $536~(100\%)$ & $536~(100\%)$ & $535~(99.8\%)$ & $533~(99.4\%)$ \\
$0.01^{\prime\prime}$\texttt{-EPL} & \cmark & $535~(99.8\%)$ & $536~(100\%)$ & $536~(100\%)$ & $536~(100\%)$ & $536~(100\%)$ & $536~(100.0\%)$ & $535~(99.8\%)$ \\
\hline
\texttt{2 mas-SIE} & \xmark  & $128~(23.9\%)$ & $45~(8.4\%)$ & $144~(26.9\%)$ & $111~(20.7\%)$ & $0~(0.0\%)$ & $99~(18.5\%)$ & $177~(33.0\%)$ \\
\texttt{2 mas-EPL} & \xmark & $250~(46.6\%)$ & $231~(43.1\%)$ & $286~(53.4\%)$ & $232~(43.3\%)$ & $291~(54.3\%)$ & $253~(47.2\%)$ & $356~(66.4\%)$ \\
\enddata\
\end{deluxetable*}

\begin{figure*}[htbp]
    \centering
    \includegraphics[width=1\linewidth]{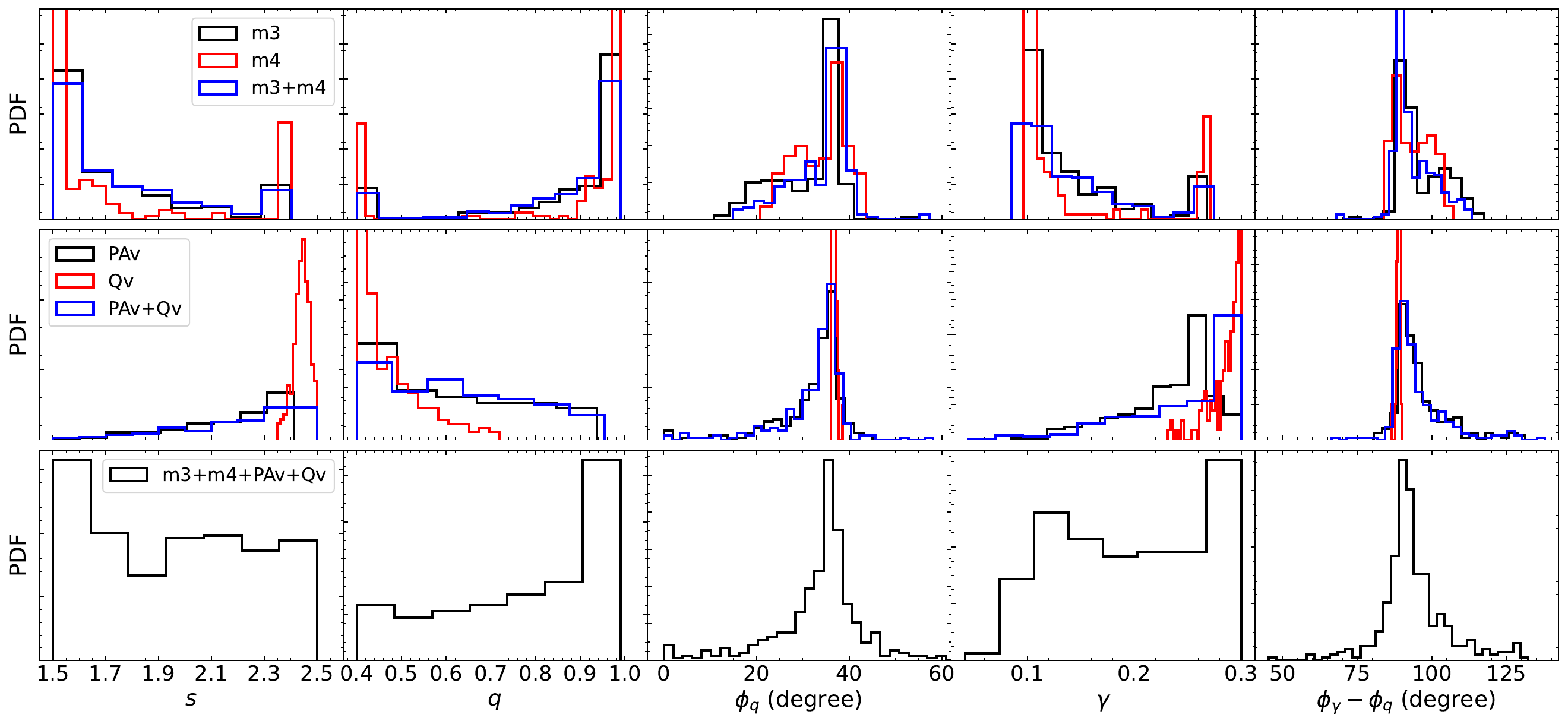}
    \caption{Distributions of best (successful) macro-model parameters, $s$, $q$, $\phi_q$, $\gamma$, $\phi_{\gamma}-\phi_q$, for all seven kinds of perturbations each added to \texttt{EPL}+$\gamma$ to fit to image positions only (assuming $\sigma_{\rm p}=2$ mas, essentially the last row in Table \ref{tab:ratepos}). We deliberately omit the $y$-axis ticks in each panel so that panels in the same row can share a common $y$-axis.  }
    \label{fig:fitpos_SP}
\end{figure*}

The differences between results from the two adopted macro-model families are easy to understand. The success rates among all perturbation scenarios are significantly higher when using \texttt{EPL}+$\gamma$ than using \texttt{SIE}+$\gamma$ (with $\sigma_{\rm p}=2$\,mas). Such differences arise essentially because \texttt{EPL}, with a further degree of freedom in the logarithmic radial slope $s$, can partially accommodate effects from yet-unaccounted macroscopic or clumpy perturbative features (even in the angular dimension). This has been clearly demonstrated and explained in \citet{Kochanek2021}: when fitting point-sources, incorrectly-inferred radial slopes may mimic the effect due to azimuthal structures if the fitting models do not already include such ingredients.

It is also worth noting that, in practice, using observational constraints from image positions {\it alone} makes it impossible to simultaneously constrain both the smooth model and the perturbations. This fundamental degeneracy is exactly why such an approach is not pursued in observational studies, and also why we fix the perturbation components by directly adopting them from our simulation setups. In light of this, we caution the reader that, despite solutions found, they are highly likely unphysical. This can be seen from Fig.\,\ref{fig:fitpos_SP}, which presents the distributions of the best (successful) macro-model parameters, $s$, $q$, $\phi_q$, $\gamma$, $\phi_{\gamma}-\phi_q$, for all seven kinds of perturbations each added to \texttt{EPL}+$\gamma$ to fit to image positions only (assuming $\sigma_{\rm p}=2$ mas, essentially the last row in Table \ref{tab:ratepos}). As can be seen, the two perturbation types -- lower-order multipoles (\texttt{m3}, \texttt{m4}, \texttt{m3+m4}) and ellipse variations (\texttt{PAv}, \texttt{Qv}, \texttt{PAv+Qv}) -- show markedly different preference in macro-model parameters. While the two kinds both agree on $\phi_q$ and the shear angle difference $\phi_{\gamma}-\phi_q$, they diverge on the radial slope $s$, axis ratio $q$, and external shear strength $\gamma$: $m_3,\,m_4$ multipoles favor shallower profiles, rounder shapes and weaker shear, whereas $\Delta q,\,\Delta\phi_q$ variations favor steeper profiles, flatter shapes and stronger shear. In either case, the inferred distributions on $s$, $q$, and $\gamma$ hit the boundary -- a classical sign of unphysical model inference. This strongly demonstrates that image positions alone, even at such high astrometric precision, cannot tightly/uniquely constrain the underlying macro-model; large degeneracies persist between different (smooth and perturbative) model components.

\section{Fitting image positions and flux ratios simultaneously} 
\label{sec:fit_fluxratio}

In this section, we present results from fitting models (again in the presence of the investigated perturbations) to both image positions and flux ratios simultaneously. A B1422-analog must be produced by the model combination that successfully predicts positions and flux ratios around the observed values for the four images $A$, $B$, $C$, and $D$ within $3\sigma$ uncertainties. For this exercise, we adopt an astrometry uncertainty of $\sigma_{\rm p}=10$ mas across all model combinations -- to guarantee that ``astrometric anomalies'' would not occur at this level (see Section~\ref{sec:fit_position}). The flux ratio uncertainty $\sigma_{\rm f}$ is assumed to take three different values -- 10\%, 5\%, and 2\% -- for the image pairs $B/A$ and $C/A$, and 15\% for $D/A$ (see Section \ref{ssec:mock} for more details).

Table \ref{tab:ratepf} presents the numbers and percentages of successful cases (out of 536) under different perturbation combinations (including none), different macro-models with different observational uncertainties. In the previous section, we already showed that when image positions alone are used (assuming $\sigma_{\rm p}=10$ mas), one can not only obtain a best-fit  \texttt{SIE}+$\gamma$ model alone, but also always find a best-fit \texttt{SIE}+$\gamma$ in the presence of all the investigated perturbation scenarios (which is also true for \texttt{EPL}+$\gamma$, see Table \ref{tab:ratepos}). Now, with constraints from both observed positions and flux ratios simultaneously, the \texttt{SIE}+$\gamma$ model alone cannot reproduce the data, and adding the investigated perturbations does not improve the fitting performance, even with $\sigma_{\rm f}$ as large as 10\% (for $B/A$ and $C/A$). This is simply a reinstatement of the ``flux-ratio anomaly problem''. Adding the investigated perturbations cannot effectively increase the chance of reproducing the observations; the combined models lack of sufficient degree of freedom to explain the data.

When the \texttt{EPL}+$\gamma$ model is adopted, the situation changes significantly. With \texttt{EPL}+$\gamma$ alone, the smooth model can simultaneously fit image positions and flux ratios at $\sigma_{\rm f}=10\%$ and 5\% (thanks to the extra degree of freedom in radial slope $s$). This means that no ``flux-ratio anomaly problem'' would be identified when using \texttt{EPL}+$\gamma$ model alone at such photometric precision\footnote{Although the successful models tend to require a fairly small axis ratio $q\sim 0.5-0.6$ and a large external shear $\gamma\sim0.3$.}.

It is interesting to note that adding the investigated perturbations in this case would actually decrease the success rates from 100\% (see the fourth and fifth rows in Table \ref{tab:ratepf}). The occurrence of failed cases when adding the investigated perturbations in fact suggests that the corresponding perturbation features (among the failed cases) introduce flux-ratio fluctuation at a level beyond what the macro-model can further adjust for, thus preventing it from explaining the observations. It is more so when adding $\Delta q, \, \Delta \phi_q$ variations than adding $m_3,\,m_4$ multipoles alone: including only $m_3,\,m_4$ multipoles can still explain the data in almost all cases ($\gtrsim 90\%$); while if $\Delta q$ and $\Delta \phi_q$ variations are added, the success rates decrease to $60-80\%$ when $\sigma_{\rm f}=10\%$ and further to $\lesssim 10\%$ when $\sigma_{\rm f}=5\%$.

\begin{figure}[htbp]
    \centering
    \includegraphics[width=1\linewidth]{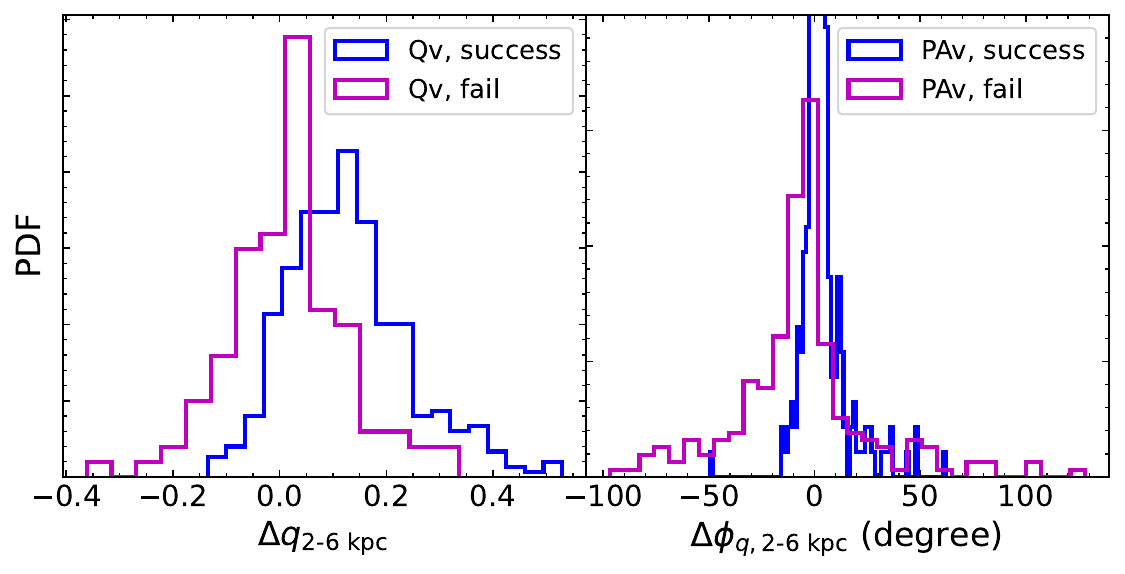}
    \caption{The left panel presents the distributions of the radial difference $\Delta q_{2-6{\rm kpc}} \equiv q_{6{\rm kpc}}-q_{2{\rm kpc}}$ for both the successful and failed cases when adopting \texttt{EPL}+$\gamma$ in the presence of perturbation \texttt{Qv} alone (assuming $\sigma_{\rm f}=10\%$, $\sigma_{\rm p}=10$ mas), and in the right panel the distributions for $\Delta \phi_{q,\,{2-6{\rm kpc}}} \equiv \phi_{q,\,{6{\rm kpc}}}-\phi_{q,\,{2{\rm kpc}}}$ when adding perturbation \texttt{PAv} alone. The samples  essentially corresponds to the fourth row in Table \ref{tab:ratepf}.}
    \label{fig:PAvQvfailvssucc}
\end{figure}

Clearly, in this case a much severer penalty is introduced by adding radial variations of iso-density ellipses. In order to gain some intuitive understanding on this, we look into the differences between the successful and failed cases in two specific perturbation scenarios where $\Delta q$ and $\Delta \phi_q$ variations alone are added, and the sample sizes are also sufficiently large in both cases. In the left panel of Fig.\,\ref{fig:PAvQvfailvssucc} we present the distributions of the radial difference $\Delta q_{2-6{\rm kpc}} \equiv q_{6{\rm kpc}}-q_{2{\rm kpc}}$ for both the successful and failed cases when adopting \texttt{EPL}+$\gamma$ in the presence of perturbation \texttt{Qv} alone (assuming $\sigma_{\rm f}=10\%$, $\sigma_{\rm p}=10$ mas), and in the right panel the distributions for $\Delta \phi_{q,\,{2-6{\rm kpc}}} \equiv \phi_{q,\,{6{\rm kpc}}}-\phi_{q,\,{2{\rm kpc}}}$ when adding perturbation \texttt{PAv} alone. As can be seen, the distributions between the successful and failed cases are markedly different. Quantitatively, on average $\Delta q_{2-6{\rm kpc}}=0.12 \pm 0.10$ and $0.02 \pm 0.10$ for the successful and failed cases, respectively. While $\Delta \phi_{q,\,{2-6{\rm kpc}}}=1.3^{\circ}\, ^{+5.6^{\circ}}_{-3.3^{\circ}}$ and $-5.7^{\circ}\,^{+16.4^{\circ}}_{-24.4^{\circ}}$ for the successful and failed cases, respectively. This may suggest some general tendency that when the iso-density ellipses become rounder at larger radii and/or are less rotated, it is easier for the macro-model to accommodate and maintain a successful fitting to data.

\begin{figure*}[htbp]
    \centering
    \includegraphics[width=1\linewidth]{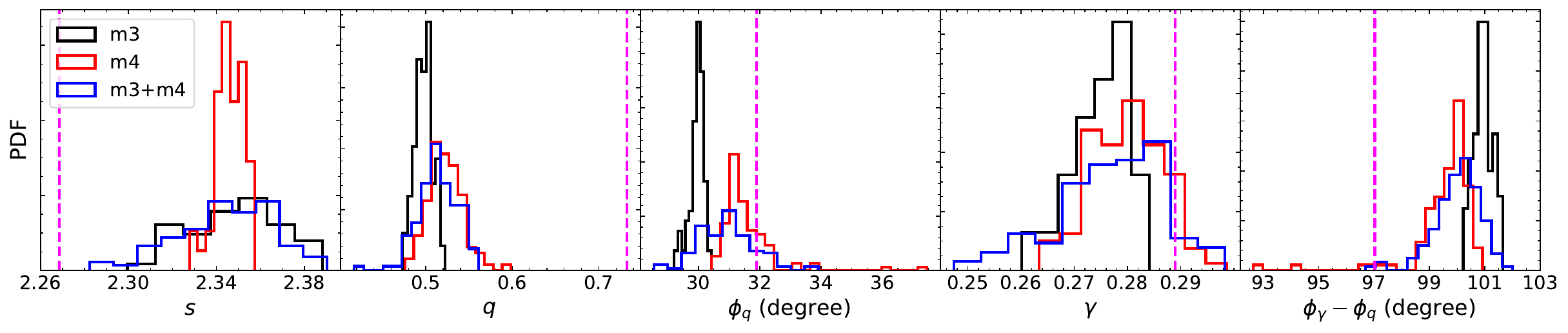}
    \caption{Distributions of the best (successful) macro-model parameters, $s$, $q$, $\phi_q$, $\gamma$, $\phi_{\gamma}-\phi_q$, for three kinds of perturbations (\texttt{m3, \,m4, \,m3+m4}), each added to \texttt{EPL}+$\gamma$ to fit both image positions and flux ratios (assuming $\sigma_{\rm f}=2\%$ and $\sigma_{\rm p}=10$ mas, essentially the last row in Table \ref{tab:ratepf}). The vertical lines indicate the parameter values of one example successful mass models from \texttt{m3+m4+PAv+Qv},for which the image configuration along with the corresponding critical curves and caustics are shown in Fig.~\ref{fig:B1422_winsol}.}
    \label{fig:PSFA_SP}
\end{figure*}

As the photometric precision progressively increases (smaller $\sigma_{\rm f}$), \texttt{EPL}+$\gamma$ alone can no longer reproduce the observed image positions and flux ratios simultaneously at $\sigma_{\rm f}=2\%$ -- ``flux-ratio anomalies'' eventually show up at the smooth model level. Adding $m_3,\,m_4$ multipoles alone can help explain the data in about a hundred ($\sim 20\%$) cases. Fig.\,\ref{fig:PSFA_SP} presents the distributions of the best (successful) macro-model parameters, $s$, $q$, $\phi_q$, $\gamma$, $\phi_{\gamma}-\phi_q$, for three kinds of perturbations (\texttt{m3,\,m4,\,m3+m4}) each added to \texttt{EPL}+$\gamma$ to fit to both image positions and flux ratios (assuming $\sigma_{\rm f}=2\%$ and $\sigma_{\rm p}=10$ mas). As can be seen, unlike in Fig.~\ref{fig:fitpos_SP}, here no inference results hit the boundary, and the ranges of all macro-model parameters are largely consistent across the different perturbation scenarios. Interestingly, the recovered parameters deviate markedly from the norm -- the radial slopes much steeper than isothermal ($s\sim 2.35$), unusually small axis ratios ($q\sim 0.5$), and fairly large external shears ($\gamma \sim 0.27$). One can ask whether such a solution remains physically plausible, given how strongly it deviates from our prior beliefs about the population. Indeed, B1422 appears to be a unique case in this regard: studies focusing on this system typically find that, in order to simultaneously explain the observed image positions and flux ratios -- without introducing dark matter subhalos, one must invoke fairly small $q\sim 0.5$ (\citealt{KSB1994B1422}) or large shear strength $\gamma \sim 0.3$ (\citealt{Sluse2012_AAduetoSatellite}), or when the slope fixed to $s=2$ then ridiculously large $m_3,\,m_4$ multipole strengths (e.g., \citealt{WynWitt2003TruthDilusion}). Here, by relaxing the slope constraint and adopting the realistic multipole distributions derived from the simulation, we find some significantly steepened density profiles that are favored to account for the distinctive observations of B1422.

\begin{figure}[htbp]
    \centering
    \includegraphics[width=1\linewidth]{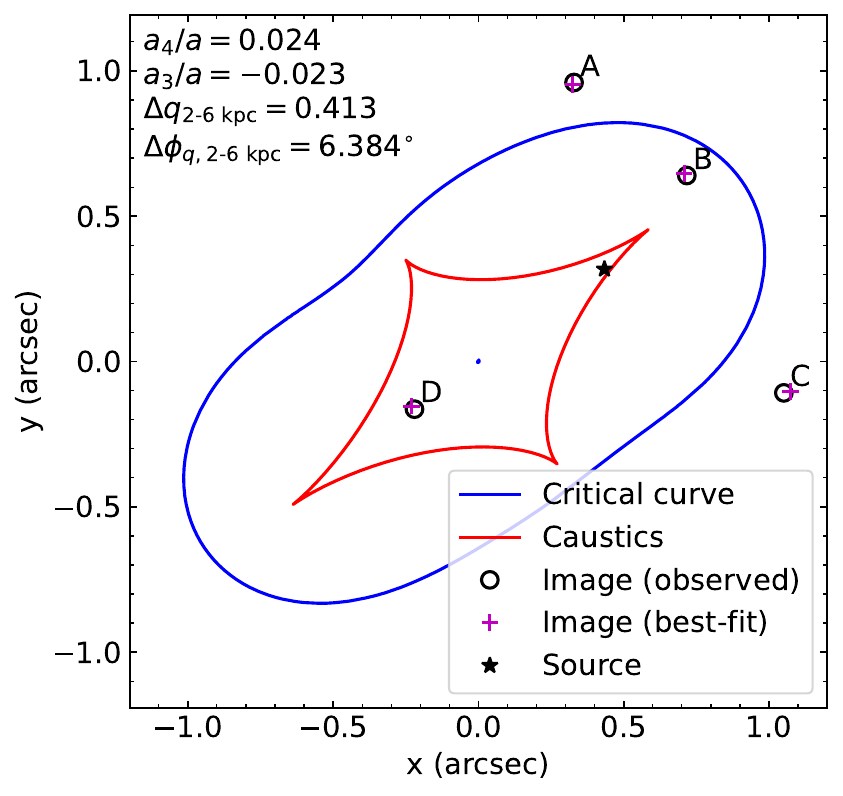}
    \caption{The critical curves and caustics together with the image and source positions predicted by one of the best-fit models in \texttt{EPL}+$\gamma$+\texttt{m3+m4+PAv+Qv} (assuming $\sigma_{\rm f}=2\%$ and  $\sigma_{\rm p}=10$ mas). The best model parameters are also marked out by the vertical lines in Fig.\,\ref{fig:PSFA_SP}. The observed image positions for B1422 are also given (circles).}
    \label{fig:B1422_winsol}
\end{figure}

When the radial variations of iso-density ellipses is taken into account, no successful case is found  when adding only $\Delta \phi_q$ (i.e., \texttt{PAv}), but successful cases (although only a few cases) do exist when adding $\Delta q$ alone (i.e., \texttt{Qv}), or $\Delta \phi_q$ and $\Delta q$ together (i.e., \texttt{PAv+Qv}), as well as if all four perturbations are added simultaneously (i.e., \texttt{m3+m4+PAv+Qv}). Fig.\,\ref{fig:B1422_winsol} present the critical curves and caustics together with the image and source positions predicted by one of the best-fit models in \texttt{EPL}+$\gamma$+\texttt{m3+m4+PAv+Qv} (assuming $\sigma_{\rm f}=2\%$ and  $\sigma_{\rm p}=10$ mas). The best model parameters are also marked out by the vertical lines in Fig.\,\ref{fig:PSFA_SP}. These results have already answered the very question that we post as the title of this paper: ``{\it Can third- and fourth-order multipoles plus radial variation of iso-density ellipses explain the observed flux ratios in B1422$+$231?}''. And the answer is YES -- without the inclusion of clumpy mass perturbations.

Regarding the low ($\lesssim 2\%$) success rates when including $\Delta q,\,\Delta \phi_q$ variations (i.e., in \texttt{PAv}, \texttt{Qv}, \texttt{PAv+Qv} and \texttt{m3+m4+PAv+Qv}) versus the relatively high ($\sim 20\%$) success rates when adding $m_3,\,m_4$ multipoles (i.e., in \texttt{m3}, \texttt{m4} and \texttt{m3+m4}), one is fair to say, from a pure model fitting perspective, that introducing global $m_3,\,m_4$ multipoles is more flexible than implementing $\Delta q,\,\Delta \phi_q$ variations (to the combined mass model) to explain the data. However, one shall also caution NOT to directly interpret this result as ``{\it $\Delta q,\,\Delta \phi_q$ variations are less important than $m_3,\,m_4$ multipoles in causing flux-ratio anomalies}'' or ``{\it $m_3,\,m_4$ multipoles are more plausible than $\Delta q,\,\Delta \phi_q$ variations to be responsible for the observed flux anomalies}''. The reason is twofold. Firstly, as already explained in Section \ref{ssec:mock}, the actual value of the success rate in fact depends on how fine one can ``sample'' the entire perturbation feature space through adopting such a cosmological simulation-based strong lens sample. Between the two perturbation types, it is not difficult to see that the parameter space for $m_3,\,m_4$ multipoles has much lower dimensions than that for the radial-dependent variations in $\Delta q,\,\Delta \phi_q$. If a larger galaxy sample is used, it would manage to sample the high-dimension perturbation space more finely, and thus have somehow larger chances (for such rare events) to pick up realizations that closely resemble the observed system (hence increased successful rates). Secondly and more importantly, a key lesson that we learn from using a realistic lens sample from cosmological simulations is that the two kinds of perturbations are subtly connected: very rarely a galaxy would only possess $m_3,\,m_4$ multipole perturbations but be free from $\Delta q,\,\Delta \phi_q$ variations. As soon as the latter is present, it introduces fluctuations to image flux ratios, which sensitively depend on the very way that the iso-density ellipses vary with radius. This explains why in the  perturbation scenario \texttt{m3+m4+PAv+Qv} (the last row and last column of Table \ref{tab:ratepf}) the success rate remains low:  $\Delta q,\,\Delta \phi_q$ variations induce specific fluctuations that prevents the multipoles (with macro-models) from successfully fitting the flux ratios, even though the multipoles alone often succeed.

\begin{deluxetable*}{ccccccccc}
\tablehead{Setup &  No-pert & \texttt{m3} & \texttt{m4} & \texttt{m3+m4} & \texttt{PAv} & \texttt{Qv} & \texttt{PAv+Qv} & \texttt{m3+m4+PAv+Qv}}
\tablecaption{Number and percentage of successful cases (out of a total of 536) which find a best-fit model that predicts image {\it positions and flux ratios} within $3\sigma$ around observed true image positions and flux ratios for all four images, given different perturbation complexity and baseline model setups. Assuming image astrometric uncertainty of 10 mas, flux ratio uncertainty in $D/A$ of 15\%, flux ratio uncertainties in $B/A$ and $C/A$ of 10\%, 5\% and 2\%.  
\label{tab:ratepf}}
\startdata
\texttt{10{\rm\%}-SIE} & \xmark &  $3~(0.6\%)$ & $2~(0.4\%)$ & $7~(1.3\%)$ & $0~(0.0\%)$ & $0~(0.0\%)$ & $0~(0.0\%)$ & $4~(0.7\%)$ \\
\texttt{5{\rm\%}-SIE} & \xmark & $0~(0.0\%)$ & $0~(0.0\%)$ & $0~(0.0\%)$ & $0~(0.0\%)$ & $0~(0.0\%)$ & $0~(0.0\%)$ & $0~(0.0\%)$ \\
\texttt{2{\rm\%}-SIE} & \xmark & $0~(0.0\%)$ & $0~(0.0\%)$ & $0~(0.0\%)$ & $0~(0.0\%)$ & $0~(0.0\%)$ & $0~(0.0\%)$ & $0~(0.0\%)$ \\
\hline
\texttt{10{\rm\%}-EPL} & \cmark & $533~(99.4\%)$ & $535~(99.8\%)$ & $532~(99.3\%)$ & $314~(58.6\%)$ & $428~(79.9\%)$ & $314~(58.6\%)$ & $292~(54.5\%)$ \\
\texttt{5{\rm\%}-EPL} & \cmark & $526~(98.1\%)$ & $479~(89.4\%)$ & $475~(88.6\%)$ & $12~(2.2\%)$ & $22~(4.1\%)$ & $66~(12.3\%)$ & $89~(16.6\%)$ \\
\texttt{2{\rm\%}-EPL} & \xmark & $90~(16.8\%)$ & $102~(19.0\%)$ & $126~(23.5\%)$ & $0~(0.0\%)$ & $2~(0.4\%)$ & $7~(1.3\%)$ & $10~(1.9\%)$ \\
\enddata
\end{deluxetable*}

\section{Conclusions and Discussion}
\label{sec:concl}

Flux-ratio anomalies have been a long-standing issue in multiply-imaged quasar lensing systems. In this study, we take B1422+231 -- one of the classical flux-anomaly systems -- for a case study. We utilize a strong lensing galaxy sample from the TNG100 simulation and extract lower-order density perturbations beyond elliptical shapes (in particular, the third- and fourth-order multipole moments $m_3$ and $m_4$, as well as radial variations of axis-ratio $\Delta q$ and position angle $\Delta \phi_q$ of iso-density ellipses, see Section \ref{ssec:samp_sel} and \ref{ssec:extract} for details). We then use the observed image positions alone, as well as both positions and flux ratios, assuming typical observational precisions, to constrain smooth macroscopic models (described by \texttt{SIE}+$\gamma$ and \texttt{EPL}+$\gamma$) in the presences of the extracted perturbations. This method essentially offers a much cheaper alternative way for searching over a large parameter space of lens models that include both smooth and perturbative components. Rather than explicitly sampling the enormous and highly complex parameter space of the perturbative features — in particular the radial variation of ellipticity and position angle, along with the $m_3,\,m_4$ moments that are subtly connected to them — the method adopts a simulation-based ``coarse-grain'' sampling to implicitly cover that space (see Sections \ref{ssec:lens1422model} and \ref{ssec:mock} for details). Through these experiments, we can now answer the three key questions posed at the outset: (1) Under what fitting conditions lensing (astrometry and flux-ratio) anomalies would appear? (2) Whether typical macroscopic perturbations (with respect to clumpy dark matter halos or subhalos) on top of the adopted smooth elliptical mass distributions may account for the observed flux ratios in this system; (3) How the results depend on the macro-model choice and on observational precisions. Below, we summarize our main findings in response to these questions.

\begin{itemize}
    \item ``Astrometric anomalies'' between the smooth model predicted and the observed image positions do not appear at astrometric uncertainty $\sigma_{\rm p}=10$ mas, either using \texttt{SIE}+$\gamma$ or \texttt{EPL}+$\gamma$. Adding the investigated perturbations in form of $m_3, \, m_4, \, \Delta q, \Delta \phi_q$ does not change the conclusion -- the model combination can always reproduce image astrometry within $3\sigma_{\rm p}$ at such an uncertainty level (see top two rows in Table \ref{tab:ratepos} and Section \ref{sec:fit_position}).

    \item ``Astrometric anomalies'' for this system occur at $\sigma_{\rm p}=2$ mas, where neither of the two macro-models {\it alone} can reproduce the observed image positions within $3\sigma_{\rm p}$. Adding the investigated perturbations can explain the data, although strong model degeneracy exists (see bottom two panels in Table \ref{tab:ratepos}, Fig.\,\ref{fig:fitpos_SP} and more discussion in Section \ref{sec:fit_position}). 
    
    \item ``Flux-ratio anomalies'' between the smooth model prediction and the observation (flux ratios) come to exist when  \texttt{SIE}+$\gamma$ alone is adopted. This is the true even with flux-ratio uncertainty as large as $\sigma_{\rm f}= 10\%$ (for image $B/A$ and $C/A$) (fixing $\sigma_{\rm p}=10$ mas). Adding the investigated perturbations does not change this result (see top three rows in Table \ref{tab:ratepf} and Section \ref{sec:fit_fluxratio}). 
    
    \item ``Flux-ratio anomalies'' do not occur when using  \texttt{EPL}+$\gamma$ alone at photometric uncertainty $\sigma_{\rm f}= 5\%$ or 10\% (also fixing $\sigma_{\rm p}=10$ mas), i.e., the smooth model alone can already well fit both positions and flux ratios simultaneously without adding any kinds of perturbations (see the fourth and fifth rows in Table \ref{tab:ratepf} and Section \ref{sec:fit_fluxratio}). 
    
    \item ``Flux-ratio anomalies'' occur for \texttt{EPL}+$\gamma$ at $\sigma_{\rm f}= 2\%$ -- the smooth model family alone can no longer simultaneously fit the observed image positions and flux ratios. {\it Adding the investigated perturbations to a smooth \texttt{EPL}+$\gamma$ can explain the observation for B1422} (see the last row in Table \ref{tab:ratepf}, Figs.\,\ref{fig:PSFA_SP} and \ref{fig:B1422_winsol} and Section \ref{sec:fit_fluxratio}). 
        
\end{itemize}

Some important lessons are learned when comparing between different macro-models. It is clear that \texttt{EPL}+$\gamma$ with a free radial slope $s$ provide much greater flexibility to fit observations than \texttt{SIE}+$\gamma$. In particular, some of the cases that are identified as ``flux ratio anomalies'' using \texttt{SIE}+$\gamma$ become ``normal'' and ``happy'' again when using \texttt{EPL}+$\gamma$ instead. The overall success rates are also much bigger when \texttt{EPL}+$\gamma$ is adopted instead of \texttt{SIE}+$\gamma$. We therefore caution the use of \texttt{SIE}+$\gamma$ in future flux-ratio anomaly studies in order to avoid drawing limited and even biased conclusions (unless sufficient evidence is present for perfect isothermality).

One may also ask between the $m_3,\,m_4$ multipoles and the radial variation of $\Delta q,\Delta\phi_q$, which kind of perturbations can better explain the B1422 image configuration (position + flux ratios)? By face value of the success rates (see, e.g., the last row of Table \ref{tab:ratepf}), the former appears more likely to explain the observation -- but this is only from a model-fitting perspective. One shall bear in mind that the two perturbations are intrinsically and subtly coupled: rarely does a galaxy host $m_3,\,m_4$ multipoles while being free from $\Delta q,\,\Delta\phi_q$ variations. Once present, flux-ratio fluctuations induced by the latter would depend sensitively on how the iso-density ellipses vary with radius, which cannot be easily compensated for by the lower-order multipoles.

In this study, we have shown that including macroscopic mass perturbations -- instead of exclusively invoking clumpy substructure -- may offer a viable alternative to explain the observation of B1422. However, it is important to clarify that the primary intent of this study is not to argue that such macroscopic perturbations are the definitive cause over subhalo populations or other alternative explanations for this particular system. Rather, we aim to highlight that such a perspective warrants serious consideration. While we acknowledge the substantial efforts and careful investigations in the field, we hope that the present analysis encourages broader and further reflection on current modeling assumptions. In particular, we wish to remind the community to caution the practice of adding a global $m_3$ or $m_4$ multipole (alongside dark matter subhalos/halos) on top of the adopted smooth macro-model, such as \texttt{EPL}+$\gamma$, to fit quadruply lensed quasar image data. This approach has been widely adopted for constraining subhalo properties and, by extension, dark matter cosmology. However, it warrants careful scrutiny. Specifically as we have demonstrated in this study, due to the intrinsic coupling between these lower-order multipoles and other macroscopic perturbation features such as radial variation of iso-density ellipses, the practice of implementing global m3, m4 alone may artificially break hidden degeneracies and thus lead to biased inference (along the same lines see \citealt{Cohen2024} but also see Appendix B of \citealt{Gilman2024_FAandArc}).

It is worth noting that B1422 is indeed classified as a flux-anomaly system, albeit exhibiting relatively milder anomalous ratios. A natural question is whether the macroscopic perturbations considered here could account for more extreme cases, such as B2045+265, which shows significantly anomalous ratios (e.g., \citealt{Fassnacht1999_B2045, DalalKochanek2002_FACDMSub, Keeton2003_Cusp,  KochanekDalal2004_PropagationMacroM34, McKean2007_B2045, Xu2015}). In a parallel work (Feng et al., in prep), we shall address this question from a statistical perspective using a larger sample of mock systems.

As a final point of potentially comparable significance, we present a brief discussion on possible caveats from a number of macro-model aspects: a radial profile beyond a single power law and an external shear field beyond a constant $\gamma$. Indeed a number of studies from both observations and simulations suggest that a single power law may not always be a good approximation for early-type galaxies (e.g., \citealt{Xu2016_PLfromSim, Cao2022_EPLfail}) and that pure power-law macro-models may even induce false detections of low-mass dark matter haloes, leading to biased inferences on dark matter models (e.g., \citealt {He2023SubDetectMainTwoComp, Nightingale2024_Imaging_VPLs, ORiordanVegetti2024}) and on $H_0$ determinations (e.g.,  \citealt{Sonnenfeld2018_EPLH0bias, Millon2020_H0EPLComposite}). The adoption of a constant shear field through $\gamma$ is also a concern. It was originally introduced to account for a cumulative deflection effect from neighboring or sightline galaxies (e.g., \citealt{Keeton1997_ExternalShear, Wong2011_Shear}), but has been shown to mostly act as a fudge factor to compensate for macroscopic models lacking sufficient complexity (see \citealt{Sluse2012_AAduetoSatellite, Etherington2024Shear}). In addition, there have also been recent discussions on potential perturbation effects from the $m=1$ mode, which have also been implemented into model fitting pipelines (\citealt{PaugnatGilman25_EM, Keeley2025_JWST_Warm, Nierenberg2026_CDM_min}).

In order to utilize more sophisticated macro-models or model-free fitting approaches (e.g., \citealt{Liesenborgs2009NonParLens, Lefor2013LensModelReview, Lubini2014FreeModLens, Birrer2021_CAB, Paugnat2025_ModelFreeFA}), one has to introduce more observations to put effective constraints on the extra degree of freedom. This essentially calls for combined analyses using both point-image position and flux ratios, as well as surface brightness distributions of arcs when available (e.g., \citealt{Spingola2018, Birrer2021_CAB, Gilman2024_FAandArc, Gilman2025_JWST_FAandArc, Paugnat2025_ModelFreeFA, oh2026}). 
Future high-precision observations from adaptive optics will also play a crucial role in this aspect (see \citealt{Zelko2024_FuturePrecisionAO}).

\begin{acknowledgments}
We thank Dr. Conor O'Riordan for very constructive comments to improve the quality of this work. We acknowledge Dr.~Yan Liang for useful discussion on sample construction, and Wenshuo Xu, Xincheng Zhu and Dr.~Xiaoyue Cao on lens mock and model fitting strategies. We thank Drs.~Frederic Courbin, Jeremy Jin Leong Lim and Alfred Amruth for insightful discussion and comments. This work is supported by the General Program from National Natural Science Foundation of China (No.~12073013), the National Natural Science Foundation of China (Grant No.~12433003), and the China Manned Space Project (No.~CMS-CSST-2025-A10). XDD and FRZ thank the Dushi program and funding from Tsinghua University. We also acknowledge the Tsinghua Astrophysics High-Performance Computing platform at Tsinghua University for providing computational and data storage resources that have contributed to the research results reported in this study. 
\end{acknowledgments}

\begin{contribution}

RZF conducted the numerical experiments, wrote the software and administers the project github repositories, and was also responsible for writing the manuscript. DDX came up with the initial research concept, supervised the graduate RZF, was responsible for formulating the conclusion, writing and submitting the manuscript. DS was responsible for formulating key conclusions and editing the manuscript. GD, AS and CH edited the manuscript and provided key technical and data support to this work. 
\end{contribution}

%

\software{\texttt{astropy} \citep{astropy:2013, astropy:2018, astropy:2022}, \texttt{lenstronomy} \citep{lenstronomy01,lenstronomy02}, \texttt{matplotlib} \citep{Hunter:2007}, \texttt{numba} \citep{numbapaper}, \texttt{numpy} \citep{harris2020array}, \texttt{photutils} \citep{photutils2.3.0}, \texttt{python} \citep{python}, \texttt{scipy} \citep{2020SciPy-NMeth}}




\bibliography{references,software,tng}{}

@ARTICLE{Cao2022_EPLfail,
       author = {{Cao}, Xiaoyue and {Li}, Ran and {Nightingale}, J.~W. and {Massey}, Richard and {Robertson}, Andrew and {Frenk}, Carlos S. and {Amvrosiadis}, Aristeidis and {Amorisco}, Nicola C. and {He}, Qiuhan and {Etherington}, Amy and {Cole}, Shaun and {Zhu}, Kai},
    title = "{Systematic Errors Induced by the Elliptical Power-law model in Galaxy-Galaxy Strong Lens Modeling}",
      journal = {Research in Astronomy and Astrophysics},
    year = 2022,
    month = feb,
       volume = {22},
       number = {2},
     eid = {025014},
    pages = {025014},
    doi = {10.1088/1674-4527/ac3f2b},
archivePrefix = {arXiv},
       eprint = {2110.14554},
 primaryClass = {astro-ph.GA},
       adsurl = {https://ui.adsabs.harvard.edu/abs/2022RAA....22b5014C}
}

@ARTICLE{Zelko2024_FuturePrecisionAO,
       author = {{Zelko}, Ioana A. and {Nierenberg}, Anna M. and {Treu}, Tommaso},
    title = "{Probing dark matter with adaptive-optics based flux ratio anomalies: photometric and astrometric precision}",
      journal = {\mnras},
    year = 2024,
    month = jun,
    volume = {531},
    number = {1},
    pages = {885-897},
    doi = {10.1093/mnras/stae970},
archivePrefix = {arXiv},
       eprint = {2311.17140},
 primaryClass = {astro-ph.CO},
       adsurl = {https://ui.adsabs.harvard.edu/abs/2024MNRAS.531..885Z}
}

@ARTICLE{Birrer2021_CAB,
       author = {{Birrer}, Simon},
    title = "{Gravitational Lensing Formalism in a Curved Arc Basis: A Continuous Description of Observables and Degeneracies from the Weak to the Strong Lensing Regime}",
      journal = {\apj},
    year = 2021,
    month = sep,
       volume = {919},
       number = {1},
    eid = {38},
    pages = {38},
    doi = {10.3847/1538-4357/ac1108},
archivePrefix = {arXiv},
       eprint = {2104.09522},
 primaryClass = {astro-ph.CO},
       adsurl = {https://ui.adsabs.harvard.edu/abs/2021ApJ...919...38B}
}

@ARTICLE{Paugnat2025_ModelFreeFA,
       author = {{Paugnat}, Hadrien and {Treu}, Tommaso and {Gilman}, Daniel},
    title = "{Macromodel-free flux-ratio prediction in quadruply imaged quasars with local constraints from lensed arcs}",
      journal = {\prd},
    year = 2025,
    month = dec,
    volume = {112},
    number = {12},
    eid = {123002},
    pages = {123002},
    doi = {10.1103/y3j4-t8cl},
archivePrefix = {arXiv},
    eprint = {2509.05416},
 primaryClass = {astro-ph.CO},
       adsurl = {https://ui.adsabs.harvard.edu/abs/2025PhRvD.112l3002P}
}

@ARTICLE{Dike2023_PBH,
       author = {{Dike}, Veronica and {Gilman}, Daniel and {Treu}, Tommaso},
        title = "{Strong lensing constraints on primordial black holes as a dark matter candidate}",
      journal = {\mnras},
         year = 2023,
        month = jul,
       volume = {522},
       number = {4},
        pages = {5434-5441},
          doi = {10.1093/mnras/stad1313},
archivePrefix = {arXiv},
       eprint = {2210.09493},
 primaryClass = {astro-ph.CO},
       adsurl = {https://ui.adsabs.harvard.edu/abs/2023MNRAS.522.5434D}
}

@misc{hou2026fluxratioanomaliescuspquasars,
      author = {{Hou}, Siyuan and {Xiang}, Shucheng and {Sming Tsai}, Yue-Lin and {Yang}, Daneng and {Shu}, Yiping and {Li}, Nan and {Dong}, Jiang and {He}, Zizhao and {Li}, Guoliang and {Fan}, Yizhong},
        title = "{Flux-ratio anomalies in cusp quasars reveal dark matter beyond CDM}",
      journal = {arXiv e-prints},
         year = 2026,
        month = jan,
          eid = {arXiv:2601.16818},
        pages = {arXiv:2601.16818},
          doi = {10.48550/arXiv.2601.16818},
archivePrefix = {arXiv},
       eprint = {2601.16818},
 primaryClass = {astro-ph.CO},
       adsurl = {https://ui.adsabs.harvard.edu/abs/2026arXiv260116818H}
}

@article{Harvey_2019_WDM,
   title={Exploiting flux ratio anomalies to probe warm dark matter in future large-scale surveys},
   volume={491},
   ISSN={1365-2966},
   url={http://dx.doi.org/10.1093/mnras/stz3305},
   DOI={10.1093/mnras/stz3305},
   number={3},
   journal={\mnras},
   publisher={Oxford University Press (OUP)},
   author={Harvey, David and Valkenburg, Wessel and Tamone, Amelie and Boyarsky, Alexey and Courbin, Frederic and Lovell, Mark},
   year={2019},
   month=nov, pages={4247–4253} 
}

@ARTICLE{Gilman2023SIDM,
       author = {{Gilman}, Daniel and {Zhong}, Yi-Ming and {Bovy}, Jo},
        title = "{Constraining resonant dark matter self-interactions with strong gravitational lenses}",
      journal = {\prd},
         year = 2023,
        month = may,
       volume = {107},
       number = {10},
          eid = {103008},
        pages = {103008},
          doi = {10.1103/PhysRevD.107.103008},
archivePrefix = {arXiv},
       eprint = {2207.13111},
 primaryClass = {astro-ph.CO},
       adsurl = {https://ui.adsabs.harvard.edu/abs/2023PhRvD.107j3008G}
}

@ARTICLE{Gilman2021SIDM,
       author = {{Gilman}, Daniel and {Bovy}, Jo and {Treu}, Tommaso and {Nierenberg}, Anna and {Birrer}, Simon and {Benson}, Andrew and {Sameie}, Omid},
        title = "{Strong lensing signatures of self-interacting dark matter in low-mass haloes}",
      journal = {\mnras},
         year = 2021,
        month = oct,
       volume = {507},
       number = {2},
        pages = {2432-2447},
          doi = {10.1093/mnras/stab2335},
archivePrefix = {arXiv},
       eprint = {2105.05259},
 primaryClass = {astro-ph.CO},
       adsurl = {https://ui.adsabs.harvard.edu/abs/2021MNRAS.507.2432G}
}

@ARTICLE{Sonnenfeld2018_EPLH0bias,
       author = {{Sonnenfeld}, Alessandro},
        title = "{On the choice of lens density profile in time delay cosmography}",
      journal = {\mnras},
         year = 2018,
        month = mar,
       volume = {474},
       number = {4},
        pages = {4648-4659},
          doi = {10.1093/mnras/stx3105},
archivePrefix = {arXiv},
       eprint = {1710.05925},
 primaryClass = {astro-ph.CO},
       adsurl = {https://ui.adsabs.harvard.edu/abs/2018MNRAS.474.4648S}
}

@ARTICLE{Millon2020_H0EPLComposite,
       author = {{Millon}, M. and {Galan}, A. and {Courbin}, F. and {Treu}, T. and {Suyu}, S.~H. and {Ding}, X. and {Birrer}, S. and {Chen}, G.~C.-F. and {Shajib}, A.~J. and {Sluse}, D. and {Wong}, K.~C. and {Agnello}, A. and {Auger}, M.~W. and {Buckley-Geer}, E.~J. and {Chan}, J.~H.~H. and {Collett}, T. and {Fassnacht}, C.~D. and {Hilbert}, S. and {Koopmans}, L.~V.~E. and {Motta}, V. and {Mukherjee}, S. and {Rusu}, C.~E. and {Sonnenfeld}, A. and {Spiniello}, C. and {Van de Vyvere}, L.},
        title = "{TDCOSMO. I. An exploration of systematic uncertainties in the inference of H$_{0}$ from time-delay cosmography}",
      journal = {\aap},
         year = 2020,
        month = jul,
       volume = {639},
          eid = {A101},
        pages = {A101},
          doi = {10.1051/0004-6361/201937351},
archivePrefix = {arXiv},
       eprint = {1912.08027},
 primaryClass = {astro-ph.CO},
       adsurl = {https://ui.adsabs.harvard.edu/abs/2020A&A...639A.101M}
}

@ARTICLE{Kochanek2021,
       author = {{Kochanek}, C.~S.},
        title = "{Overconstrained models of time delay lenses redux: how the angular tail wags the radial dog}",
      journal = {\mnras},
         year = 2021,
        month = mar,
       volume = {501},
       number = {4},
        pages = {5021-5028},
          doi = {10.1093/mnras/staa4033},
archivePrefix = {arXiv},
       eprint = {2003.08395},
 primaryClass = {astro-ph.CO},
       adsurl = {https://ui.adsabs.harvard.edu/abs/2021MNRAS.501.5021K}
}

@ARTICLE{Lubini2014FreeModLens,
       author = {{Lubini}, M. and {Sereno}, M. and {Coles}, J. and {Jetzer}, Ph. and {Saha}, P.},
        title = "{Cosmological parameter determination in free-form strong gravitational lens modelling}",
      journal = {\mnras},
         year = 2014,
        month = jan,
       volume = {437},
       number = {3},
        pages = {2461-2470},
          doi = {10.1093/mnras/stt2057},
archivePrefix = {arXiv},
       eprint = {1310.7578},
 primaryClass = {astro-ph.CO},
       adsurl = {https://ui.adsabs.harvard.edu/abs/2014MNRAS.437.2461L}
}

@ARTICLE{Lefor2013LensModelReview,
       author = {{Lefor}, Alan T. and {Futamase}, Toshifumi and {Akhlaghi}, Mohammad},
        title = "{A systematic review of strong gravitational lens modeling software}",
      journal = {\nar},
         year = 2013,
        month = jul,
       volume = {57},
       number = {1-2},
        pages = {1-13},
          doi = {10.1016/j.newar.2013.05.001},
archivePrefix = {arXiv},
       eprint = {1206.4382},
 primaryClass = {astro-ph.IM},
       adsurl = {https://ui.adsabs.harvard.edu/abs/2013NewAR..57....1L}
}

@ARTICLE{Liesenborgs2009NonParLens,
       author = {{Liesenborgs}, J. and {de Rijcke}, S. and {Dejonghe}, H. and {Bekaert}, P.},
        title = "{Non-parametric strong lens inversion of SDSS J1004+4112}",
      journal = {\mnras},
         year = 2009,
        month = jul,
       volume = {397},
       number = {1},
        pages = {341-349},
          doi = {10.1111/j.1365-2966.2009.14912.x},
archivePrefix = {arXiv},
       eprint = {0904.2382},
 primaryClass = {astro-ph.CO},
       adsurl = {https://ui.adsabs.harvard.edu/abs/2009MNRAS.397..341L}
}

@ARTICLE{Schramm1994LensingSheet,
       author = {{Schramm}, T.},
        title = "{A toolbox for general elliptical gravitational lenses.}",
      journal = {\aap},
         year = 1994,
        month = apr,
       volume = {284},
        pages = {44-50},
          doi = {10.48550/arXiv.astro-ph/9311021},
archivePrefix = {arXiv},
       eprint = {astro-ph/9311021},
 primaryClass = {astro-ph},
       adsurl = {https://ui.adsabs.harvard.edu/abs/1994A&A...284...44S}
}

@ARTICLE{Michard1985CetrRndOutEllp,
       author = {{Michard}, R.},
        title = "{Detailed surface photometry of 36 E-SO galaxies.}",
      journal = {\aaps},
         year = 1985,
        month = feb,
       volume = {59},
        pages = {205-228},
       adsurl = {https://ui.adsabs.harvard.edu/abs/1985A&AS...59..205M}
}

@ARTICLE{Vigroux1988TwistEC,
       author = {{Vigroux}, L. and {Souviron}, J. and {Lachieze-Rey}, M. and {Vader}, J.~P.},
        title = "{3-color surface photometry of a selected sample of early-type galaxies. I. Observations and data reduction.}",
      journal = {\aaps},
         year = 1988,
        month = apr,
       volume = {73},
        pages = {1-36},
       adsurl = {https://ui.adsabs.harvard.edu/abs/1988A&AS...73....1V}
}

@ARTICLE{Jedrzejewski1987Discy,
       author = {{Jedrzejewski}, Robert I.},
        title = "{CCD surface photometry of elliptical galaxies - I. Observations, reduction and results.}",
      journal = {\mnras},
         year = 1987,
        month = jun,
       volume = {226},
        pages = {747-768},
          doi = {10.1093/mnras/226.4.747},
       adsurl = {https://ui.adsabs.harvard.edu/abs/1987MNRAS.226..747J}
}

@ARTICLE{Lauer1985BoxyDiscy,
       author = {{Lauer}, T.~R.},
        title = "{Boxy isophotes, discs and dust lanes in elliptical galaxies.}",
      journal = {\mnras},
         year = 1985,
        month = sep,
       volume = {216},
        pages = {429-438},
          doi = {10.1093/mnras/216.2.429},
       adsurl = {https://ui.adsabs.harvard.edu/abs/1985MNRAS.216..429L}
}

@ARTICLE{Fadely2010DC0957,
       author = {{Fadely}, R. and {Keeton}, C.~R. and {Nakajima}, R. and {Bernstein}, G.~M.},
        title = "{Improved Constraints on the Gravitational Lens Q0957+561. II. Strong Lensing}",
      journal = {\apj},
         year = 2010,
        month = mar,
       volume = {711},
       number = {1},
        pages = {246-267},
          doi = {10.1088/0004-637X/711/1/246},
archivePrefix = {arXiv},
       eprint = {0909.1807},
 primaryClass = {astro-ph.CO},
       adsurl = {https://ui.adsabs.harvard.edu/abs/2010ApJ...711..246F}
}

@ARTICLE{BernsteinFischer1999DPL0957,
       author = {{Bernstein}, Gary and {Fischer}, Philippe},
        title = "{Values of H\_0 from Models of the Gravitational Lens 0957+561}",
      journal = {\aj},
         year = 1999,
        month = jul,
       volume = {118},
       number = {1},
        pages = {14-34},
          doi = {10.1086/300949},
archivePrefix = {arXiv},
       eprint = {astro-ph/9903274},
 primaryClass = {astro-ph},
       adsurl = {https://ui.adsabs.harvard.edu/abs/1999AJ....118...14B}
}

@ARTICLE{Bernstein1997Host0957TwistEC,
       author = {{Bernstein}, Gary and {Fischer}, Philippe and {Tyson}, J. Anthony and {Rhee}, George},
        title = "{Improved Parameters and New Lensed Features for Q0957+561 from WFPC2 Imaging}",
      journal = {\apjl},
         year = 1997,
        month = jul,
       volume = {483},
       number = {2},
        pages = {L79-L82},
          doi = {10.1086/310747},
archivePrefix = {arXiv},
       eprint = {astro-ph/9705024},
 primaryClass = {astro-ph},
       adsurl = {https://ui.adsabs.harvard.edu/abs/1997ApJ...483L..79B}
}

@ARTICLE{Keeton2000Host0957TwistModel,
       author = {{Keeton}, C.~R. and {Falco}, E.~E. and {Impey}, C.~D. and {Kochanek}, C.~S. and {Leh{\'a}r}, J. and {McLeod}, B.~A. and {Rix}, H. -W. and {Mu{\~n}oz}, J.~A. and {Peng}, C.~Y.},
        title = "{The Host Galaxy of the Lensed Quasar Q0957+561}",
      journal = {\apj},
         year = 2000,
        month = oct,
       volume = {542},
       number = {1},
        pages = {74-93},
          doi = {10.1086/309517},
archivePrefix = {arXiv},
       eprint = {astro-ph/0001500},
 primaryClass = {astro-ph},
       adsurl = {https://ui.adsabs.harvard.edu/abs/2000ApJ...542...74K}
}

@ARTICLE{FasanoBonoli1989Isophotwist,
       author = {{Fasano}, G. and {Bonoli}, C.},
        title = "{Isophotal twisting in isolated elliptical galaxies.}",
      journal = {\aaps},
         year = 1989,
        month = aug,
       volume = {79},
        pages = {291-312},
       adsurl = {https://ui.adsabs.harvard.edu/abs/1989A&AS...79..291F}
}

@ARTICLE{Etherington2024Shear,
       author = {{Etherington}, Amy and {Nightingale}, James W. and {Massey}, Richard and {Tam}, Sut-Ieng and {Cao}, XiaoYue and {Niemiec}, Anna and {He}, Qiuhan and {Robertson}, Andrew and {Li}, Ran and {Amvrosiadis}, Aristeidis and {Cole}, Shaun and {Diego}, Jose M. and {Frenk}, Carlos S. and {Frye}, Brenda L. and {Harvey}, David and {Jauzac}, Mathilde and {Koekemoer}, Anton M. and {Lagattuta}, David J. and {Lange}, Samuel and {Limousin}, Marceau and {Mahler}, Guillaume and {Sirks}, Ellen and {Steinhardt}, Charles L.},
        title = "{Strong gravitational lensing's 'external shear' is not shear}",
      journal = {\mnras},
         year = 2024,
        month = jul,
       volume = {531},
       number = {3},
        pages = {3684-3697},
          doi = {10.1093/mnras/stae1375},
archivePrefix = {arXiv},
       eprint = {2301.05244},
 primaryClass = {astro-ph.CO},
       adsurl = {https://ui.adsabs.harvard.edu/abs/2024MNRAS.531.3684E}
}

@ARTICLE{InoueTakahashi2012LOS,
       author = {{Inoue}, Kaiki Taro and {Takahashi}, Ryuichi},
        title = "{Weak lensing by line-of-sight haloes as the origin of flux-ratio anomalies in quadruply lensed QSOs}",
      journal = {\mnras},
         year = 2012,
        month = nov,
       volume = {426},
       number = {4},
        pages = {2978-2993},
          doi = {10.1111/j.1365-2966.2012.21915.x},
archivePrefix = {arXiv},
       eprint = {1207.2139},
 primaryClass = {astro-ph.CO},
       adsurl = {https://ui.adsabs.harvard.edu/abs/2012MNRAS.426.2978I}
}

@ARTICLE{Amorisco2022,
       author = {{Amorisco}, Nicola C. and {Nightingale}, James and {He}, Qiuhan and {Amvrosiadis}, Aristeidis and {Cao}, Xiaoyue and {Cole}, Shaun and {Etherington}, Amy and {Frenk}, Carlos S. and {Li}, Ran and {Massey}, Richard and {Robertson}, Andrew},
        title = "{Halo concentration strengthens dark matter constraints in galaxy-galaxy strong lensing analyses}",
      journal = {\mnras},
         year = 2022,
        month = feb,
       volume = {510},
       number = {2},
        pages = {2464-2479},
          doi = {10.1093/mnras/stab3527},
archivePrefix = {arXiv},
       eprint = {2109.00018},
 primaryClass = {astro-ph.CO},
       adsurl = {https://ui.adsabs.harvard.edu/abs/2022MNRAS.510.2464A}
}

@ARTICLE{Despali2018LOS,
       author = {{Despali}, Giulia and {Vegetti}, Simona and {White}, Simon D.~M. and {Giocoli}, Carlo and {van den Bosch}, Frank C.},
        title = "{Modelling the line-of-sight contribution in substructure lensing}",
      journal = {\mnras},
         year = 2018,
        month = apr,
       volume = {475},
       number = {4},
        pages = {5424-5442},
          doi = {10.1093/mnras/sty159},
archivePrefix = {arXiv},
       eprint = {1710.05029},
 primaryClass = {astro-ph.CO},
       adsurl = {https://ui.adsabs.harvard.edu/abs/2018MNRAS.475.5424D}
}

@ARTICLE{Moller2003DiskAffectRatio,
       author = {{M{\"o}ller}, Ole and {Hewett}, Paul and {Blain}, A.~W.},
        title = "{Discs in early-type lensing galaxies: effects on magnification ratios and measurements of H$_{0}$}",
      journal = {\mnras},
         year = 2003,
        month = oct,
       volume = {345},
       number = {1},
        pages = {1-15},
          doi = {10.1046/j.1365-8711.2003.06758.x},
archivePrefix = {arXiv},
       eprint = {astro-ph/0212467},
 primaryClass = {astro-ph},
       adsurl = {https://ui.adsabs.harvard.edu/abs/2003MNRAS.345....1M}
}

@ARTICLE{Shajib2019UniMod13Quads,
       author = {{Shajib}, A.~J. and {Birrer}, S. and {Treu}, T. and {Auger}, M.~W. and {Agnello}, A. and {Anguita}, T. and {Buckley-Geer}, E.~J. and {Chan}, J.~H.~H. and {Collett}, T.~E. and {Courbin}, F. and {Fassnacht}, C.~D. and {Frieman}, J. and {Kayo}, I. and {Lemon}, C. and {Lin}, H. and {Marshall}, P.~J. and {McMahon}, R. and {More}, A. and {Morgan}, N.~D. and {Motta}, V. and {Oguri}, M. and {Ostrovski}, F. and {Rusu}, C.~E. and {Schechter}, P.~L. and {Shanks}, T. and {Suyu}, S.~H. and {Meylan}, G. and {Abbott}, T.~M.~C. and {Allam}, S. and {Annis}, J. and {Avila}, S. and {Bertin}, E. and {Brooks}, D. and {Carnero Rosell}, A. and {Carrasco Kind}, M. and {Carretero}, J. and {Cunha}, C.~E. and {da Costa}, L.~N. and {De Vicente}, J. and {Desai}, S. and {Doel}, P. and {Flaugher}, B. and {Fosalba}, P. and {Garc{\'\i}a-Bellido}, J. and {Gerdes}, D.~W. and {Gruen}, D. and {Gruendl}, R.~A. and {Gutierrez}, G. and {Hartley}, W.~G. and {Hollowood}, D.~L. and {Hoyle}, B. and {James}, D.~J. and {Kuehn}, K. and {Kuropatkin}, N. and {Lahav}, O. and {Lima}, M. and {Maia}, M.~A.~G. and {March}, M. and {Marshall}, J.~L. and {Melchior}, P. and {Menanteau}, F. and {Miquel}, R. and {Plazas}, A.~A. and {Sanchez}, E. and {Scarpine}, V. and {Sevilla-Noarbe}, I. and {Smith}, M. and {Soares-Santos}, M. and {Sobreira}, F. and {Suchyta}, E. and {Swanson}, M.~E.~C. and {Tarle}, G. and {Walker}, A.~R.},
        title = "{Is every strong lens model unhappy in its own way? Uniform modelling of a sample of 13 quadruply+ imaged quasars}",
      journal = {\mnras},
         year = 2019,
        month = mar,
       volume = {483},
       number = {4},
        pages = {5649-5671},
          doi = {10.1093/mnras/sty3397},
archivePrefix = {arXiv},
       eprint = {1807.09278},
 primaryClass = {astro-ph.GA},
       adsurl = {https://ui.adsabs.harvard.edu/abs/2019MNRAS.483.5649S}
}

@ARTICLE{Kormendy2009,
       author = {{Kormendy}, John and {Fisher}, David B. and {Cornell}, Mark E. and {Bender}, Ralf},
        title = "{Structure and Formation of Elliptical and Spheroidal Galaxies}",
      journal = {\apjs},
         year = 2009,
        month = may,
       volume = {182},
       number = {1},
        pages = {216-309},
          doi = {10.1088/0067-0049/182/1/216},
archivePrefix = {arXiv},
       eprint = {0810.1681},
 primaryClass = {astro-ph},
       adsurl = {https://ui.adsabs.harvard.edu/abs/2009ApJS..182..216K}
}

@ARTICLE{Bender1989,
       author = {{Bender}, R. and {Surma}, P. and {Doebereiner}, S. and {Moellenhoff}, C. and {Madejsky}, R.},
        title = "{Isophote shapes of elliptical galaxies. II. Correlations with global optical, radio and X-ray properties.}",
      journal = {\aap},
         year = 1989,
        month = jun,
       volume = {217},
        pages = {35-43},
       adsurl = {https://ui.adsabs.harvard.edu/abs/1989A&A...217...35B}
}

@ARTICLE{Bender1988,
       author = {{Bender}, R. and {Doebereiner}, S. and {Moellenhoff}, C.},
        title = "{Isophote shapes of elliptical galaxies. I. The data.}",
      journal = {\aaps},
         year = 1988,
        month = sep,
       volume = {74},
        pages = {385-426},
       adsurl = {https://ui.adsabs.harvard.edu/abs/1988A&AS...74..385B}
}

@ARTICLE{Mitsuda2017,
       author = {{Mitsuda}, Kazuma and {Doi}, Mamoru and {Morokuma}, Tomoki and {Suzuki}, Nao and {Yasuda}, Naoki and {Perlmutter}, Saul and {Aldering}, Greg and {Meyers}, Joshua},
        title = "{Isophote Shapes of Early-type Galaxies in Massive Clusters at z {\ensuremath{\sim}} 1 and 0}",
      journal = {\apj},
         year = 2017,
        month = jan,
       volume = {834},
       number = {2},
          eid = {109},
        pages = {109},
          doi = {10.3847/1538-4357/834/2/109},
archivePrefix = {arXiv},
       eprint = {1611.05870},
 primaryClass = {astro-ph.GA},
       adsurl = {https://ui.adsabs.harvard.edu/abs/2017ApJ...834..109M}
}

@ARTICLE{Krajnovic2013,
       author = {{Krajnovi{\'c}}, Davor and {Alatalo}, Katherine and {Blitz}, Leo and {Bois}, Maxime and {Bournaud}, Fr{\'e}d{\'e}ric and {Bureau}, Martin and {Cappellari}, Michele and {Davies}, Roger L. and {Davis}, Timothy A. and {de Zeeuw}, P.~T. and {Duc}, Pierre-Alain and {Emsellem}, Eric and {Khochfar}, Sadegh and {Kuntschner}, Harald and {McDermid}, Richard M. and {Morganti}, Raffaella and {Naab}, Thorsten and {Oosterloo}, Tom and {Sarzi}, Marc and {Scott}, Nicholas and {Serra}, Paolo and {Weijmans}, Anne-Marie and {Young}, Lisa M.},
        title = "{The ATLAS$^{3D}$ project - XVII. Linking photometric and kinematic signatures of stellar discs in early-type galaxies}",
      journal = {\mnras},
         year = 2013,
        month = jul,
       volume = {432},
       number = {3},
        pages = {1768-1795},
          doi = {10.1093/mnras/sts315},
archivePrefix = {arXiv},
       eprint = {1210.8167},
 primaryClass = {astro-ph.CO},
       adsurl = {https://ui.adsabs.harvard.edu/abs/2013MNRAS.432.1768K}
}

@ARTICLE{Pasquali2006,
       author = {{Pasquali}, A. and {Ferreras}, I. and {Panagia}, N. and {Daddi}, E. and {Malhotra}, S. and {Rhoads}, J.~E. and {Pirzkal}, N. and {Windhorst}, R.~A. and {Koekemoer}, A.~M. and {Moustakas}, L. and {Xu}, C. and {Gronwall}, C.},
        title = "{The Structure and Star Formation History of Early-Type Galaxies in the Ultra Deep Field/GRAPES Survey}",
      journal = {\apj},
         year = 2006,
        month = jan,
       volume = {636},
       number = {1},
        pages = {115-133},
          doi = {10.1086/497290},
archivePrefix = {arXiv},
       eprint = {astro-ph/0504264},
 primaryClass = {astro-ph},
       adsurl = {https://ui.adsabs.harvard.edu/abs/2006ApJ...636..115P}
}

@ARTICLE{Rest2001,
       author = {{Rest}, Armin and {van den Bosch}, Frank C. and {Jaffe}, Walter and {Tran}, Hien and {Tsvetanov}, Zlatan and {Ford}, Holland C. and {Davies}, James and {Schafer}, Joanna},
        title = "{WFPC2 Images of the Central Regions of Early-Type Galaxies. I. The Data}",
      journal = {\aj},
         year = 2001,
        month = may,
       volume = {121},
       number = {5},
        pages = {2431-2482},
          doi = {10.1086/320370},
archivePrefix = {arXiv},
       eprint = {astro-ph/0102286},
 primaryClass = {astro-ph},
       adsurl = {https://ui.adsabs.harvard.edu/abs/2001AJ....121.2431R}
}

@ARTICLE{Hao2006,
       author = {{Hao}, C.~N. and {Mao}, S. and {Deng}, Z.~G. and {Xia}, X.~Y. and {Wu}, Hong},
        title = "{Isophotal shapes of elliptical/lenticular galaxies from the Sloan Digital Sky Survey}",
      journal = {\mnras},
         year = 2006,
        month = aug,
       volume = {370},
       number = {3},
        pages = {1339-1350},
          doi = {10.1111/j.1365-2966.2006.10545.x},
archivePrefix = {arXiv},
       eprint = {astro-ph/0605319},
 primaryClass = {astro-ph},
       adsurl = {https://ui.adsabs.harvard.edu/abs/2006MNRAS.370.1339H}
}

@ARTICLE{Hao2006_Erratum,
       author = {{Hao}, C.~N. and {Mao}, S. and {Deng}, Z.~G. and {Xia}, X.~Y. and {Wu}, Hong},
        title = "{Erratum: Isophotal shapes of elliptical/lenticular galaxies from the Sloan Digital Sky Survey}",
      journal = {\mnras},
         year = 2006,
        month = dec,
       volume = {373},
       number = {3},
        pages = {1264-1264},
          doi = {10.1111/j.1365-2966.2006.11137.x},
       adsurl = {https://ui.adsabs.harvard.edu/abs/2006MNRAS.373.1264H}
}

@ARTICLE{He2023SubDetectMainTwoComp,
       author = {{He}, Qiuhan and {Nightingale}, James and {Robertson}, Andrew and {Amvrosiadis}, Aristeidis and {Cole}, Shaun and {Frenk}, Carlos S. and {Massey}, Richard and {Li}, Ran and {Amorisco}, Nicola C. and {Metcalf}, R. Benton and {Cao}, Xiaoyue and {Etherington}, Amy},
        title = "{Testing strong lensing subhalo detection with a cosmological simulation}",
      journal = {\mnras},
         year = 2023,
        month = jan,
       volume = {518},
       number = {1},
        pages = {220-239},
          doi = {10.1093/mnras/stac2779},
archivePrefix = {arXiv},
       eprint = {2202.10191},
 primaryClass = {astro-ph.CO},
       adsurl = {https://ui.adsabs.harvard.edu/abs/2023MNRAS.518..220H}
}

@ARTICLE{Gilman2024_FAandArc,
       author = {{Gilman}, Daniel and {Birrer}, Simon and {Nierenberg}, Anna and {Oh}, Maverick S.~H.},
        title = "{Turbocharging constraints on dark matter substructure through a synthesis of strong lensing flux ratios and extended lensed arcs}",
      journal = {\mnras},
         year = 2024,
        month = sep,
       volume = {533},
       number = {2},
        pages = {1687-1713},
          doi = {10.1093/mnras/stae1810},
archivePrefix = {arXiv},
       eprint = {2403.03253},
 primaryClass = {astro-ph.CO},
       adsurl = {https://ui.adsabs.harvard.edu/abs/2024MNRAS.533.1687G}
}

@ARTICLE{Stacey2024,
       author = {{Stacey}, H.~R. and {Powell}, D.~M. and {Vegetti}, S. and {McKean}, J.~P. and {Fassnacht}, C.~D. and {Wen}, D. and {O'Riordan}, C.~M.},
        title = "{Complex angular structure of three elliptical galaxies from high-resolution ALMA observations of strong gravitational lenses}",
      journal = {\aap},
         year = 2024,
        month = aug,
       volume = {688},
          eid = {A110},
        pages = {A110},
          doi = {10.1051/0004-6361/202449710},
archivePrefix = {arXiv},
       eprint = {2403.04850},
 primaryClass = {astro-ph.GA},
       adsurl = {https://ui.adsabs.harvard.edu/abs/2024A&A...688A.110S}
}

@ARTICLE{Fassnacht1999_B2045,
       author = {{Fassnacht}, C.~D. and {Blandford}, R.~D. and {Cohen}, J.~G. and {Matthews}, K. and {Pearson}, T.~J. and {Readhead}, A.~C.~S. and {Womble}, D.~S. and {Myers}, S.~T. and {Browne}, I.~W.~A. and {Jackson}, N.~J. and {Marlow}, D.~R. and {Wilkinson}, P.~N. and {Koopmans}, L.~V.~E. and {de Bruyn}, A.~G. and {Schilizzi}, R.~T. and {Bremer}, M. and {Miley}, G.},
        title = "{B2045+265: A New Four-Image Gravitational Lens from CLASS}",
      journal = {\aj},
         year = 1999,
        month = feb,
       volume = {117},
       number = {2},
        pages = {658-670},
          doi = {10.1086/300724},
archivePrefix = {arXiv},
       eprint = {astro-ph/9811167},
 primaryClass = {astro-ph},
       adsurl = {https://ui.adsabs.harvard.edu/abs/1999AJ....117..658F}
}

@ARTICLE{Cohen2024,
       author = {{Cohen}, Jacob S. and {Fassnacht}, Christopher D. and {O'Riordan}, Conor M. and {Vegetti}, Simona},
        title = "{General multipoles and their implications for dark matter inference}",
      journal = {\mnras},
         year = 2024,
        month = jul,
       volume = {531},
       number = {3},
        pages = {3431-3443},
          doi = {10.1093/mnras/stae1228},
archivePrefix = {arXiv},
       eprint = {2403.08895},
 primaryClass = {astro-ph.CO},
       adsurl = {https://ui.adsabs.harvard.edu/abs/2024MNRAS.531.3431C}
}

@ARTICLE{EvansWitt2003,
       author = {{Evans}, N. Wyn and {Witt}, Hans J.},
        title = "{Fitting gravitational lenses: truth or delusion}",
      journal = {\mnras},
         year = 2003,
        month = nov,
       volume = {345},
       number = {4},
        pages = {1351-1364},
          doi = {10.1046/j.1365-2966.2003.07057.x},
archivePrefix = {arXiv},
       eprint = {astro-ph/0212013},
 primaryClass = {astro-ph},
       adsurl = {https://ui.adsabs.harvard.edu/abs/2003MNRAS.345.1351E}
}

@ARTICLE{CongdonKeeton2005,
       author = {{Congdon}, Arthur B. and {Keeton}, Charles R.},
        title = "{Multipole models of four-image gravitational lenses with anomalous flux ratios}",
      journal = {\mnras},
         year = 2005,
        month = dec,
       volume = {364},
       number = {4},
        pages = {1459-1466},
          doi = {10.1111/j.1365-2966.2005.09699.x},
archivePrefix = {arXiv},
       eprint = {astro-ph/0510232},
 primaryClass = {astro-ph},
       adsurl = {https://ui.adsabs.harvard.edu/abs/2005MNRAS.364.1459C}
}

@ARTICLE{ORiordanVegetti2024,
       author = {{O'Riordan}, Conor M. and {Vegetti}, Simona},
        title = "{Angular complexity in strong lens substructure detection}",
      journal = {\mnras},
         year = 2024,
        month = feb,
       volume = {528},
       number = {2},
        pages = {1757-1768},
          doi = {10.1093/mnras/stae153},
archivePrefix = {arXiv},
       eprint = {2310.10714},
 primaryClass = {astro-ph.CO},
       adsurl = {https://ui.adsabs.harvard.edu/abs/2024MNRAS.528.1757O}
}

@ARTICLE{ORiordan2025_NGC6065,
       author = {{O'Riordan}, C.~M. and {Oldham}, L.~J. and {Nersesian}, A. and {Li}, T. and {Collett}, T.~E. and {Sluse}, D. and {Altieri}, B. and {Cl{\'e}ment}, B. and {Vasan}, K.~G.~C. and {Rhoades}, S. and {Chen}, Y. and {Jones}, T. and {Adami}, C. and {Gavazzi}, R. and {Vegetti}, S. and {Powell}, D.~M. and {Acevedo Barroso}, J.~A. and {Andika}, I.~T. and {Bhatawdekar}, R. and {Cooray}, A.~R. and {Despali}, G. and {Diego}, J.~M. and {Ecker}, L.~R. and {Galan}, A. and {G{\'o}mez-Alvarez}, P. and {Leuzzi}, L. and {Meneghetti}, M. and {Metcalf}, R.~B. and {Schirmer}, M. and {Serjeant}, S. and {Tortora}, C. and {Vaccari}, M. and {Vernardos}, G. and {Walmsley}, M. and {Amara}, A. and {Andreon}, S. and {Auricchio}, N. and {Aussel}, H. and {Baccigalupi}, C. and {Baldi}, M. and {Balestra}, A. and {Bardelli}, S. and {Basset}, A. and {Battaglia}, P. and {Bender}, R. and {Bonino}, D. and {Branchini}, E. and {Brescia}, M. and {Brinchmann}, J. and {Caillat}, A. and {Camera}, S. and {Capobianco}, V. and {Carbone}, C. and {Carretero}, J. and {Casas}, S. and {Castander}, F.~J. and {Castellano}, M. and {Castignani}, G. and {Cavuoti}, S. and {Cimatti}, A. and {Colodro-Conde}, C. and {Congedo}, G. and {Conselice}, C.~J. and {Conversi}, L. and {Copin}, Y. and {Corcione}, L. and {Courbin}, F. and {Courtois}, H.~M. and {Cropper}, M. and {Da Silva}, A. and {Degaudenzi}, H. and {De Lucia}, G. and {Di Giorgio}, A.~M. and {Dinis}, J. and {Dubath}, F. and {Duncan}, C.~A.~J. and {Dupac}, X. and {Dusini}, S. and {Farina}, M. and {Farrens}, S. and {Faustini}, F. and {Ferriol}, S. and {Fourmanoit}, N. and {Frailis}, M. and {Franceschi}, E. and {Fumana}, M. and {Galeotta}, S. and {Gillard}, W. and {Gillis}, B. and {Giocoli}, C. and {Granett}, B.~R. and {Grazian}, A. and {Grupp}, F. and {Guzzo}, L. and {Haugan}, S.~V.~H. and {Hoar}, J. and {Hoekstra}, H. and {Holmes}, W. and {Hook}, I. and {Hormuth}, F. and {Hornstrup}, A. and {Hudelot}, P. and {Jahnke}, K. and {Jhabvala}, M. and {Joachimi}, B. and {Keih{\"a}nen}, E. and {Kermiche}, S. and {Kiessling}, A. and {Kilbinger}, M. and {Kohley}, R. and {Kubik}, B. and {K{\"u}mmel}, M. and {Kunz}, M. and {Kurki-Suonio}, H. and {Lahav}, O. and {Laureijs}, R. and {Le Mignant}, D. and {Ligori}, S. and {Lilje}, P.~B. and {Lindholm}, V. and {Lloro}, I. and {Mainetti}, G. and {Maiorano}, E. and {Mansutti}, O. and {Marggraf}, O. and {Markovic}, K. and {Martinelli}, M. and {Martinet}, N. and {Marulli}, F. and {Massey}, R. and {Medinaceli}, E. and {Mei}, S. and {Melchior}, M. and {Mellier}, Y. and {Merlin}, E. and {Meylan}, G. and {Moresco}, M. and {Moscardini}, L. and {Nakajima}, R. and {Nichol}, R.~C. and {Niemi}, S.-M. and {Nightingale}, J.~W. and {Padilla}, C. and {Paltani}, S. and {Pasian}, F. and {Pedersen}, K. and {Percival}, W.~J. and {Pettorino}, V. and {Pires}, S. and {Polenta}, G. and {Poncet}, M. and {Popa}, L.~A. and {Pozzetti}, L. and {Raison}, F. and {Rebolo}, R. and {Renzi}, A. and {Rhodes}, J. and {Riccio}, G. and {Rix}, H.-W. and {Romelli}, E. and {Roncarelli}, M. and {Rossetti}, E. and {Rusholme}, B. and {Saglia}, R. and {Sakr}, Z. and {S{\'a}nchez}, A.~G. and {Sapone}, D. and {Sartoris}, B. and {Schneider}, P. and {Schrabback}, T. and {Secroun}, A. and {Seidel}, G. and {Serrano}, S. and {Sirignano}, C. and {Sirri}, G. and {Stanco}, L. and {Steinwagner}, J. and {Tallada-Cresp{\'\i}}, P. and {Tereno}, I. and {Toledo-Moreo}, R. and {Torradeflot}, F. and {Tutusaus}, I. and {Valenziano}, L. and {Vassallo}, T. and {Verdoes Kleijn}, G. and {Veropalumbo}, A. and {Wang}, Y. and {Weller}, J. and {Zacchei}, A. and {Zamorani}, G. and {Zucca}, E. and {Burigana}, C. and {Casenove}, P. and {Mora}, A. and {Scottez}, V. and {Viel}, M. and {Jauzac}, M. and {Dannerbauer}, H.},
        title = "{Euclid: A complete Einstein ring in NGC 6505}",
      journal = {\aap},
         year = 2025,
        month = feb,
       volume = {694},
          eid = {A145},
        pages = {A145},
          doi = {10.1051/0004-6361/202453014},
archivePrefix = {arXiv},
       eprint = {2502.06505},
 primaryClass = {astro-ph.GA},
       adsurl = {https://ui.adsabs.harvard.edu/abs/2025A&A...694A.145O}
}

@ARTICLE{SchneiderWeiss1992,
       author = {{Schneider}, P. and {Weiss}, A.},
        title = "{The gravitational lens equation near cusps}",
      journal = {\aap},
         year = 1992,
        month = jul,
       volume = {260},
       number = {1-2},
        pages = {1-13},
       adsurl = {https://ui.adsabs.harvard.edu/abs/1992A&A...260....1S}
}

@ARTICLE{Zakharov1995,
       author = {{Zakharov}, A.~F.},
        title = "{On the magnification of gravitational lens images near cusps.}",
      journal = {\aap},
         year = 1995,
        month = jan,
       volume = {293},
        pages = {1-4},
       adsurl = {https://ui.adsabs.harvard.edu/abs/1995A&A...293....1Z}
}

@ARTICLE{BlandfordNarayan1986,
       author = {{Blandford}, Roger and {Narayan}, Ramesh},
        title = "{Fermat's Principle, Caustics, and the Classification of Gravitational Lens Images}",
      journal = {\apj},
         year = 1986,
        month = nov,
       volume = {310},
        pages = {568},
          doi = {10.1086/164709},
       adsurl = {https://ui.adsabs.harvard.edu/abs/1986ApJ...310..568B}
}

@ARTICLE{Schechter_Wambsganss_2002_Parity,
       author = {{Schechter}, Paul L. and {Wambsganss}, Joachim},
        title = "{Quasar Microlensing at High Magnification and the Role of Dark Matter: Enhanced Fluctuations and Suppressed Saddle Points}",
      journal = {\apj},
         year = 2002,
        month = dec,
       volume = {580},
       number = {2},
        pages = {685-695},
          doi = {10.1086/343856},
archivePrefix = {arXiv},
       eprint = {astro-ph/0204425},
 primaryClass = {astro-ph},
       adsurl = {https://ui.adsabs.harvard.edu/abs/2002ApJ...580..685S}
}

@ARTICLE{Jackson2000_JVASCLASS,
       author = {{Jackson}, N. and {Xanthopoulos}, E. and {Browne}, I.~W.~A.},
        title = "{NICMOS images of JVAS/CLASS gravitational lens systems}",
      journal = {\mnras},
         year = 2000,
        month = jan,
       volume = {311},
       number = {2},
        pages = {389-396},
          doi = {10.1046/j.1365-8711.2000.03115.x},
archivePrefix = {arXiv},
       eprint = {astro-ph/9909474},
 primaryClass = {astro-ph},
       adsurl = {https://ui.adsabs.harvard.edu/abs/2000MNRAS.311..389J}
}

@ARTICLE{Koopmans2003_JVASCLASS,
       author = {{Koopmans}, L.~V.~E. and {Biggs}, A. and {Blandford}, R.~D. and {Browne}, I.~W.~A. and {Jackson}, N.~J. and {Mao}, S. and {Wilkinson}, P.~N. and {de Bruyn}, A.~G. and {Wambsganss}, J.},
        title = "{Extrinsic Radio Variability of JVAS/CLASS Gravitational Lenses}",
      journal = {\apj},
         year = 2003,
        month = oct,
       volume = {595},
       number = {2},
        pages = {712-718},
          doi = {10.1086/377434},
archivePrefix = {arXiv},
       eprint = {astro-ph/0302189},
 primaryClass = {astro-ph},
       adsurl = {https://ui.adsabs.harvard.edu/abs/2003ApJ...595..712K}
}

@ARTICLE{Metcalf2012_FARcuspSimSys,
       author = {{Metcalf}, R. Benton and {Amara}, Adam},
        title = "{Small-scale structures of dark matter and flux anomalies in quasar gravitational lenses}",
      journal = {\mnras},
         year = 2012,
        month = feb,
       volume = {419},
       number = {4},
        pages = {3414-3425},
          doi = {10.1111/j.1365-2966.2011.19982.x},
archivePrefix = {arXiv},
       eprint = {1007.1599},
 primaryClass = {astro-ph.CO},
       adsurl = {https://ui.adsabs.harvard.edu/abs/2012MNRAS.419.3414M}
}

@ARTICLE{Sluse2012_AAduetoSatellite,
       author = {{Sluse}, D. and {Chantry}, V. and {Magain}, P. and {Courbin}, F. and {Meylan}, G.},
        title = "{COSMOGRAIL: the COSmological MOnitoring of GRAvItational Lenses. X. Modeling based on high-precision astrometry of a sample of 25 lensed quasars: consequences for ellipticity, shear, and astrometric anomalies}",
      journal = {\aap},
         year = 2012,
        month = feb,
       volume = {538},
          eid = {A99},
        pages = {A99},
          doi = {10.1051/0004-6361/201015844},
archivePrefix = {arXiv},
       eprint = {1112.0005},
 primaryClass = {astro-ph.CO},
       adsurl = {https://ui.adsabs.harvard.edu/abs/2012A&A...538A..99S}
}

@ARTICLE{Keeton2003_Cusp,
       author = {{Keeton}, Charles R. and {Gaudi}, B. Scott and {Petters}, A.~O.},
        title = "{Identifying Lenses with Small-Scale Structure. I. Cusp Lenses}",
      journal = {\apj},
         year = 2003,
        month = nov,
       volume = {598},
       number = {1},
        pages = {138-161},
          doi = {10.1086/378934},
archivePrefix = {arXiv},
       eprint = {astro-ph/0210318},
 primaryClass = {astro-ph},
       adsurl = {https://ui.adsabs.harvard.edu/abs/2003ApJ...598..138K}
}

@ARTICLE{Keeton2005_fold,
       author = {{Keeton}, Charles R. and {Gaudi}, B. Scott and {Petters}, A.~O.},
        title = "{Identifying Lenses with Small-Scale Structure. II. Fold Lenses}",
      journal = {\apj},
         year = 2005,
        month = dec,
       volume = {635},
       number = {1},
        pages = {35-59},
          doi = {10.1086/497324},
archivePrefix = {arXiv},
       eprint = {astro-ph/0503452},
 primaryClass = {astro-ph},
       adsurl = {https://ui.adsabs.harvard.edu/abs/2005ApJ...635...35K}
}

@ARTICLE{MetcalfMadau2001_SimCDM,
       author = {{Metcalf}, R. Benton and {Madau}, Piero},
        title = "{Compound Gravitational Lensing as a Probe of Dark Matter Substructure within Galaxy Halos}",
      journal = {\apj},
         year = 2001,
        month = dec,
       volume = {563},
       number = {1},
        pages = {9-20},
          doi = {10.1086/323695},
archivePrefix = {arXiv},
       eprint = {astro-ph/0108224},
 primaryClass = {astro-ph},
       adsurl = {https://ui.adsabs.harvard.edu/abs/2001ApJ...563....9M}
}

@ARTICLE{Metacalf2002_FRDsub,
       author = {{Metcalf}, R. Benton and {Zhao}, HongSheng},
        title = "{Flux Ratios as a Probe of Dark Substructures in Quadruple-Image Gravitational Lenses}",
      journal = {\apjl},
         year = 2002,
        month = mar,
       volume = {567},
       number = {1},
        pages = {L5-L8},
          doi = {10.1086/339798},
archivePrefix = {arXiv},
       eprint = {astro-ph/0111427},
 primaryClass = {astro-ph},
       adsurl = {https://ui.adsabs.harvard.edu/abs/2002ApJ...567L...5M}
}

@ARTICLE{Wong2011_Shear,
       author = {{Wong}, Kenneth C. and {Keeton}, Charles R. and {Williams}, Kurtis A. and {Momcheva}, Ivelina G. and {Zabludoff}, Ann I.},
        title = "{The Effect of Environment on Shear in Strong Gravitational Lenses}",
      journal = {\apj},
         year = 2011,
        month = jan,
       volume = {726},
       number = {2},
          eid = {84},
        pages = {84},
          doi = {10.1088/0004-637X/726/2/84},
archivePrefix = {arXiv},
       eprint = {1011.2504},
 primaryClass = {astro-ph.CO},
       adsurl = {https://ui.adsabs.harvard.edu/abs/2011ApJ...726...84W}
}

@ARTICLE{Keeton1997_ExternalShear,
       author = {{Keeton}, C.~R. and {Kochanek}, C.~S. and {Seljak}, U.},
        title = "{Shear and Ellipticity in Gravitational Lenses}",
      journal = {\apj},
         year = 1997,
        month = jun,
       volume = {482},
       number = {2},
        pages = {604-620},
          doi = {10.1086/304172},
archivePrefix = {arXiv},
       eprint = {astro-ph/9610163},
 primaryClass = {astro-ph},
       adsurl = {https://ui.adsabs.harvard.edu/abs/1997ApJ...482..604K}
}

@ARTICLE{DalalKochanek2002_FACDMSub,
       author = {{Dalal}, N. and {Kochanek}, C.~S.},
        title = "{Direct Detection of Cold Dark Matter Substructure}",
      journal = {\apj},
         year = 2002,
        month = jun,
       volume = {572},
       number = {1},
        pages = {25-33},
          doi = {10.1086/340303},
archivePrefix = {arXiv},
       eprint = {astro-ph/0111456},
 primaryClass = {astro-ph},
       adsurl = {https://ui.adsabs.harvard.edu/abs/2002ApJ...572...25D}
}

@ARTICLE{KochanekDalal2004_PropagationMacroM34,
       author = {{Kochanek}, C.~S. and {Dalal}, N.},
        title = "{Tests for Substructure in Gravitational Lenses}",
      journal = {\apj},
         year = 2004,
        month = jul,
       volume = {610},
       number = {1},
        pages = {69-79},
          doi = {10.1086/421436},
archivePrefix = {arXiv},
       eprint = {astro-ph/0302036},
 primaryClass = {astro-ph},
       adsurl = {https://ui.adsabs.harvard.edu/abs/2004ApJ...610...69K}
}

@ARTICLE{VdV2022_LackAF,
       author = {{Van de Vyvere}, Lyne and {Sluse}, Dominique and {Gomer}, Matthew R. and {Mukherjee}, Sampath},
        title = "{Consequences of the lack of azimuthal freedom in the modeling of lensing galaxies}",
      journal = {\aap},
         year = 2022,
        month = jul,
       volume = {663},
          eid = {A179},
        pages = {A179},
          doi = {10.1051/0004-6361/202243382},
archivePrefix = {arXiv},
       eprint = {2206.00022},
 primaryClass = {astro-ph.CO},
       adsurl = {https://ui.adsabs.harvard.edu/abs/2022A&A...663A.179V}
}

@ARTICLE{Nierenberg2024_JWST_WarmDustFA,
       author = {{Nierenberg}, A.~M. and {Keeley}, R.~E. and {Sluse}, D. and {Gilman}, D. and {Birrer}, S. and {Treu}, T. and {Abazajian}, K.~N. and {Anguita}, T. and {Benson}, A.~J. and {Bennert}, V.~N. and {Djorgovski}, S.~G. and {Du}, X. and {Fassnacht}, C.~D. and {Hoenig}, S.~F. and {Kusenko}, A. and {Lemon}, C. and {Malkan}, M. and {Motta}, V. and {Moustakas}, L.~A. and {Stern}, D. and {Wechsler}, R.~H.},
        title = "{JWST lensed quasar dark matter survey - I. Description and first results}",
      journal = {\mnras},
         year = 2024,
        month = may,
       volume = {530},
       number = {3},
        pages = {2960-2971},
          doi = {10.1093/mnras/stae499},
archivePrefix = {arXiv},
       eprint = {2309.10101},
 primaryClass = {astro-ph.CO},
       adsurl = {https://ui.adsabs.harvard.edu/abs/2024MNRAS.530.2960N}
}

@ARTICLE{Schmidt2023STRIDES,
       author = {{Schmidt}, T. and {Treu}, T. and {Birrer}, S. and {Shajib}, A.~J. and {Lemon}, C. and {Millon}, M. and {Sluse}, D. and {Agnello}, A. and {Anguita}, T. and {Auger-Williams}, M.~W. and {McMahon}, R.~G. and {Motta}, V. and {Schechter}, P. and {Spiniello}, C. and {Kayo}, I. and {Courbin}, F. and {Ertl}, S. and {Fassnacht}, C.~D. and {Frieman}, J.~A. and {More}, A. and {Schuldt}, S. and {Suyu}, S.~H. and {Aguena}, M. and {Andrade-Oliveira}, F. and {Annis}, J. and {Bacon}, D. and {Bertin}, E. and {Brooks}, D. and {Burke}, D.~L. and {Carnero Rosell}, A. and {Carrasco Kind}, M. and {Carretero}, J. and {Conselice}, C. and {Costanzi}, M. and {da Costa}, L.~N. and {Pereira}, M.~E.~S. and {De Vicente}, J. and {Desai}, S. and {Doel}, P. and {Everett}, S. and {Ferrero}, I. and {Friedel}, D. and {Garc{\'\i}a-Bellido}, J. and {Gaztanaga}, E. and {Gruen}, D. and {Gruendl}, R.~A. and {Gschwend}, J. and {Gutierrez}, G. and {Hinton}, S.~R. and {Hollowood}, D.~L. and {Honscheid}, K. and {James}, D.~J. and {Kuehn}, K. and {Lahav}, O. and {Menanteau}, F. and {Miquel}, R. and {Palmese}, A. and {Paz-Chinch{\'o}n}, F. and {Pieres}, A. and {Plazas Malag{\'o}n}, A.~A. and {Prat}, J. and {Rodriguez-Monroy}, M. and {Romer}, A.~K. and {Sanchez}, E. and {Scarpine}, V. and {Sevilla-Noarbe}, I. and {Smith}, M. and {Suchyta}, E. and {Tarle}, G. and {To}, C. and {Varga}, T.~N. and {DES Collaboration}},
        title = "{STRIDES: automated uniform models for 30 quadruply imaged quasars}",
      journal = {\mnras},
         year = 2023,
        month = jan,
       volume = {518},
       number = {1},
        pages = {1260-1300},
          doi = {10.1093/mnras/stac2235},
archivePrefix = {arXiv},
       eprint = {2206.04696},
 primaryClass = {astro-ph.CO},
       adsurl = {https://ui.adsabs.harvard.edu/abs/2023MNRAS.518.1260S}
}

@ARTICLE{Keeley2023_mixCDMWDM,
       author = {{Keeley}, Ryan E. and {Nierenberg}, Anna M. and {Gilman}, Daniel and {Birrer}, Simon and {Benson}, Andrew and {Treu}, Tommaso},
        title = "{Pushing the limits of detectability: mixed dark matter from strong gravitational lenses}",
      journal = {\mnras},
         year = 2023,
        month = oct,
       volume = {524},
       number = {4},
        pages = {6159-6166},
          doi = {10.1093/mnras/stad2251},
archivePrefix = {arXiv},
       eprint = {2301.07265},
 primaryClass = {astro-ph.CO},
       adsurl = {https://ui.adsabs.harvard.edu/abs/2023MNRAS.524.6159K}
}

@ARTICLE{Keeley2024_JWST_WarmDustFA,
       author = {{Keeley}, Ryan E. and {Nierenberg}, A.~M. and {Gilman}, D. and {Gannon}, C. and {Birrer}, S. and {Treu}, T. and {Benson}, A.~J. and {Du}, X. and {Abazajian}, K.~N. and {Anguita}, T. and {Bennert}, V.~N. and {Djorgovski}, S.~G. and {Gupta}, K.~K. and {Hoenig}, S.~F. and {Kusenko}, A. and {Lemon}, C. and {Malkan}, M. and {Motta}, V. and {Moustakas}, L.~A. and {Oh}, Maverick S.~H. and {Sluse}, D. and {Stern}, D. and {Wechsler}, R.~H.},
        title = "{JWST lensed quasar dark matter survey - II. Strongest gravitational lensing limit on the dark matter free streaming length to date}",
      journal = {\mnras},
         year = 2024,
        month = dec,
       volume = {535},
       number = {2},
        pages = {1652-1671},
          doi = {10.1093/mnras/stae2458},
archivePrefix = {arXiv},
       eprint = {2405.01620},
 primaryClass = {astro-ph.CO},
       adsurl = {https://ui.adsabs.harvard.edu/abs/2024MNRAS.535.1652K}
}

@ARTICLE{Keeley2025_JWST_Warm,
       author = {{Keeley}, R.~E. and {Nierenberg}, A.~M. and {Gilman}, D. and {Treu}, T. and {Du}, X. and {Gannon}, C. and {Mozumdar}, P. and {Wong}, K.~C. and {Paugnat}, H. and {Birrer}, S. and {Malkan}, M. and {Benson}, A.~J. and {Abazajian}, K.~N. and {Anguita}, T. and {Bennert}, V.~N. and {Djorgovski}, S.~G. and {Hoenig}, S.~F. and {Kusenko}, A. and {Larsson}, H.~R. and {Morishita}, T. and {Motta}, V. and {Moustakas}, L.~A. and {Sheu}, W. and {Sluse}, D. and {Stern}, D. and {Stiavelli}, M. and {Williams}, D.},
        title = "{JWST Lensed Quasar Dark Matter Survey III: Dark Matter Sensitive Flux Ratios and Warm Dark Matter Constraint from the Full Sample}",
      journal = {arXiv e-prints},
         year = 2025,
        month = nov,
          eid = {arXiv:2511.07765},
        pages = {arXiv:2511.07765},
          doi = {10.48550/arXiv.2511.07765},
archivePrefix = {arXiv},
       eprint = {2511.07765},
 primaryClass = {astro-ph.CO},
       adsurl = {https://ui.adsabs.harvard.edu/abs/2025arXiv251107765K}
}

@ARTICLE{Chen2003_FALOS,
       author = {{Chen}, Jacqueline and {Kravtsov}, Andrey V. and {Keeton}, Charles R.},
        title = "{Lensing Optical Depths for Substructure and Isolated Dark Matter Halos}",
      journal = {\apj},
         year = 2003,
        month = jul,
       volume = {592},
       number = {1},
        pages = {24-31},
          doi = {10.1086/375639},
archivePrefix = {arXiv},
       eprint = {astro-ph/0302005},
 primaryClass = {astro-ph},
       adsurl = {https://ui.adsabs.harvard.edu/abs/2003ApJ...592...24C}
}

@ARTICLE{Inoue2016_FAlos,
       author = {{Inoue}, Kaiki Taro},
        title = "{On the origin of the flux ratio anomaly in quadruple lens systems}",
      journal = {\mnras},
         year = 2016,
        month = sep,
       volume = {461},
       number = {1},
        pages = {164-175},
          doi = {10.1093/mnras/stw1270},
archivePrefix = {arXiv},
       eprint = {1601.04414},
 primaryClass = {astro-ph.CO},
       adsurl = {https://ui.adsabs.harvard.edu/abs/2016MNRAS.461..164I}
}

@ARTICLE{Chan2020_FRFuzzyDM,
       author = {{Chan}, James H.~H. and {Schive}, Hsi-Yu and {Wong}, Shing-Kwong and {Chiueh}, Tzihong and {Broadhurst}, Tom},
        title = "{Multiple Images and Flux Ratio Anomaly of Fuzzy Gravitational Lenses}",
      journal = {\prl},
         year = 2020,
        month = sep,
       volume = {125},
       number = {11},
          eid = {111102},
        pages = {111102},
          doi = {10.1103/PhysRevLett.125.111102},
archivePrefix = {arXiv},
       eprint = {2002.10473},
 primaryClass = {astro-ph.GA},
       adsurl = {https://ui.adsabs.harvard.edu/abs/2020PhRvL.125k1102C}
}

@ARTICLE{Kamada2016PRD_FRWDM,
       author = {{Kamada}, Ayuki and {Inoue}, Kaiki Taro and {Takahashi}, Tomo},
        title = "{Constraints on mixed dark matter from anomalous strong lens systems}",
      journal = {\prd},
         year = 2016,
        month = jul,
       volume = {94},
       number = {2},
          eid = {023522},
        pages = {023522},
          doi = {10.1103/PhysRevD.94.023522},
archivePrefix = {arXiv},
       eprint = {1604.01489},
 primaryClass = {astro-ph.CO},
       adsurl = {https://ui.adsabs.harvard.edu/abs/2016PhRvD..94b3522K}
}

@ARTICLE{Inoue2015_FRWDM,
       author = {{Inoue}, Kaiki Taro and {Takahashi}, Ryuichi and {Takahashi}, Tomo and {Ishiyama}, Tomoaki},
        title = "{Constraints on warm dark matter from weak lensing in anomalous quadruple lenses}",
      journal = {\mnras},
         year = 2015,
        month = apr,
       volume = {448},
       number = {3},
        pages = {2704-2716},
          doi = {10.1093/mnras/stv194},
archivePrefix = {arXiv},
       eprint = {1409.1326},
 primaryClass = {astro-ph.CO},
       adsurl = {https://ui.adsabs.harvard.edu/abs/2015MNRAS.448.2704I}
}

@ARTICLE{Gilman2020_CDMMC,
       author = {{Gilman}, Daniel and {Du}, Xiaolong and {Benson}, Andrew and {Birrer}, Simon and {Nierenberg}, Anna and {Treu}, Tommaso},
        title = "{Constraints on the mass-concentration relation of cold dark matter halos with 11 strong gravitational lenses}",
      journal = {\mnras},
         year = 2020,
        month = feb,
       volume = {492},
       number = {1},
        pages = {L12-L16},
          doi = {10.1093/mnrasl/slz173},
archivePrefix = {arXiv},
       eprint = {1909.02573},
 primaryClass = {astro-ph.CO},
       adsurl = {https://ui.adsabs.harvard.edu/abs/2020MNRAS.492L..12G}
}

@ARTICLE{Nierenberg2026_CDM_min,
       author = {{Nierenberg}, A.~M. and {Gilman}, D. and {Treu}, T. and {Du}, X. and {Gannon}, C. and {Paugnat}, H. and {Birrer}, S. and {Benson}, A.~J. and {Abazajian}, K.~N. and {Anguita}, T. and {Djorgovski}, S.~G. and {Hoenig}, S.~F. and {Keeley}, R.~E. and {Kusenko}, A. and {Larsson}, H.~R. and {Moustakas}, L.~A. and {Mozumdar}, P. and {Sheu}, W. and {Sluse}, D. and {Stern}, D. and {Williams}, D. and {Wong}, K.~C.},
        title = "{JWST Lensed Quasar Dark Matter Survey V: Measuring the minimum halo mass with strong gravitational lensing}",
      journal = {arXiv e-prints},
         year = 2026,
        month = apr,
          eid = {arXiv:2604.05237},
        pages = {arXiv:2604.05237},
          doi = {10.48550/arXiv.2604.05237},
archivePrefix = {arXiv},
       eprint = {2604.05237},
 primaryClass = {astro-ph.CO},
       adsurl = {https://ui.adsabs.harvard.edu/abs/2026arXiv260405237N}
}

@ARTICLE{Nierenberg2020_NLR_WFC3grism,
       author = {{Nierenberg}, A.~M. and {Gilman}, D. and {Treu}, T. and {Brammer}, G. and {Birrer}, S. and {Moustakas}, L. and {Agnello}, A. and {Anguita}, T. and {Fassnacht}, C.~D. and {Motta}, V. and {Peter}, A.~H.~G. and {Sluse}, D.},
        title = "{Double dark matter vision: twice the number of compact-source lenses with narrow-line lensing and the WFC3 grism}",
      journal = {\mnras},
         year = 2020,
        month = mar,
       volume = {492},
       number = {4},
        pages = {5314-5335},
          doi = {10.1093/mnras/stz3588},
archivePrefix = {arXiv},
       eprint = {1908.06344},
 primaryClass = {astro-ph.GA},
       adsurl = {https://ui.adsabs.harvard.edu/abs/2020MNRAS.492.5314N}
}

@ARTICLE{Gilman2019_WD_withLOS,
       author = {{Gilman}, Daniel and {Birrer}, Simon and {Treu}, Tommaso and {Nierenberg}, Anna and {Benson}, Andrew},
        title = "{Probing dark matter structure down to {}10$^{7}$ solar masses: flux ratio statistics in gravitational lenses with line-of-sight haloes}",
      journal = {\mnras},
         year = 2019,
        month = aug,
       volume = {487},
       number = {4},
        pages = {5721-5738},
          doi = {10.1093/mnras/stz1593},
archivePrefix = {arXiv},
       eprint = {1901.11031},
 primaryClass = {astro-ph.CO},
       adsurl = {https://ui.adsabs.harvard.edu/abs/2019MNRAS.487.5721G}
}

@ARTICLE{Hsueh2020_SHARP-VII_DMFS,
       author = {{Hsueh}, J. -W. and {Enzi}, W. and {Vegetti}, S. and {Auger}, M.~W. and {Fassnacht}, C.~D. and {Despali}, G. and {Koopmans}, L.~V.~E. and {McKean}, J.~P.},
        title = "{SHARP - VII. New constraints on the dark matter free-streaming properties and substructure abundance from gravitationally lensed quasars}",
      journal = {\mnras},
         year = 2020,
        month = feb,
       volume = {492},
       number = {2},
        pages = {3047-3059},
          doi = {10.1093/mnras/stz3177},
archivePrefix = {arXiv},
       eprint = {1905.04182},
 primaryClass = {astro-ph.CO},
       adsurl = {https://ui.adsabs.harvard.edu/abs/2020MNRAS.492.3047H}
}

@ARTICLE{Gilman2018_FR_DMNature,
       author = {{Gilman}, Daniel and {Birrer}, Simon and {Treu}, Tommaso and {Keeton}, Charles R. and {Nierenberg}, Anna},
        title = "{Probing the nature of dark matter by forward modelling flux ratios in strong gravitational lenses}",
      journal = {\mnras},
         year = 2018,
        month = nov,
       volume = {481},
       number = {1},
        pages = {819-834},
          doi = {10.1093/mnras/sty2261},
archivePrefix = {arXiv},
       eprint = {1712.04945},
 primaryClass = {astro-ph.CO},
       adsurl = {https://ui.adsabs.harvard.edu/abs/2018MNRAS.481..819G}
}

@ARTICLE{Gilman2020_WDM,
       author = {{Gilman}, Daniel and {Birrer}, Simon and {Nierenberg}, Anna and {Treu}, Tommaso and {Du}, Xiaolong and {Benson}, Andrew},
        title = "{Warm dark matter chills out: constraints on the halo mass function and the free-streaming length of dark matter with eight quadruple-image strong gravitational lenses}",
      journal = {\mnras},
         year = 2020,
        month = feb,
       volume = {491},
       number = {4},
        pages = {6077-6101},
          doi = {10.1093/mnras/stz3480},
archivePrefix = {arXiv},
       eprint = {1908.06983},
 primaryClass = {astro-ph.CO},
       adsurl = {https://ui.adsabs.harvard.edu/abs/2020MNRAS.491.6077G}
}

@ARTICLE{Gilman2017,
       author = {{Gilman}, Daniel and {Agnello}, Adriano and {Treu}, Tommaso and {Keeton}, Charles R. and {Nierenberg}, Anna M.},
        title = "{Strong lensing signatures of luminous structure and substructure in early-type galaxies}",
      journal = {\mnras},
         year = 2017,
        month = jun,
       volume = {467},
       number = {4},
        pages = {3970-3992},
          doi = {10.1093/mnras/stx158},
archivePrefix = {arXiv},
       eprint = {1610.08525},
 primaryClass = {astro-ph.CO},
       adsurl = {https://ui.adsabs.harvard.edu/abs/2017MNRAS.467.3970G}
}

@ARTICLE{AlfredLim2023Nat,
       author = {{Amruth}, Alfred and {Broadhurst}, Tom and {Lim}, Jeremy and {Oguri}, Masamune and {Smoot}, George F. and {Diego}, Jose M. and {Leung}, Enoch and {Emami}, Razieh and {Li}, Juno and {Chiueh}, Tzihong and {Schive}, Hsi-Yu and {Yeung}, Michael C.~H. and {Li}, Sung Kei},
        title = "{Einstein rings modulated by wavelike dark matter from anomalies in gravitationally lensed images}",
      journal = {Nature Astronomy},
         year = 2023,
        month = jun,
       volume = {7},
        pages = {736-747},
          doi = {10.1038/s41550-023-01943-9},
archivePrefix = {arXiv},
       eprint = {2304.09895},
 primaryClass = {astro-ph.CO},
       adsurl = {https://ui.adsabs.harvard.edu/abs/2023NatAs...7..736A}
}

@ARTICLE{Xu2016_PLfromSim,
       author = {{Xu}, Dandan and {Sluse}, Dominique and {Schneider}, Peter and {Springel}, Volker and {Vogelsberger}, Mark and {Nelson}, Dylan and {Hernquist}, Lars},
        title = "{Lens galaxies in the Illustris simulation: power-law models and the bias of the Hubble constant from time delays}",
      journal = {\mnras},
         year = 2016,
        month = feb,
       volume = {456},
       number = {1},
        pages = {739-755},
          doi = {10.1093/mnras/stv2708},
archivePrefix = {arXiv},
       eprint = {1507.07937},
 primaryClass = {astro-ph.GA},
       adsurl = {https://ui.adsabs.harvard.edu/abs/2016MNRAS.456..739X}
}

@article{Xu2015,
        author = {{Xu}, Dandan and {Sluse}, Dominique and {Gao}, Liang and {Wang}, Jie and {Frenk}, Carlos and {Mao}, Shude and {Schneider}, Peter and {Springel}, Volker},
        title = "{How well can cold dark matter substructures account for the observed radio flux-ratio anomalies}",
      journal = {\mnras},
         year = 2015,
        month = mar,
       volume = {447},
       number = {4},
        pages = {3189-3206},
          doi = {10.1093/mnras/stu2673},
archivePrefix = {arXiv},
       eprint = {1410.3282},
 primaryClass = {astro-ph.CO},
       adsurl = {https://ui.adsabs.harvard.edu/abs/2015MNRAS.447.3189X}
}

@ARTICLE{Xu2012,
       author = {{Xu}, Dandan and {Mao}, Shude and {Cooper}, Andrew P. and {Gao}, Liang and {Frenk}, Carlos S. and {Angulo}, Raul E. and {Helly}, John},
        title = "{On the effects of line-of-sight structures on lensing flux-ratio anomalies in a {\ensuremath{\Lambda}}CDM universe}",
      journal = {\mnras},
         year = 2012,
        month = apr,
       volume = {421},
       number = {3},
        pages = {2553-2567},
          doi = {10.1111/j.1365-2966.2012.20484.x},
archivePrefix = {arXiv},
       eprint = {1110.1185},
 primaryClass = {astro-ph.CO},
       adsurl = {https://ui.adsabs.harvard.edu/abs/2012MNRAS.421.2553X}
}

@ARTICLE{Xu2010,
       author = {{Xu}, Dandan and {Mao}, Shude and {Cooper}, Andrew P. and {Wang}, Jie and {Gao}, Liang and {Frenk}, Carlos S. and {Springel}, V.},
        title = "{Substructure lensing: effects of galaxies, globular clusters and satellite streams}",
      journal = {\mnras},
         year = 2010,
        month = nov,
       volume = {408},
       number = {3},
        pages = {1721-1729},
          doi = {10.1111/j.1365-2966.2010.17235.x},
archivePrefix = {arXiv},
       eprint = {1004.3094},
 primaryClass = {astro-ph.CO},
       adsurl = {https://ui.adsabs.harvard.edu/abs/2010MNRAS.408.1721X}
}

@ARTICLE{Xu2009,
       author = {{Xu}, Dandan and {Mao}, Shude and {Wang}, Jie and {Springel}, V. and {Gao}, Liang and {White}, S.~D.~M. and {Frenk}, Carlos S. and {Jenkins}, Adrian and {Li}, Guoliang and {Navarro}, Julio F.},
        title = "{Effects of dark matter substructures on gravitational lensing: results from the Aquarius simulations}",
      journal = {\mnras},
         year = 2009,
        month = sep,
       volume = {398},
       number = {3},
        pages = {1235-1253},
          doi = {10.1111/j.1365-2966.2009.15230.x},
archivePrefix = {arXiv},
       eprint = {0903.4559},
 primaryClass = {astro-ph.CO},
       adsurl = {https://ui.adsabs.harvard.edu/abs/2009MNRAS.398.1235X}
}

@ARTICLE{Nightingale2024_Imaging_VPLs,
       author = {{Nightingale}, James W. and {He}, Qiuhan and {Cao}, Xiaoyue and {Amvrosiadis}, Aristeidis and {Etherington}, Amy and {Frenk}, Carlos S. and {Hayes}, Richard G. and {Robertson}, Andrew and {Cole}, Shaun and {Lange}, Samuel and {Li}, Ran and {Massey}, Richard},
        title = "{Scanning for dark matter subhaloes in Hubble Space Telescope imaging of 54 strong lenses}",
      journal = {\mnras},
         year = 2024,
        month = feb,
       volume = {527},
       number = {4},
        pages = {10480-10506},
          doi = {10.1093/mnras/stad3694},
archivePrefix = {arXiv},
       eprint = {2209.10566},
 primaryClass = {astro-ph.CO},
       adsurl = {https://ui.adsabs.harvard.edu/abs/2024MNRAS.52710480N}
}

@ARTICLE{SHARP-IV_Hsueh0712_2017,
       author = {{Hsueh}, J. -W. and {Oldham}, L. and {Spingola}, C. and {Vegetti}, S. and {Fassnacht}, C.~D. and {Auger}, M.~W. and {Koopmans}, L.~V.~E. and {McKean}, J.~P. and {Lagattuta}, D.~J.},
        title = "{SHARP - IV. An apparent flux-ratio anomaly resolved by the edge-on disc in B0712+472}",
      journal = {\mnras},
         year = 2017,
        month = aug,
       volume = {469},
       number = {3},
        pages = {3713-3721},
          doi = {10.1093/mnras/stx1082},
archivePrefix = {arXiv},
       eprint = {1701.06575},
 primaryClass = {astro-ph.GA},
       adsurl = {https://ui.adsabs.harvard.edu/abs/2017MNRAS.469.3713H}
}

@ARTICLE{Spingola2020_B0712_VLBI,
       author = {{Spingola}, Cristiana and {Barnacka}, Anna},
        title = "{Constraining VLBI-optical offsets in high redshift galaxies using strong gravitational lensing}",
      journal = {\mnras},
         year = 2020,
        month = may,
       volume = {494},
       number = {2},
        pages = {2312-2326},
          doi = {10.1093/mnras/staa870},
archivePrefix = {arXiv},
       eprint = {2003.11551},
 primaryClass = {astro-ph.CO},
       adsurl = {https://ui.adsabs.harvard.edu/abs/2020MNRAS.494.2312S}
}

@article{SHARP-II_Hsueh2016_B1555,
  author     = {Hsueh, J.-W. and Fassnacht, C. D. and Vegetti, S. and McKean, J. P. and Spingola, C. and others},
  title      = {{SHARP - II. M}ass structure in strong lenses is not necessarily dark matter substructure: A flux ratio anomaly from an edge-on disc},
  journal    = {MNRAS},
  year       = {2016},
  volume     = {463},
  number     = {1},
  pages      = {L51--L55},
}

@ARTICLE{Hsueh2018_IllustrisDisk,
       author = {{Hsueh}, Jen-Wei and {Despali}, Giulia and {Vegetti}, Simona and {Xu}, Dandan and {Fassnacht}, Christopher D. and {Metcalf}, R. Benton},
        title = "{Flux-ratio anomalies from discs and other baryonic structures in the Illustris simulation}",
      journal = {\mnras},
         year = 2018,
        month = apr,
       volume = {475},
       number = {2},
        pages = {2438-2451},
          doi = {10.1093/mnras/stx3320},
archivePrefix = {arXiv},
       eprint = {1707.07680},
 primaryClass = {astro-ph.GA},
       adsurl = {https://ui.adsabs.harvard.edu/abs/2018MNRAS.475.2438H}
}

@ARTICLE{Wen_2024_PBH_B1422,
       author = {{Wen}, Di and {Kemball}, Athol J.},
        title = "{Testing Primordial Black Hole Dark Matter with Atacama Large Millimeter Array Observations of the Gravitational Lens B1422+231}",
      journal = {Universe},
         year = 2024,
        month = jan,
       volume = {10},
       number = {1},
          eid = {37},
        pages = {37},
          doi = {10.3390/universe10010037},
       adsurl = {https://ui.adsabs.harvard.edu/abs/2024Univ...10...37W}
}

@ARTICLE{Nierenberg2014_B1422_FA,
       author = {{Nierenberg}, A.~M. and {Treu}, T. and {Wright}, S.~A. and {Fassnacht}, C.~D. and {Auger}, M.~W.},
        title = "{Detection of substructure with adaptive optics integral field spectroscopy of the gravitational lens B1422+231}",
      journal = {\mnras},
         year = 2014,
        month = aug,
       volume = {442},
       number = {3},
        pages = {2434-2445},
          doi = {10.1093/mnras/stu862},
archivePrefix = {arXiv},
       eprint = {1402.1496},
 primaryClass = {astro-ph.GA},
       adsurl = {https://ui.adsabs.harvard.edu/abs/2014MNRAS.442.2434N}
}

@ARTICLE{Bradac2002_B1422_FA,
       author = {{Brada{\v{c}}}, M. and {Schneider}, P. and {Steinmetz}, M. and {Lombardi}, M. and {King}, L.~J. and {Porcas}, R.},
        title = "{B1422+231: The influence of mass substructure on strong lensing}",
      journal = {\aap},
         year = 2002,
        month = jun,
       volume = {388},
        pages = {373-382},
          doi = {10.1051/0004-6361:20020559},
archivePrefix = {arXiv},
       eprint = {astro-ph/0112038},
 primaryClass = {astro-ph},
       adsurl = {https://ui.adsabs.harvard.edu/abs/2002A&A...388..373B}
}

@ARTICLE{Shan2026_BarFA,
       author = {{Shan}, Xikai and {Jin}, Yunpeng and {Mao}, Shude},
        title = "{Gravitational lensing by a spiral galaxy ─ I. The influence from bar's structure to the flux ratio anomaly}",
      journal = {\mnras},
         year = 2026,
        month = jun,
       volume = {549},
       number = {2},
          eid = {stag883},
        pages = {stag883},
          doi = {10.1093/mnras/stag883},
archivePrefix = {arXiv},
       eprint = {2510.02805},
 primaryClass = {astro-ph.CO},
       adsurl = {https://ui.adsabs.harvard.edu/abs/2026MNRAS.549ag883S}
}

@ARTICLE{Bradac2004_Sim,
       author = {{Brada{\v{c}}}, M. and {Schneider}, P. and {Lombardi}, M. and {Steinmetz}, M. and {Koopmans}, L.~V.~E. and {Navarro}, J.~F.},
        title = "{The signature of substructure on gravitational lensing in the {\ensuremath{\Lambda}}CDM cosmological model}",
      journal = {\aap},
         year = 2004,
        month = sep,
       volume = {423},
        pages = {797-809},
          doi = {10.1051/0004-6361:20040168},
archivePrefix = {arXiv},
       eprint = {astro-ph/0306238},
 primaryClass = {astro-ph},
       adsurl = {https://ui.adsabs.harvard.edu/abs/2004A&A...423..797B}
}

@ARTICLE{Chiba2005_B1422_FA,
       author = {{Chiba}, Masashi and {Minezaki}, Takeo and {Kashikawa}, Nobunari and {Kataza}, Hirokazu and {Inoue}, Kaiki Taro},
        title = "{Subaru Mid-Infrared Imaging of the Quadruple Lenses PG 1115+080 and B1422+231: Limits on Substructure Lensing}",
      journal = {\apj},
         year = 2005,
        month = jul,
       volume = {627},
       number = {1},
        pages = {53-61},
          doi = {10.1086/430403},
archivePrefix = {arXiv},
       eprint = {astro-ph/0503487},
 primaryClass = {astro-ph},
       adsurl = {https://ui.adsabs.harvard.edu/abs/2005ApJ...627...53C}
}

@article{Chiba2002,
  title = {Probing {{Dark Matter Substructure}} in {{Lens Galaxies}}},
  journal = {\apj},
  author = {Chiba, Masashi},
  year = 2002,
  journaltitle = {The Astrophysical Journal},
  volume = {565},
  pages = {17--23},
  publisher = {IOP},
  issn = {0004-637X},
  doi = {10.1086/324493},
  url = {https://ui.adsabs.harvard.edu/abs/2002ApJ...565...17C},
  urldate = {2025-12-08},
}

@ARTICLE{DoblerKeeton2006,
       author = {{Dobler}, Gregory and {Keeton}, Charles R.},
        title = "{Finite source effects in strong lensing: implications for the substructure mass scale}",
      journal = {\mnras},
         year = 2006,
        month = feb,
       volume = {365},
       number = {4},
        pages = {1243-1262},
          doi = {10.1111/j.1365-2966.2005.09809.x},
archivePrefix = {arXiv},
       eprint = {astro-ph/0502436},
 primaryClass = {astro-ph},
       adsurl = {https://ui.adsabs.harvard.edu/abs/2006MNRAS.365.1243D}
}

@ARTICLE{Impey_1996_B1422,
       author = {{Impey}, C.~D. and {Foltz}, C.~B. and {Petry}, C.~E. and {Browne}, I.~W.~A. and {Patnaik}, A.~R.},
        title = "{Hubble Space Telescope Observations of the Gravitational Lens System B1422+231}",
      journal = {\apjl},
         year = 1996,
        month = may,
       volume = {462},
        pages = {L53},
          doi = {10.1086/310035},
       adsurl = {https://ui.adsabs.harvard.edu/abs/1996ApJ...462L..53I}
}

@ARTICLE{Patnaik_1992_B1422,
       author = {{Patnaik}, A.~R. and {Browne}, I.~W.~A. and {Walsh}, D. and {Chaffee}, F.~H. and {Foltz}, C.~B.},
        title = "{B 1422+231 : a new gravitationally lensed system at Z = 3.62.}",
      journal = {\mnras},
         year = 1992,
        month = nov,
       volume = {259},
        pages = {1P-4},
          doi = {10.1093/mnras/259.1.1P},
       adsurl = {https://ui.adsabs.harvard.edu/abs/1992MNRAS.259P...1P}
}

@ARTICLE{Lawrence1992B1422,
       author = {{Lawrence}, C.~R. and {Neugebauer}, G. and {Weir}, N. and {Matthews}, K. and {Patnaik}, A.~R.},
        title = "{Infrared observations of the gravitational lens system B 1422+231.}",
      journal = {\mnras},
         year = 1992,
        month = nov,
       volume = {259},
        pages = {5P-7},
          doi = {10.1093/mnras/259.1.5P},
       adsurl = {https://ui.adsabs.harvard.edu/abs/1992MNRAS.259P...5L}
}

@ARTICLE{HB1994B1422,
       author = {{Hogg}, D.~W. and {Blandford}, R.~D.},
        title = "{The gravitational lens system B 1422+231 : dark matter, superluminal expansion and the Hubble constant.}",
      journal = {\mnras},
         year = 1994,
        month = jun,
       volume = {268},
        pages = {889-893},
          doi = {10.1093/mnras/268.4.889},
archivePrefix = {arXiv},
       eprint = {astro-ph/9311077},
 primaryClass = {astro-ph},
       adsurl = {https://ui.adsabs.harvard.edu/abs/1994MNRAS.268..889H}
}

@ARTICLE{KSB1994B1422,
       author = {{Kormann}, R. and {Schneider}, P. and {Bartelmann}, M.},
        title = "{A gravitational lens model for B1422+231.}",
      journal = {\aap},
         year = 1994,
        month = jun,
       volume = {286},
        pages = {357-364},
          doi = {10.48550/arXiv.astro-ph/9311011},
archivePrefix = {arXiv},
       eprint = {astro-ph/9311011},
 primaryClass = {astro-ph},
       adsurl = {https://ui.adsabs.harvard.edu/abs/1994A&A...286..357K}
}

@ARTICLE{WynWitt2003TruthDilusion,
       author = {{Evans}, N. Wyn and {Witt}, Hans J.},
        title = "{Fitting gravitational lenses: truth or delusion}",
      journal = {\mnras},
         year = 2003,
        month = nov,
       volume = {345},
       number = {4},
        pages = {1351-1364},
          doi = {10.1046/j.1365-2966.2003.07057.x},
archivePrefix = {arXiv},
       eprint = {astro-ph/0212013},
 primaryClass = {astro-ph},
       adsurl = {https://ui.adsabs.harvard.edu/abs/2003MNRAS.345.1351E}
}

@ARTICLE{McKean2007_B2045,
       author = {{McKean}, J.~P. and {Koopmans}, L.~V.~E. and {Flack}, C.~E. and {Fassnacht}, C.~D. and {Thompson}, D. and {Matthews}, K. and {Blandford}, R.~D. and {Readhead}, A.~C.~S. and {Soifer}, B.~T.},
        title = "{High-resolution imaging of the anomalous flux ratio gravitational lens system CLASS B2045+265: dark or luminous satellites?}",
      journal = {\mnras},
         year = 2007,
        month = jun,
       volume = {378},
       number = {1},
        pages = {109-118},
          doi = {10.1111/j.1365-2966.2007.11744.x},
archivePrefix = {arXiv},
       eprint = {astro-ph/0611215},
 primaryClass = {astro-ph},
       adsurl = {https://ui.adsabs.harvard.edu/abs/2007MNRAS.378..109M}
}

@ARTICLE{FadelyKeeton2012HE0435,
       author = {{Fadely}, Ross and {Keeton}, Charles R.},
        title = "{Substructure in the lens HE 0435-1223}",
      journal = {\mnras},
         year = 2012,
        month = jan,
       volume = {419},
       number = {2},
        pages = {936-951},
          doi = {10.1111/j.1365-2966.2011.19729.x},
archivePrefix = {arXiv},
       eprint = {1109.0548},
 primaryClass = {astro-ph.CO},
       adsurl = {https://ui.adsabs.harvard.edu/abs/2012MNRAS.419..936F}
}

@ARTICLE{Spingola2018,
       author = {{Spingola}, C. and {McKean}, J.~P. and {Auger}, M.~W. and {Fassnacht}, C.~D. and {Koopmans}, L.~V.~E. and {Lagattuta}, D.~J. and {Vegetti}, S.},
        title = "{SHARP - V. Modelling gravitationally lensed radio arcs imaged with global VLBI observations}",
      journal = {\mnras},
         year = 2018,
        month = aug,
       volume = {478},
       number = {4},
        pages = {4816-4829},
          doi = {10.1093/mnras/sty1326},
archivePrefix = {arXiv},
       eprint = {1807.05566},
 primaryClass = {astro-ph.GA},
       adsurl = {https://ui.adsabs.harvard.edu/abs/2018MNRAS.478.4816S}
}

@ARTICLE{Spingola2019VLBA,
       author = {{Spingola}, C. and {McKean}, J.~P. and {Lee}, M. and {Deller}, A. and {Moldon}, J.},
        title = "{A novel search for gravitationally lensed radio sources in wide-field VLBI imaging from the mJIVE-20 survey}",
      journal = {\mnras},
         year = 2019,
        month = feb,
       volume = {483},
       number = {2},
        pages = {2125-2153},
          doi = {10.1093/mnras/sty3189},
archivePrefix = {arXiv},
       eprint = {1811.09152},
 primaryClass = {astro-ph.GA},
       adsurl = {https://ui.adsabs.harvard.edu/abs/2019MNRAS.483.2125S}
}

@article{Mao1998,
  author     = {Mao, S. and Schneider, P.},
  title      = {Evidence for substructure in lens galaxies?},
  journal    = {MNRAS},
  year       = {1998},
  volume     = {295},
  number     = {3},
  pages      = {587--594},
}

@ARTICLE{Planck_Collaboration_2016,
       author = {{Planck Collaboration} and others},
        title = "{Planck 2015 results. XIII. Cosmological parameters}",
      journal = {\aap},
         year = 2016,
        month = sep,
       volume = {594},
          eid = {A13},
        pages = {A13},
          doi = {10.1051/0004-6361/201525830},
archivePrefix = {arXiv},
       eprint = {1502.01589},
 primaryClass = {astro-ph.CO},
       adsurl = {https://ui.adsabs.harvard.edu/abs/2016A&A...594A..13P}
}

@ARTICLE{Sonnenfeld2013a,
       author = {{Sonnenfeld}, Alessandro and {Gavazzi}, Rapha{\"e}l and {Suyu}, Sherry H. and {Treu}, Tommaso and {Marshall}, Philip J.},
        title = "{The SL2S Galaxy-scale Lens Sample. III. Lens Models, Surface Photometry, and Stellar Masses for the Final Sample}",
      journal = {\apj},
         year = 2013,
        month = nov,
       volume = {777},
       number = {2},
          eid = {97},
        pages = {97},
          doi = {10.1088/0004-637X/777/2/97},
archivePrefix = {arXiv},
       eprint = {1307.4764},
 primaryClass = {astro-ph.CO},
       adsurl = {https://ui.adsabs.harvard.edu/abs/2013ApJ...777...97S}
}

@ARTICLE{Sonnenfeld2013b,
       author = {{Sonnenfeld}, Alessandro and {Treu}, Tommaso and {Gavazzi}, Rapha{\"e}l and {Suyu}, Sherry H. and {Marshall}, Philip J. and {Auger}, Matthew W. and {Nipoti}, Carlo},
        title = "{The SL2S Galaxy-scale Lens Sample. IV. The Dependence of the Total Mass Density Profile of Early-type Galaxies on Redshift, Stellar Mass, and Size}",
      journal = {\apj},
         year = 2013,
        month = nov,
       volume = {777},
       number = {2},
          eid = {98},
        pages = {98},
          doi = {10.1088/0004-637X/777/2/98},
archivePrefix = {arXiv},
       eprint = {1307.4759},
 primaryClass = {astro-ph.CO},
       adsurl = {https://ui.adsabs.harvard.edu/abs/2013ApJ...777...98S}
}

@ARTICLE{patnaik1999,
       author = {{Patnaik}, A.~R. and {Kemball}, A.~J. and {Porcas}, R.~W. and {Garrett}, M.~A.},
        title = "{Milliarcsec-scale polarization observations of the gravitational lens B1422+231}",
      journal = {\mnras},
         year = 1999,
        month = jul,
       volume = {307},
       number = {1},
        pages = {L1-L5},
          doi = {10.1046/j.1365-8711.1999.02813.x},
archivePrefix = {arXiv},
       eprint = {astro-ph/9905311},
 primaryClass = {astro-ph},
       adsurl = {https://ui.adsabs.harvard.edu/abs/1999MNRAS.307L...1P}
}

@ARTICLE{2005MNRAS.362...41G,
       author = {{Gallazzi}, Anna and {Charlot}, St{\'e}phane and {Brinchmann}, Jarle and {White}, Simon D.~M. and {Tremonti}, Christy A.},
        title = "{The ages and metallicities of galaxies in the local universe}",
      journal = {\mnras},
         year = 2005,
        month = sep,
       volume = {362},
       number = {1},
        pages = {41-58},
          doi = {10.1111/j.1365-2966.2005.09321.x},
archivePrefix = {arXiv},
       eprint = {astro-ph/0506539},
 primaryClass = {astro-ph},
       adsurl = {https://ui.adsabs.harvard.edu/abs/2005MNRAS.362...41G}
}

@ARTICLE{oh2026,
       author = {{Oh}, Maverick S.~H. and {Nierenberg}, Anna and {Gilman}, Daniel and {Birrer}, Simon},
        title = "{Joint semi-analytic multipole priors from galaxy isophotes and constraints from lensed arcs}",
      journal = {\jcap},
         year = 2026,
        month = mar,
       volume = {2026},
       number = {3},
          eid = {039},
        pages = {039},
          doi = {10.1088/1475-7516/2026/03/039},
archivePrefix = {arXiv},
       eprint = {2404.17124},
 primaryClass = {astro-ph.CO},
       adsurl = {https://ui.adsabs.harvard.edu/abs/2026JCAP...03..039O}
}

@ARTICLE{Gilman2025_JWST_FAandArc,
       author = {{Gilman}, D. and {Nierenberg}, A.~M. and {Treu}, T. and {Gannon}, C. and {Du}, X. and {Paugnat}, H. and {Birrer}, S. and {Benson}, A.~J. and {Mozumdar}, P. and {Wong}, K.~C. and {Williams}, D. and {Keeley}, R.~E. and {Abazajian}, K.~N. and {Anguita}, T. and {Bennert}, V.~N. and {Djorgovski}, S.~G. and {Hoenig}, S.~H. and {Kusenko}, A. and {Malkan}, M. and {Morishita}, T. and {Motta}, V. and {Moustakas}, L.~A. and {Sheu}, W. and {Sluse}, D. and {Stern}, D. and {Stiavelli}, M.},
        title = "{JWST lensed quasar dark matter survey IV: Stringent warm dark matter constraints from the joint reconstruction of extended lensed arcs and quasar flux ratios}",
      journal = {arXiv e-prints},
         year = 2025,
        month = nov,
          eid = {arXiv:2511.07513},
        pages = {arXiv:2511.07513},
          doi = {10.48550/arXiv.2511.07513},
archivePrefix = {arXiv},
       eprint = {2511.07513},
 primaryClass = {astro-ph.CO},
       adsurl = {https://ui.adsabs.harvard.edu/abs/2025arXiv251107513G}
}

@ARTICLE{PaugnatGilman25_EM,
       author = {{Paugnat}, Hadrien and {Gilman}, Daniel},
        title = "{Elliptical multipoles for gravitational lenses}",
      journal = {\prd},
         year = 2025,
        month = jun,
       volume = {111},
       number = {12},
          eid = {123014},
        pages = {123014},
          doi = {10.1103/d14h-f5mn},
archivePrefix = {arXiv},
       eprint = {2502.03530},
 primaryClass = {astro-ph.CO},
       adsurl = {https://ui.adsabs.harvard.edu/abs/2025PhRvD.111l3014P}
}

@ARTICLE{keeton2001,
       author = {{Keeton}, Charles R.},
        title = "{Computational Methods for Gravitational Lensing}",
      journal = {arXiv e-prints},
         year = 2001,
        month = feb,
          eid = {astro-ph/0102340},
        pages = {astro-ph/0102340},
          doi = {10.48550/arXiv.astro-ph/0102340},
archivePrefix = {arXiv},
       eprint = {astro-ph/0102340},
 primaryClass = {astro-ph},
       adsurl = {https://ui.adsabs.harvard.edu/abs/2001astro.ph..2340K}
}

@ARTICLE{1992ARA&A..30..543M,
       author = {{Monaghan}, J.~J.},
        title = "{Smoothed particle hydrodynamics.}",
      journal = {\araa},
         year = 1992,
        month = jan,
       volume = {30},
        pages = {543-574},
          doi = {10.1146/annurev.aa.30.090192.002551},
       adsurl = {https://ui.adsabs.harvard.edu/abs/1992ARA&A..30..543M}
}

@ARTICLE{2010ApJ...724..511A,
       author = {{Auger}, M.~W. and {Treu}, T. and {Bolton}, A.~S. and {Gavazzi}, R. and {Koopmans}, L.~V.~E. and {Marshall}, P.~J. and {Moustakas}, L.~A. and {Burles}, S.},
        title = "{The Sloan Lens ACS Survey. X. Stellar, Dynamical, and Total Mass Correlations of Massive Early-type Galaxies}",
      journal = {\apj},
         year = 2010,
        month = nov,
       volume = {724},
       number = {1},
        pages = {511-525},
          doi = {10.1088/0004-637X/724/1/511},
archivePrefix = {arXiv},
       eprint = {1007.2880},
 primaryClass = {astro-ph.CO},
       adsurl = {https://ui.adsabs.harvard.edu/abs/2010ApJ...724..511A}
}

@ARTICLE{2012ApJ...744...41B,
       author = {{Brownstein}, Joel R. and {Bolton}, Adam S. and {Schlegel}, David J. and {Eisenstein}, Daniel J. and {Kochanek}, Christopher S. and {Connolly}, Natalia and {Maraston}, Claudia and {Pandey}, Parul and {Seitz}, Stella and {Wake}, David A. and {Wood-Vasey}, W. Michael and {Brinkmann}, Jon and {Schneider}, Donald P. and {Weaver}, Benjamin A.},
        title = "{The BOSS Emission-Line Lens Survey (BELLS). I. A Large Spectroscopically Selected Sample of Lens Galaxies at Redshift \raisebox{-0.5ex}\textasciitilde0.5}",
      journal = {\apj},
         year = 2012,
        month = jan,
       volume = {744},
       number = {1},
          eid = {41},
        pages = {41},
          doi = {10.1088/0004-637X/744/1/41},
archivePrefix = {arXiv},
       eprint = {1112.3683},
 primaryClass = {astro-ph.CO},
       adsurl = {https://ui.adsabs.harvard.edu/abs/2012ApJ...744...41B}
}

@ARTICLE{2004ApJ...611..739T,
       author = {{Treu}, Tommaso and {Koopmans}, L{\'e}on V.~E.},
        title = "{Massive Dark Matter Halos and Evolution of Early-Type Galaxies to z \raisebox{-0.5ex}\textasciitilde 1}",
      journal = {\apj},
         year = 2004,
        month = aug,
       volume = {611},
       number = {2},
        pages = {739-760},
          doi = {10.1086/422245},
archivePrefix = {arXiv},
       eprint = {astro-ph/0401373},
 primaryClass = {astro-ph},
       adsurl = {https://ui.adsabs.harvard.edu/abs/2004ApJ...611..739T}
}

@ARTICLE{bc03,
       author = {{Bruzual}, G. and {Charlot}, S.},
        title = "{Stellar population synthesis at the resolution of 2003}",
      journal = {\mnras},
         year = 2003,
        month = oct,
       volume = {344},
       number = {4},
        pages = {1000-1028},
          doi = {10.1046/j.1365-8711.2003.06897.x},
archivePrefix = {arXiv},
       eprint = {astro-ph/0309134},
 primaryClass = {astro-ph},
       adsurl = {https://ui.adsabs.harvard.edu/abs/2003MNRAS.344.1000B}
}

@ARTICLE{Cappellari06ApertureCorr,
       author = {{Cappellari}, Michele and {Bacon}, R. and {Bureau}, M. and {Damen}, M.~C. and {Davies}, Roger L. and {de Zeeuw}, P.~T. and {Emsellem}, Eric and {Falc{\'o}n-Barroso}, Jes{\'u}s and {Krajnovi{\'c}}, Davor and {Kuntschner}, Harald and {McDermid}, Richard M. and {Peletier}, Reynier F. and {Sarzi}, Marc and {van den Bosch}, Remco C.~E. and {van de Ven}, Glenn},
        title = "{The SAURON project - IV. The mass-to-light ratio, the virial mass estimator and the Fundamental Plane of elliptical and lenticular galaxies}",
      journal = {\mnras},
         year = 2006,
        month = mar,
       volume = {366},
       number = {4},
        pages = {1126-1150},
          doi = {10.1111/j.1365-2966.2005.09981.x},
archivePrefix = {arXiv},
       eprint = {astro-ph/0505042},
 primaryClass = {astro-ph},
       adsurl = {https://ui.adsabs.harvard.edu/abs/2006MNRAS.366.1126C}
}

@software{feng_2026_22537345,
  author       = {Feng, Ruizhe and
                  Xu, Dandan and
                  Sluse, Dominique and
                  Despali, Giulia and
                  Shajib, Anowar Jaman and
                  Cai-Na, Hao},
  title        = {Can third- and fourth-order multipoles plus radial
                   variation of iso-density ellipses explain the
                   observed flux ratios in B1422+231? YES, and a
                   lesson learned from a TNG100 lensing galaxy sample
                  },
  month        = sep,
  year         = 2026,
  publisher    = {Zenodo},
  version      = {v1.0.0},
  doi          = {10.5281/zenodo.22537345},
  url          = {https://doi.org/10.5281/zenodo.22537345},
}

@article{astropy:2013,
Adsurl = {http://adsabs.harvard.edu/abs/2013A%26A...558A..33A},
Archiveprefix = {arXiv},
Author = {{Astropy Collaboration} and others},
Doi = {10.1051/0004-6361/201322068},
Eid = {A33},
Eprint = {1307.6212},
Journal = {\aap},
Month = oct,
Pages = {A33},
Primaryclass = {astro-ph.IM},
Title = {{Astropy: A community Python package for astronomy}},
Volume = 558,
Year = 2013}

@ARTICLE{astropy:2018,
       author = {{Astropy Collaboration} and others},
        title = "{The Astropy Project: Building an Open-science Project and Status of the v2.0 Core Package}",
      journal = {\aj},
         year = 2018,
        month = sep,
       volume = {156},
       number = {3},
          eid = {123},
        pages = {123},
          doi = {10.3847/1538-3881/aabc4f},
archivePrefix = {arXiv},
       eprint = {1801.02634},
 primaryClass = {astro-ph.IM},
       adsurl = {https://ui.adsabs.harvard.edu/abs/2018AJ....156..123A}
}

@ARTICLE{astropy:2022,
       author = {{Astropy Collaboration} and others},
        title = "{The Astropy Project: Sustaining and Growing a Community-oriented Open-source Project and the Latest Major Release (v5.0) of the Core Package}",
      journal = {\apj},
         year = 2022,
        month = aug,
       volume = {935},
       number = {2},
          eid = {167},
        pages = {167},
          doi = {10.3847/1538-4357/ac7c74},
archivePrefix = {arXiv},
       eprint = {2206.14220},
 primaryClass = {astro-ph.IM},
       adsurl = {https://ui.adsabs.harvard.edu/abs/2022ApJ...935..167A}
}

@ARTICLE{2020SciPy-NMeth,
  author  = {Virtanen, Pauli and Gommers, Ralf and Oliphant, Travis E. and
            Haberland, Matt and Reddy, Tyler and Cournapeau, David and
            Burovski, Evgeni and Peterson, Pearu and Weckesser, Warren and
            Bright, Jonathan and {van der Walt}, St{\'e}fan J. and
            Brett, Matthew and Wilson, Joshua and Millman, K. Jarrod and
            Mayorov, Nikolay and Nelson, Andrew R. J. and Jones, Eric and
            Kern, Robert and Larson, Eric and Carey, C J and
            Polat, {\.I}lhan and Feng, Yu and Moore, Eric W. and
            {VanderPlas}, Jake and Laxalde, Denis and Perktold, Josef and
            Cimrman, Robert and Henriksen, Ian and Quintero, E. A. and
            Harris, Charles R. and Archibald, Anne M. and
            Ribeiro, Ant{\^o}nio H. and Pedregosa, Fabian and
            {van Mulbregt}, Paul and {SciPy 1.0 Contributors}},
  title   = {{{SciPy} 1.0: Fundamental Algorithms for Scientific
            Computing in Python}},
  journal = {Nature Methods},
  year    = {2020},
  volume  = {17},
  pages   = {261--272},
  adsurl  = {https://rdcu.be/b08Wh},
  doi     = {10.1038/s41592-019-0686-2},
}

@Article{         harris2020array,
 title         = {Array programming with {NumPy}},
 author        = {Charles R. Harris and K. Jarrod Millman and St{\'{e}}fan J.
                 van der Walt and Ralf Gommers and Pauli Virtanen and David
                 Cournapeau and Eric Wieser and Julian Taylor and Sebastian
                 Berg and Nathaniel J. Smith and Robert Kern and Matti Picus
                 and Stephan Hoyer and Marten H. van Kerkwijk and Matthew
                 Brett and Allan Haldane and Jaime Fern{\'{a}}ndez del
                 R{\'{i}}o and Mark Wiebe and Pearu Peterson and Pierre
                 G{\'{e}}rard-Marchant and Kevin Sheppard and Tyler Reddy and
                 Warren Weckesser and Hameer Abbasi and Christoph Gohlke and
                 Travis E. Oliphant},
 year          = {2020},
 month         = sep,
 journal       = {\nat},
 volume        = {585},
 number        = {7825},
 pages         = {357--362},
 doi           = {10.1038/s41586-020-2649-2},
 publisher     = {Springer Science and Business Media {LLC}},
 url           = {https://doi.org/10.1038/s41586-020-2649-2}
}

@Article{Hunter:2007,
  Author    = {Hunter, J. D.},
  Title     = {Matplotlib: A 2D graphics environment},
  Journal   = {Computing in Science \& Engineering},
  Volume    = {9},
  Number    = {3},
  Pages     = {90--95},
  publisher = {IEEE COMPUTER SOC},
  doi       = {10.1109/MCSE.2007.55},
  year      = 2007
}

@inproceedings{numbapaper,
author = {Lam, Siu Kwan and Pitrou, Antoine and Seibert, Stanley},
title = {Numba: a LLVM-based Python JIT compiler},
year = {2015},
isbn = {9781450340052},
publisher = {Association for Computing Machinery},
address = {New York, NY, USA},
url = {https://doi.org/10.1145/2833157.2833162},
doi = {10.1145/2833157.2833162},
booktitle = {Proceedings of the Second Workshop on the LLVM Compiler Infrastructure in HPC},
articleno = {7},
numpages = {6},
location = {Austin, Texas},
series = {LLVM '15}
}

@ARTICLE{lenstronomy01,
       author = {{Birrer}, Simon and {Amara}, Adam},
        title = "{lenstronomy: Multi-purpose gravitational lens modelling software package}",
      journal = {Physics of the Dark Universe},
         year = 2018,
        month = dec,
       volume = {22},
        pages = {189-201},
          doi = {10.1016/j.dark.2018.11.002},
archivePrefix = {arXiv},
       eprint = {1803.09746},
 primaryClass = {astro-ph.CO},
       adsurl = {https://ui.adsabs.harvard.edu/abs/2018PDU....22..189B}
}

@book{python,
 author = {Van Rossum, Guido and Drake, Fred L.},
 title = {Python 3 Reference Manual},
 year = {2009},
 isbn = {1441412697},
 publisher = {CreateSpace},
 address = {Scotts Valley, CA}
}
\bibliographystyle{aasjournalv7}



\end{document}